\documentclass[12pt]{iopart}

\usepackage{amssymb}
\usepackage{theorem}
\usepackage{graphicx}
\usepackage{xcolor}
\usepackage{cite}

\usepackage{url} 

\newcommand{\ddx}[1]{\frac{\partial #1}{\partial x}}
\newcommand{\ddt}[1]{\frac{\partial #1}{\partial t}}
\newcommand{\ddz}[1]{\frac{\partial #1}{\partial z}}
\newcommand{\ddzz}[1]{\frac{\partial^2 #1}{\partial z^2}}

\newcommand{\R}{\mathbb{R}}

\renewcommand{\vec}[1]{\mathbf{#1}}

\newcommand{\Poincare}{Poincar\'{e} }

\newcommand{\HAin}{H_{\mathrm{in}}^{A}}
\newcommand{\HAout}{H_{\mathrm{out}}^{A}}
\newcommand{\HBin}{H_{\mathrm{in}}^{B}}

\newcommand{\xa}{\bi{x}^A}
\newcommand{\xb}{\bi{x}^B}
\newcommand{\xia}{\xi_A}
\newcommand{\xib}{\xi_B}
\newcommand{\xic}{\xi_C}
\newcommand{\xih}{\xi_H}

\newcommand{\xea}{x_e^A}
\newcommand{\yea}{y_e^A}
\newcommand{\xca}{x_c^A}
\newcommand{\yca}{y_c^A}
\newcommand{\rea}{r_e^A}
\newcommand{\thea}{\theta_e^A}

\newcommand{\xeb}{x_e^B}
\newcommand{\yeb}{y_e^B}
\newcommand{\xcb}{x_c^B}
\newcommand{\ycb}{y_c^B}

\newcommand{\pil}{\Pi_{\mathrm{loc}}}
\newcommand{\pig}{\Pi_{\mathrm{glo}}}

\newcommand{\gab}{\gamma_{AB}}

\newcommand{\hxa}{\hat{\bi{x}}^A}
\newcommand{\hxb}{\hat{\bi{x}}^B}

\newcommand{\hxea}{\hat{x}_e^A}
\newcommand{\hyea}{\hat{y}_e^A}
\newcommand{\hthea}{\hat{\theta}_e^A}

\newtheorem{defn}{Definition}

\newcommand{\vectortwo}[2]{\left( \begin{array}{cc} #1 \\ #2 \end{array}\right)}
\newcommand{\matrixtwo}[4]{\left( \begin{array}{cc} #1 & #2 \\ #3 & #4 \end{array}\right)}

\newcommand{\blue}[1]{{{#1}}}

\newcommand{\purple}[1]{{\color{black}{#1}}}

\begin{document}

\title[Heteroclinic bifurcations in Rock--Paper--Scissors]{A trio of heteroclinic bifurcations arising from a model of spatially-extended Rock--Paper--Scissors}%

\author{Claire M Postlethwaite\dag\footnote[3]{Corresponding author (c.postlethwaite@auckland.ac.nz)} and Alastair M Rucklidge\ddag}

\address{\dag Department of Mathematics, University of Auckland, Private Bag 92019, Auckland 1142, New Zealand}
\address{\ddag School of Mathematics, University of Leeds, Leeds LS2 9JT, UK}



\begin{abstract}
One of the simplest examples of a robust heteroclinic cycle involves three
saddle equilibria: each one is unstable to the next in turn, and connections
from one to the next occur within invariant subspaces. Such a situation can be
described by a third-order ordinary differential equation (ODE), and typical
trajectories approach each equilibrium point in turn, spending progressively
longer to cycle around the three points but never stopping. This cycle has been
invoked as a model of cyclic competition between populations adopting three
strategies, characterised as Rock, Paper and Scissors. When spatial
distribution and mobility of the populations is taken into account, waves of
Rock can invade regions of Scissors, only to be invaded by Paper in turn. The
dynamics is described by a set of partial differential equations (PDEs) that
has travelling wave (in one dimension) and spiral (in two dimensions)
solutions. In this paper, we explore how the robust heteroclinic cycle in the
ODE manifests itself in the PDEs. Taking the wavespeed as a parameter, and
moving into a travelling frame, the PDEs reduce to a sixth-order set of ODEs,
in which travelling waves are created in a Hopf bifurcation and are destroyed
in three different heteroclinic bifurcations, depending on parameters, as the
travelling wave approaches the heteroclinic cycle. 
We explore the three different heteroclinic bifurcations, none of
which have been observed in the context of robust heteroclinic cycles
previously. These results are an important step towards a full
understanding of the spiral patterns found in two dimensions, with possible 
application to travelling waves and spirals in other population dynamics 
models.
\end{abstract}


\noindent{Nonlinearity {\bf 32} (2019) 1375--1407. \url{https://doi.org/10.1088/1361-6544/aaf530}}


\section{Introduction}

The Rock--Paper--Scissors game, in which Rock blunts Scissors, Scissors cut
Paper, and Paper wraps Rock, provides an appealing simple model of cyclic
competition between different strategies or species in evolutionary game theory
and biology~\cite{May1975,Szolnoki2014}. The game has been invoked as a
description of three competing species of {\em E.~coli}~\cite{Kerr2002} and of
three colour-variants of side-blotched lizards~\cite{Sinervo1996}, but the idea
of cyclic competition has arisen also in rotating convection~\cite{Busse1980a}
and as the simplest example of a heteroclinic cycle~\cite{Guckenheimer1988}.

Imagine a group of people repeatedly playing Rock--Paper--Scissors, with each
person favouring one of the three choices, and let $A(t)$, $B(t)$ and $C(t)$ be
the number of people playing Rock, Paper or Scissors at any
moment of time~$t$. Pairs of people are drawn at random and when they play, either
it is a tie (if they are drawn from the same group), or one beats the other. In this
case, the loser can either adopt the strategy of the winner
(dominance--replacement) or the loser can withdraw from the game
(dominance--removal). Once removed, players are replaced (up to a maximum
number~$N$) and are assigned to Rock, Paper or Scissors with probabilities
proportional to the number of Rock, Paper or Scissors players. With these
dynamics, if all individuals who are playing (for example) Rock are eliminated
(through a random fluctuation when the number of Rock players is small), they
can never return, which means that Scissors would have no competitors and would
eventually wipe out Paper~\cite{Kerr2002}. This process
is known as {\em fixation}, and since it involves an absorbing state, is 
guaranteed (in a discrete stochastic model) to happen eventually~\cite{Kimura1969}.

In the limit of large~$N$, the discrete process becomes continuous and is modelled
by three ordinary differential equations
(ODEs)~\cite{May1975,Frey2010,Szczesny2014}:
 \begin{eqnarray}
 {\dot a} &= a (1 - (a+b+c) - (\sigma+\zeta)b + \zeta c), \nonumber\\
 {\dot b} &= b (1 - (a+b+c) - (\sigma+\zeta)c + \zeta a), 
 \label{eq:RPS_ODEs} \\
 {\dot c} &= c (1 - (a+b+c) - (\sigma+\zeta)a + \zeta b), \nonumber
 \end{eqnarray} 
where $a(t)=A/N$, $b(t)=B/N$, $c(t)=C/N$, and $\sigma$ and $\zeta$ are 
non-negative parameters
that control the rates of dominance--removal and dominance--replacement
respectively, scaled to the rate of replacement. We have assumed symmetry between
Rock, Paper and Scissors. $A$, $B$ and $C$ are numbers of individuals, so $a$, 
$b$ and $c$ are non-negative.

The ODEs~\eref{eq:RPS_ODEs} have five equilibria with
non-negative components: the trivial solution $(a,b,c)=(0,0,0)$, three on-axis
equilibria $(1,0,0)$, $(0,1,0)$ and $(0,0,1)$, and a coexistence point with
$(a,b,c)=\frac{1}{3+\sigma}(1,1,1)$. For $\sigma>0$ and $\zeta>0$, this system of
ODEs has solutions that approach each of three on-axis equilibria in turn, taking
progressively longer to cycle around the three points but never
stopping~\cite{May1975} (in contrast to eventual fixation in the discrete case).

This gradual slowing down of trajectories as they spend longer and longer near a
sequence of equilibria is a characteristic of asymptotically stable heteroclinic
cycles. The rate of slowing down is controlled by the ratio of two of the
eigenvalues of the on-axis equilibria: these are $\zeta$ and $-(\sigma+\zeta)$,
and the amount of time taken for each cycle is a factor of
$\frac{\sigma+\zeta}{\zeta}$ longer than the previous
one~\cite{Hofbauer1998}. In this expression it
is apparent that allowing either $\zeta=0$ or $\sigma=0$ requires special
attention. The situation where the eigenvalue ratio is equal to~$1$ ($\sigma=0$, 
$\zeta>0$) is normally 
called a {\em resonance bifurcation} from the heteroclinic cycle, associated with 
the creation of a long-period periodic 
orbit~\cite{Scheel1992,Postlethwaite2010}. However, in the ODEs~\eref{eq:RPS_ODEs}, 
letting $\sigma=0$ is degenerate, in that the coexistence equilibrium has 
pure imaginary eigenvalues and the ODEs have an invariant plane $a+b+c=1$ on which 
there is a continuous family of nested periodic orbits parameterised by 
$abc={}$constant.

In the last decade there has been considerable interest in the dynamics of the
Rock--Paper--Scissors game where the players are distributed in space and allowed 
to move, for example on a two-dimensional square lattice, interacting only with
their neighbours. In this case, in the limit of large~$N$ and small lattice
spacing, the dynamics is described by the partial differential equations
(PDEs)~\cite{Frey2010,Szczesny2014}:
 \begin{eqnarray}
 {\dot a} &= a (1 - (a+b+c) - (\sigma+\zeta)b + \zeta c) + \nabla^2 a, \nonumber\\
 {\dot b} &= b (1 - (a+b+c) - (\sigma+\zeta)c + \zeta a) + \nabla^2 b, 
 \label{eq:RPS_PDEs} \\
 {\dot c} &= c (1 - (a+b+c) - (\sigma+\zeta)a + \zeta b) + \nabla^2 c, \nonumber
 \end{eqnarray} 
where the spatial coordinates $(x,y)$ are scaled so that the diffusion constants 
(assumed to be equal) are equal to~$1$. 
Typically the PDEs are solved with periodic boundary conditions.
The spatial mobility allows 
for persistent spiral-like or turbulent patterns of Rock, Paper and 
Scissors~\cite{Reichenbach2008}, in which regions dominated by Rock invade regions 
of Scissors, which invade regions of Paper, which in turn invade regions of Rock.
In the case of spirals, these have a rotating core, with a point where $a=b=c$
at (or close to) the centre, and spiral arms that, far from the core, look like
they are one-dimensional periodic travelling wave (TW) solutions of the
PDEs~\eref{eq:RPS_PDEs}~\cite{Postlethwaite2017}.

The central question we address in this paper is: what is the connection between 
travelling waves in the PDEs~\eref{eq:RPS_PDEs} and heteroclinic cycles?
 The TWs are periodic orbits in a moving frame of 
reference, and, taking the wavespeed as a parameter, these periodic orbits 
originate in a Hopf bifurcation and end when they collide with a heteroclinic 
cycle~\cite{Postlethwaite2017}. In this paper we find conditions under which TWs with 
arbitrarily long wavelength can exist as solutions of~\eref{eq:RPS_PDEs}, close 
to a heteroclinic cycle in the sixth-order ODEs that describe the dynamics in the
travelling frame. We find 
that there are three different ways in which this can happen:
 \begin{itemize}
 \item there can be 
a {\em resonance bifurcation} from the heteroclinic cycle in the sixth-order ODEs,
at which a positive and a negative eigenvalue have equal magnitude; or
 \item there can be a bifurcation of {\em Belyakov--Devaney} type, at which the 
imaginary part of a 
pair of complex eigenvalues vanishes; or
 \item there can be a bifurcation of {\em orbit flip} type, at which there is a 
change in the way in which the trajectories between equilibria are oriented.
 \end{itemize}
Although our analysis proceeds along reasonably standard lines, there are
several unusual aspects, and the calculations are challenging, not least because the unstable manifolds of the equilibria in the heteroclinic cycles are of high dimension.
It turns out that each of these three bifurcations
is non-standard and, to our knowledge, has not been observed in the context of
heteroclinic cycles before. We are able to find conditions under which each of
these three bifurcations occurs, and, to some extent, how the transition from
one type to the next occurs. Our results give a much clearer picture of the
origin of the one-dimensional TW solutions of the PDEs~\eref{eq:RPS_PDEs}, a
first and necessary step in understanding their stability, which in turn is
necessary for understanding the stability of the two-dimensional spiral
solutions of~\eref{eq:RPS_PDEs}.

The remainder of this paper is structured as follows. \blue{We begin in section~\ref{sec:sims} by reviewing some
numerical results from~\cite{Postlethwaite2017} and showing some simulations of the PDEs~\eref{eq:RPS_PDEs}, both in one and two spatial dimensions. We also relate properties of the travelling wave solutions of the PDEs to periodic solutions of a related set of ODEs}. In section~\ref{sec:het}
we review the definitions of heteroclinic cycles and summarise what is already
known about ways in which they can bifurcate. We also compare these
bifurcations with those seen near homoclinic orbits, and relate these to the
new bifurcations we have found. In section~\ref{sec:setup} we describe the derivation of the ODEs we will be studying for the remainder of
the paper. Then in section~\ref{sec:Pmap} we derive a \Poincare map which
describes the flow close to the heteroclinic cycle in the ODEs. This section
contains a lot of calculation but the results are summarised at the start and
end of the section for the reader who doesn't wish to delve into too many of
the gritty details. \blue{In section~\ref{sec:moresims} we give some further
numerical results from simulation of the PDEs for a range of parameter values, and finally in section~\ref{sec:smallsig} we look at numerically computed bifurcation diagrams as the parameter $\sigma$ is varied and discuss the limit $\sigma\rightarrow 0$. }Section~\ref{sec:disc}
concludes.

\section{\blue{PDE simulations}}
\label{sec:sims}

\begin{figure}
\setlength{\unitlength}{1mm}

\begin{center}
\begin{picture}(140,40)(0,0)
\put(6,0){\includegraphics[trim= 0cm 0cm 0cm 0cm,clip=true,width=40mm]{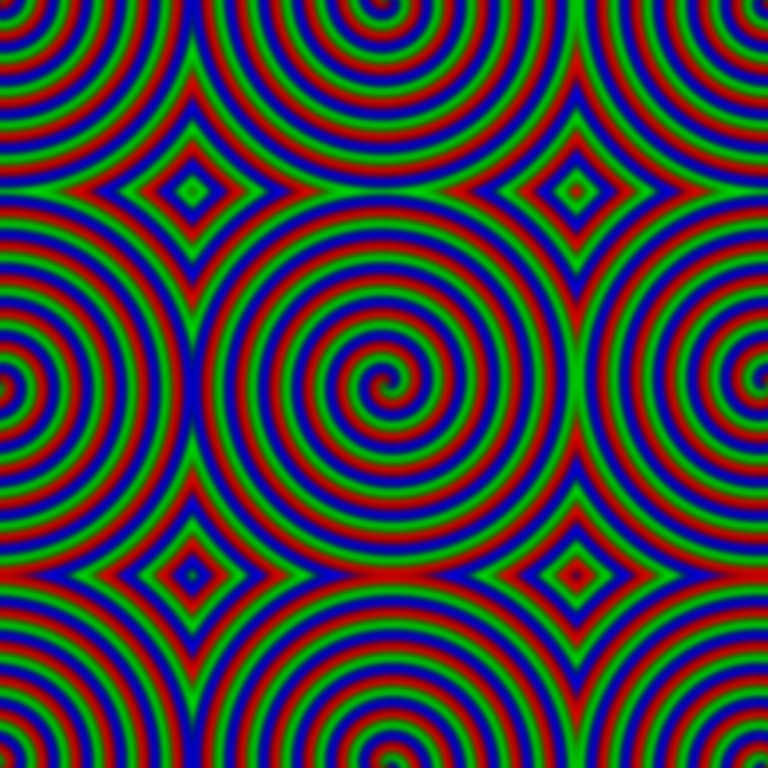}}
\put(53,0){\includegraphics[trim= 0cm 0cm 0cm 0cm,clip=true,width=40mm]{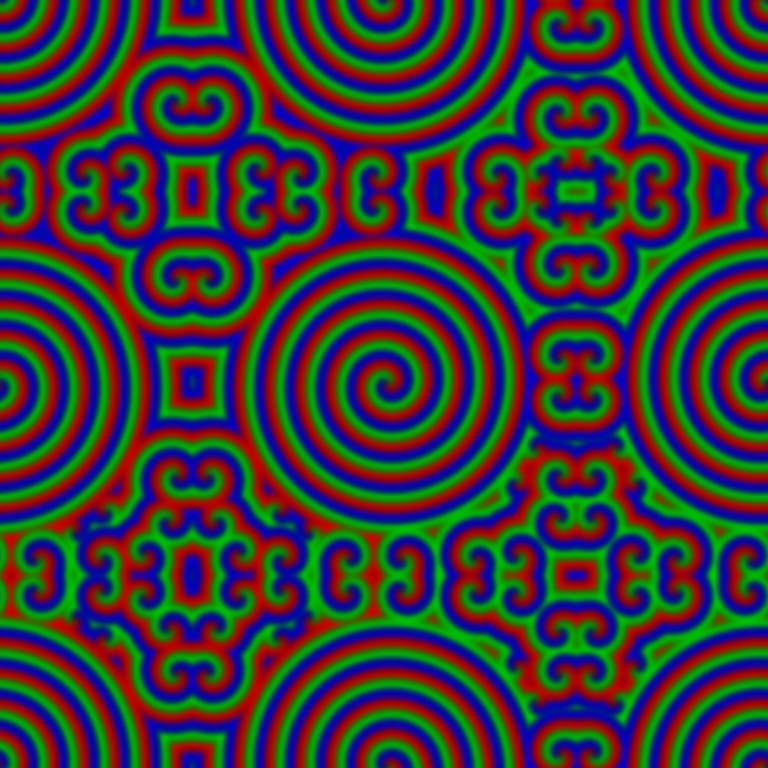}}
\put(100,0){\includegraphics[trim= 0cm 0cm 0cm 0cm,clip=true,width=40mm]{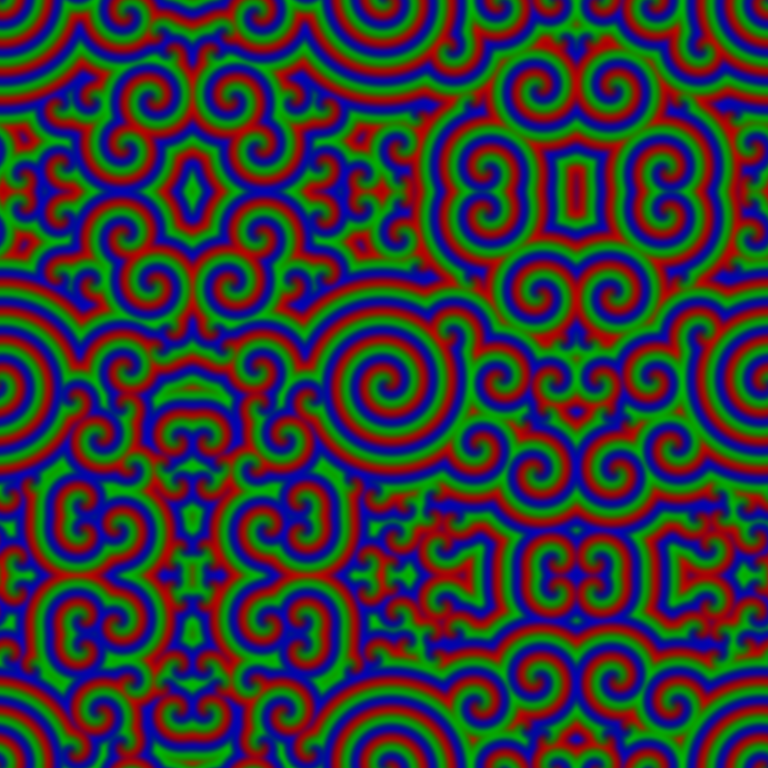}}

\put(0,37){(a)}
\put(47,37){(b)}
\put(94,37){(c)}

\end{picture}
\end{center}
 \caption{Snapshots of numerical solutions of equations~\eref{eq:RPS_PDEs}, in
two spatial dimensions with parameters $\sigma=3.2$, and (a) $\zeta=1.0$, (b)
$\zeta=2.0$, (c) $\zeta=3.0$. The domain size for the integrations was
$500\times500$. Areas in which $a$, $b$ and $c$ are dominant are shown in red,
green and blue respectively. The central spiral rotates clockwise with the three 
colours moving outwards.
 \label{fig:2d}}
 \end{figure}
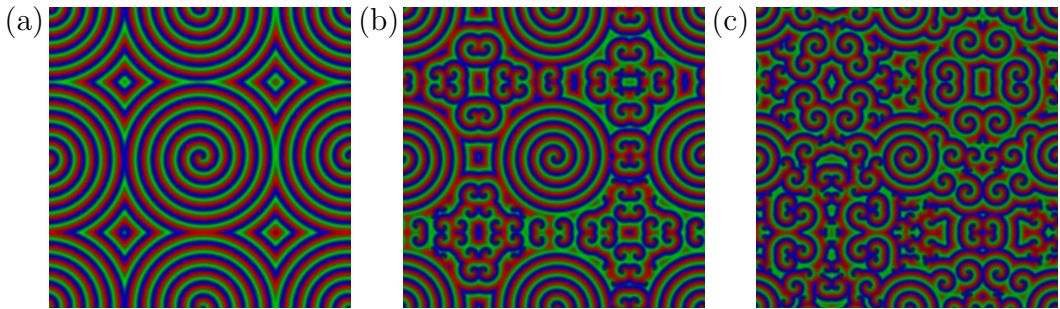

We begin with the PDEs for the spatially-extended Rock--Paper--Scissors model
as given in equations~\eref{eq:RPS_PDEs}. In figure~\ref{fig:2d} we show
numerical results from the integration of equations~\eref{eq:RPS_PDEs} in two
spatial dimensions, from~\cite{Postlethwaite2017}. A variety of behaviours can
be observed, but of particular interest are the spiral-type solutions. When a
slice is taken radially through the centre of a spiral, the profile of the
solution in the outer part of the spiral resembles a travelling wave in one spatial dimension.

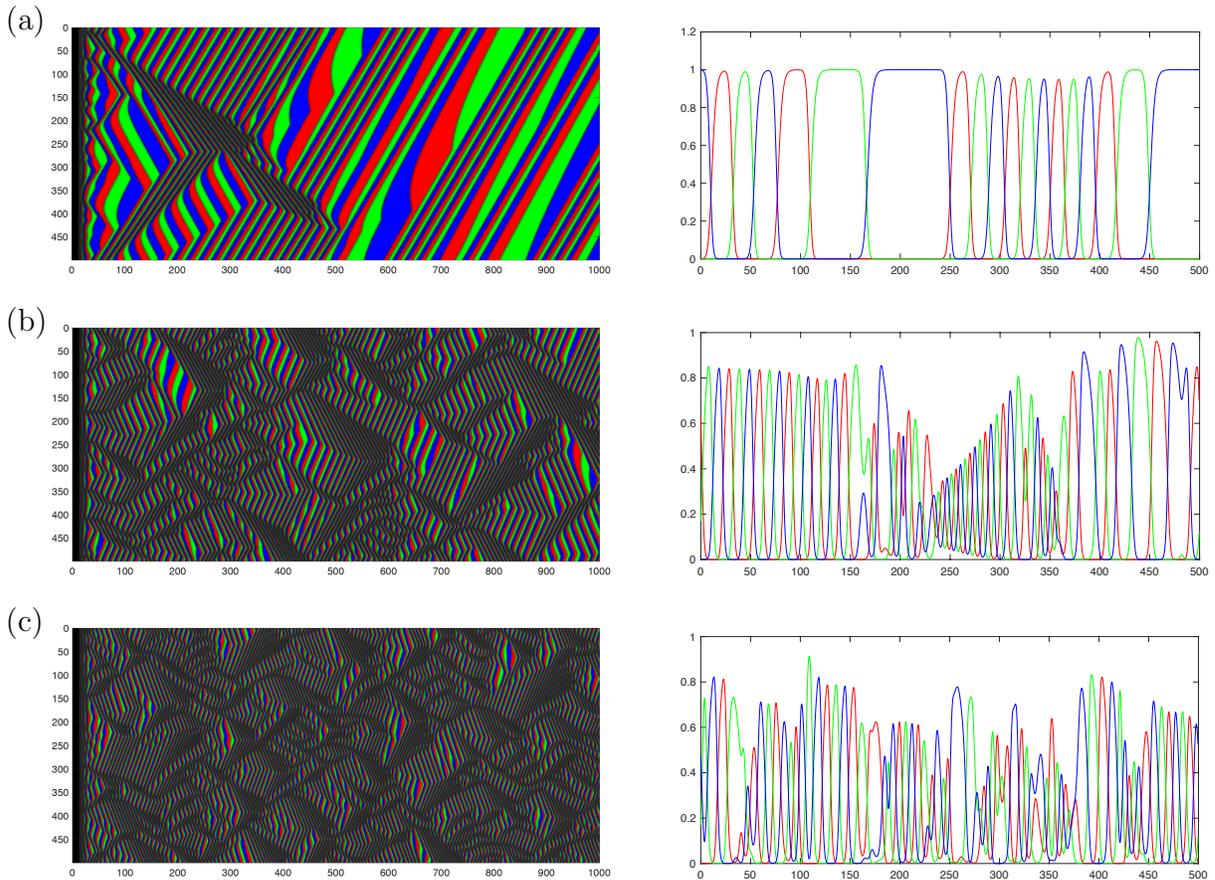
\begin{figure}
\setlength{\unitlength}{1mm}

\begin{center}
\begin{picture}(145,120)(0,0)
\put(5,2){\includegraphics[trim= 0cm 0cm 0cm 0cm,clip=true,height=35mm]{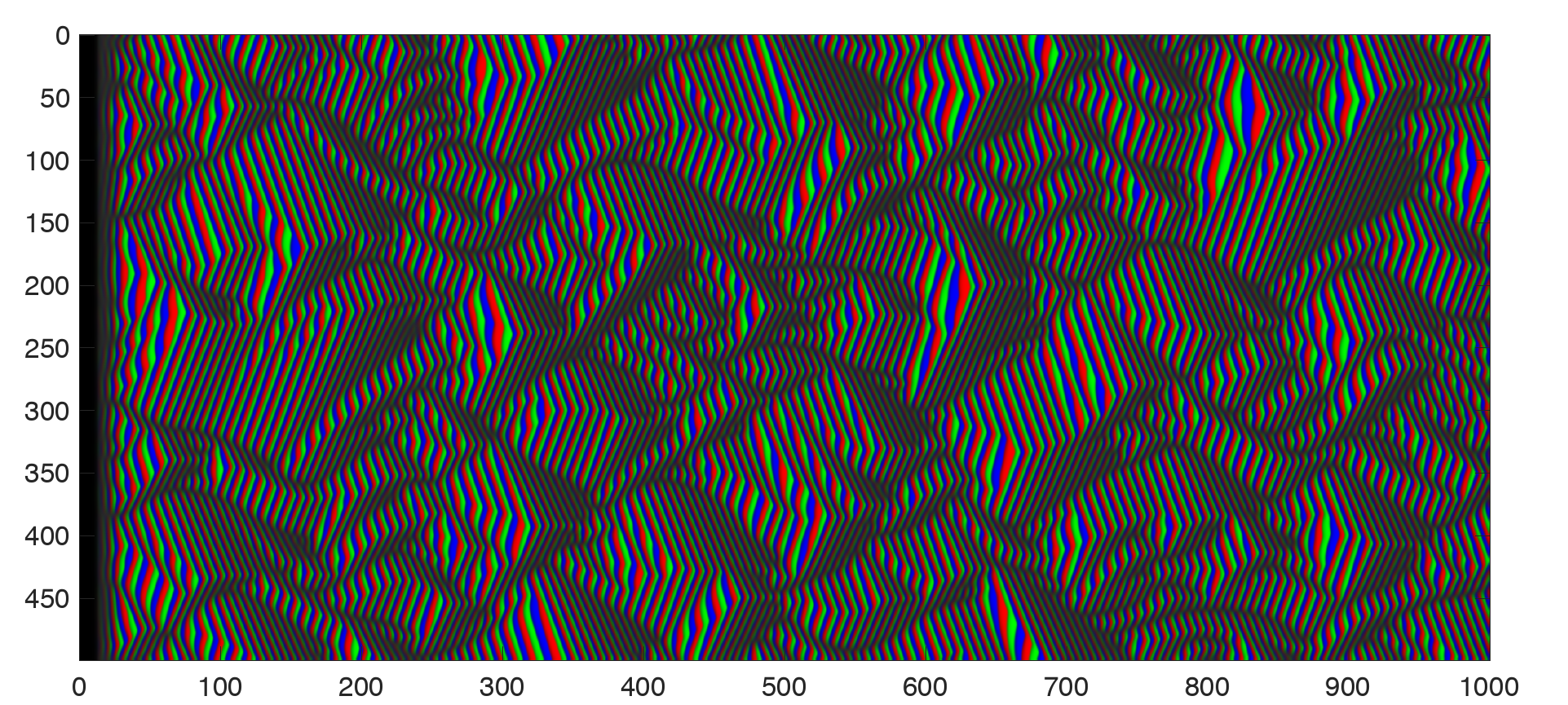}}
\put(85,0){\includegraphics[trim= 0cm 0cm 0cm 0cm,clip=true,height=37mm]{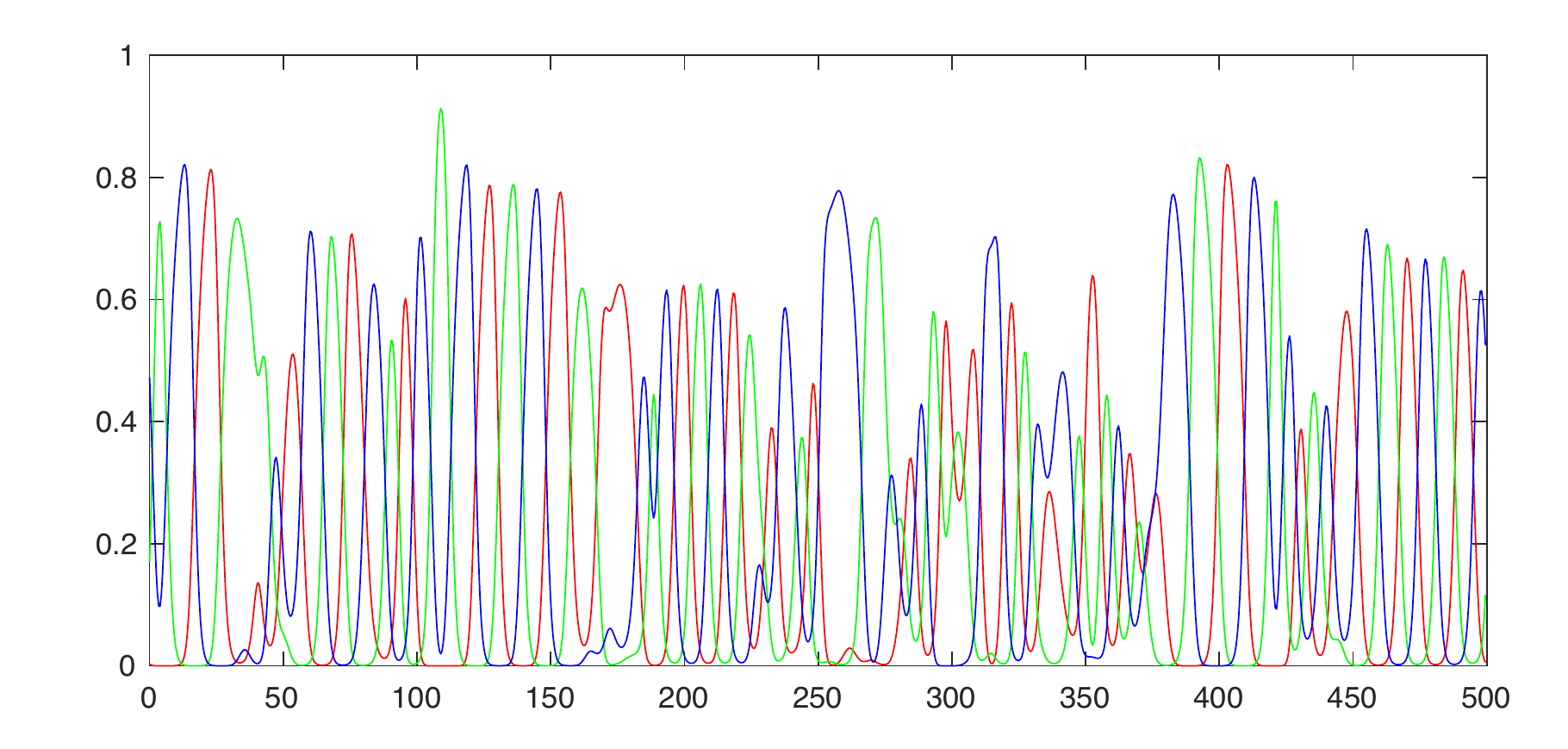}}
\put(0,35){(c)}

\put(5,42){\includegraphics[trim= 0cm 0cm 0cm 0cm,clip=true,height=35mm]{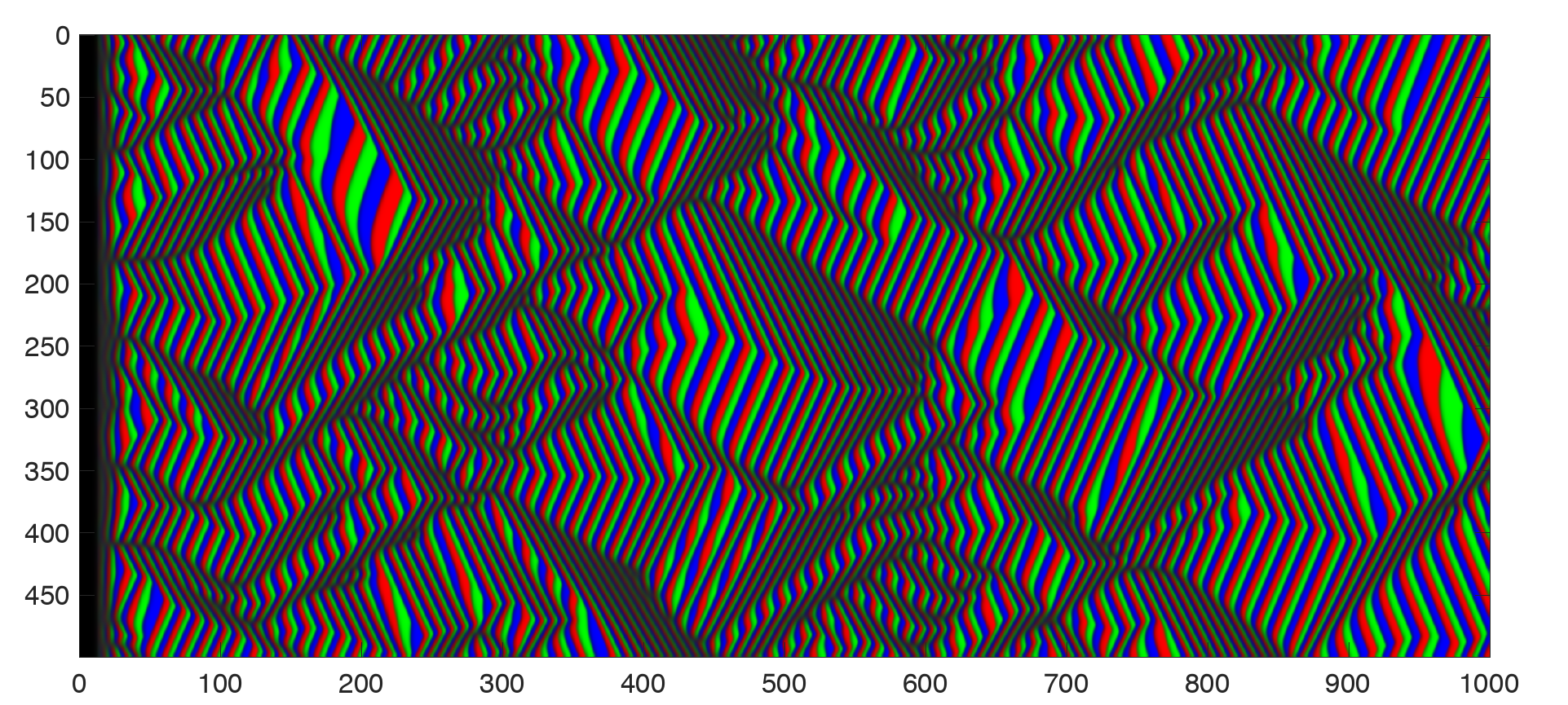}}
\put(85,40){\includegraphics[trim= 0cm 0cm 0cm 0cm,clip=true,height=37mm]{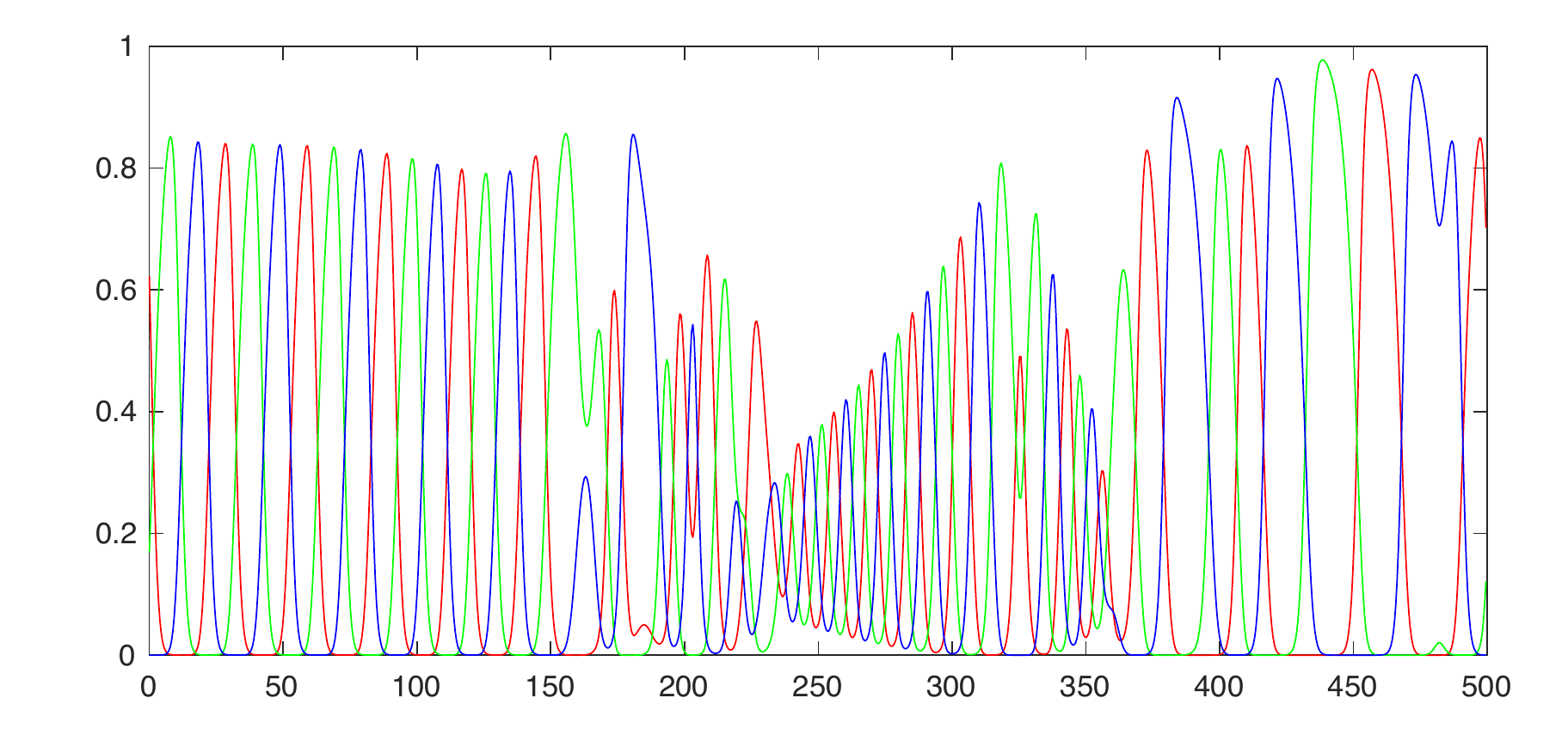}}
\put(0,75){(b)}

\put(5,82){\includegraphics[trim= 0cm 0cm 0cm 0cm,clip=true,height=35mm]{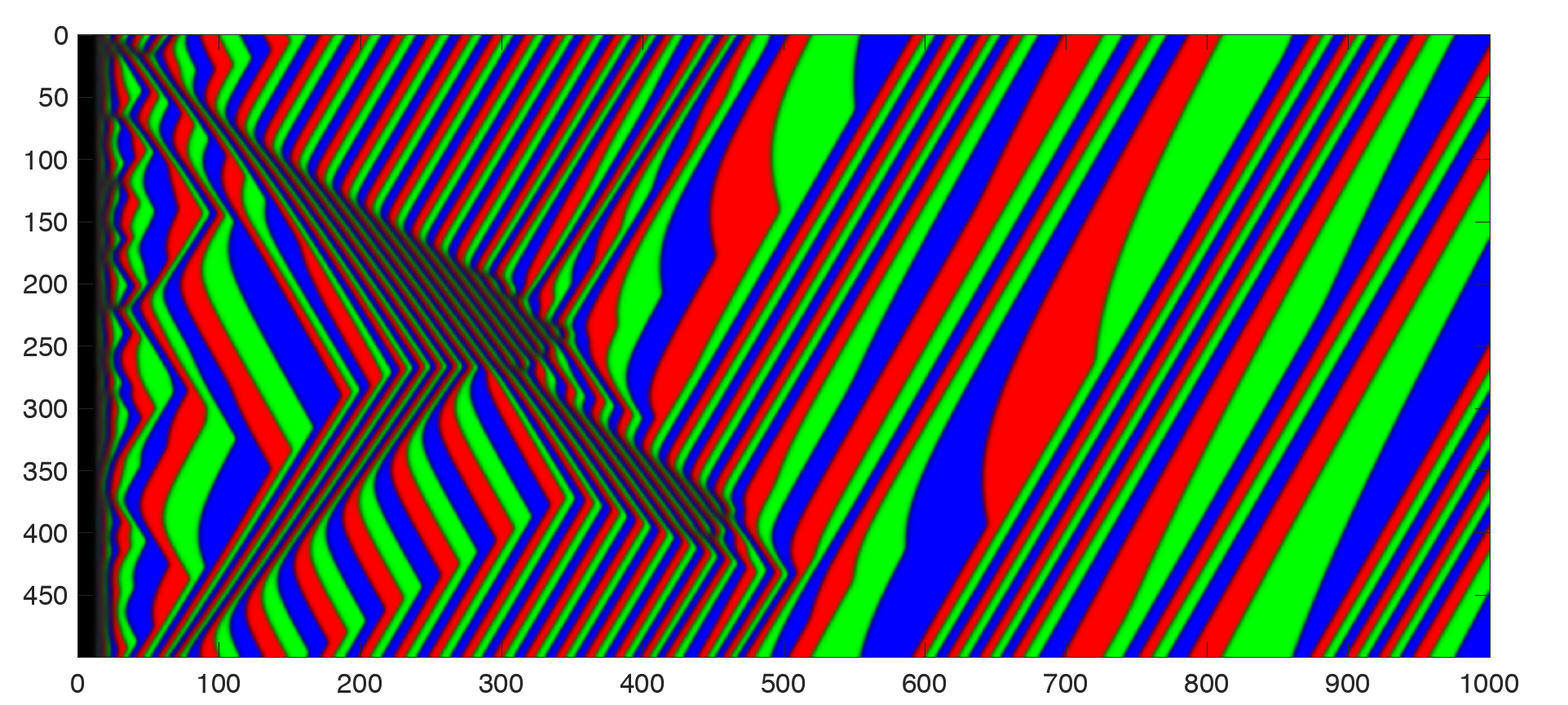}}
\put(85,80){\includegraphics[trim= 0cm 0cm 0cm 0cm,clip=true,height=37mm]{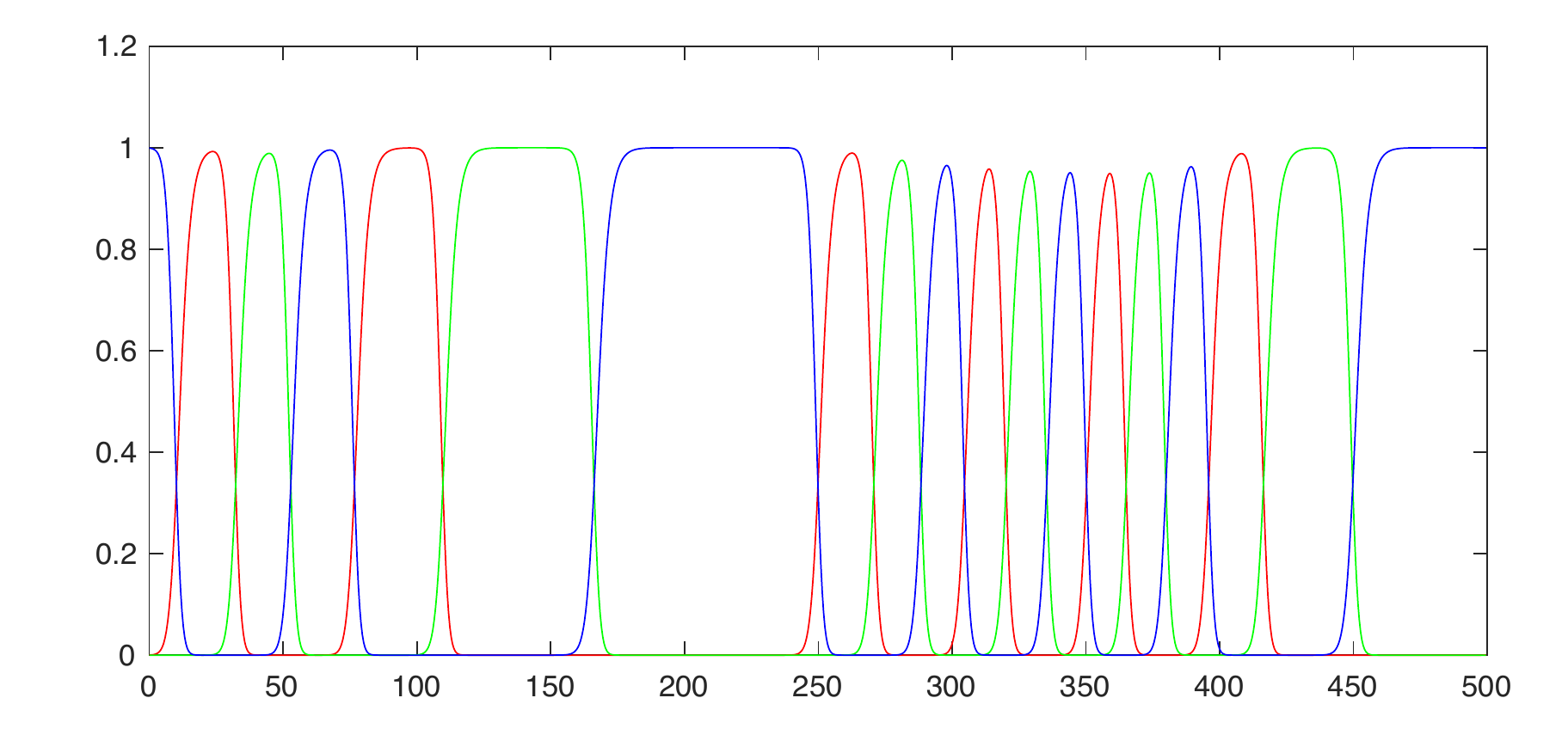}}
\put(0,115){(a)}

\end{picture}
\end{center}
 \caption{\label{fig:1dsims}\blue{The figures show results from numerical integration of equations~\eref{eq:RPS_PDEs} in one spatial dimension, in a box of size 500 with periodic boundary conditions. The left hand column shows time-space plots: time is plotted horizontally and space vertically. Areas in which $a$, $b$ and $c$ are dominant are shown in red,
green and blue respectively. The right hand column shows snapshots at $t=1000$. Parameters are $\sigma=3.2$, and (a) $\zeta=1.0$, (b)
$\zeta=2.0$, (c) $\zeta=3.0$.}}
 \end{figure}

\blue{
Figure~\ref{fig:1dsims} shows the results of numerical integration of equations~\eref{eq:RPS_PDEs} in one spatial dimension, in a large box of size 500, for $\sigma=3.2$ and $\zeta=1.0, 2.0, 3.0$. Initial conditions are of small amplitude and randomly generated, and boundary conditions are periodic. The time-space plots show clearly that multiple travelling waves arise from the initial conditions after a short transient. For all three values of $\zeta$, travelling waves of different directions, wavespeeds and wavelengths are evident. In the simulation for $\zeta=1.0$, after about $t=500$, waves consistently travel to the left, and eventually (after being integrated for a longer time period than shown here), this solution has six waves of equal wavelengths (and equal wavespeed) fitting in the periodic box. For the larger values of $\zeta$, the solutions appear more complicated, in particular, faster wavespeeds and smaller wavelengths are evident. We attempt to quantify this further in section~\ref{sec:moresims}.

We would like, ultimately, to be able to predict the behaviour of solutions to the PDEs~\eref{eq:RPS_PDEs}; that is, we would like to be able to say whether solutions will eventually asymptote onto a single travelling wave, and what the wavespeed and wavelength of that travelling wave will be. In order to do this, we would need to know both existence and stability criteria, as well as have information about the basins of attraction of the travelling waves. The latter two are difficult problems, and are beyond the scope of this paper, but in order to answer both of those questions, we first need to understand the existence problem, and that is what is addressed in this paper. 

Specifically, by relating travelling waves of the PDEs to periodic solutions of a related set of ODEs, we are able to give existence criteria for the travelling waves, and associated with that are minimum and maximum wavespeeds. For a wide range of parameters (those associated with Belyakov--Devaney-type and resonance heteroclinic bifurcations), the wavelength increases monotonically with the wavelength, and the wavelength asymptotes to infinity as the maximum wavespeed is approached. In the parameter regime for which orbit-flip heteroclinic bifurcations are observed, the dispersion relation relating wavelength and wavespeed is non-monotonic, but we are still able to identify a minimum wavespeed and wavelength, and a wavespeed which is approached asymptotically as the wavelength goes to infinity.

}

\section{Review of heteroclinic cycles and bifurcations}
\label{sec:het}

\blue{Before we begin the calculations, in this section we first} include a review of heteroclinic cycles, and the definitions
used by Krupa and Melbourne~\cite{Krupa2004} of contracting, expanding, radial
and transverse eigenvalues. In this paper, we abuse their nomenclature
slightly, and give labels to eigenvalues that don't quite fit with these
definitions, but we find that this is useful nonetheless.

Consider a system of ordinary differential equations
\begin{equation}
\dot{x}=f(x),\quad x\in\R^N.\label{eq:genf}
\end{equation}
Then we have:
\begin{defn}
A \emph{heteroclinic cycle} 
is a finite collection of equilibria
$\{\xi_1, \dots , \xi_n\}$ of~\eref{eq:genf}, together with a set of heteroclinic connections
$\{\phi_1(t), \dots , \phi_n(t) \}$, where $\phi_j(t)$ is a solution of~\eref{eq:genf} such that $\phi_j(t) \to \xi_j$
as $t \to -\infty$ and $\phi_j(t) \to \xi_{j+1}$
as $t \to \infty$, and where $\xi_{n+1} \equiv \xi_1$.
\end{defn}
In generic systems, heteroclinic connections between saddles are of high
codimension, but if a system contains invariant subspaces they can exist for
open sets of parameter values, that is, they are of codimension zero, and are
referred to as `robust'~\cite{Field1996a,Guckenheimer1988,Krupa1997}. In the
work of Krupa and Melbourne~\cite{Krupa1995,Krupa2004} and
others~(e.g., \cite{Field1991,Chossat1997,Jones1987,Melbourne1989b,
Postlethwaite2006,Kirk1994,Ashwin1999d,Driesse2009}), robust heteroclinic cycles arise due to
invariant subspaces which are a by-product of symmetry in the ODEs. In this
paper, we show that for the ODEs we are studying, heteroclinic connections
exist for open sets of parameter values due to a combination of invariant
subspaces and the dimensions of stable and unstable manifolds of equilibria
\emph{for the flow restricted to these invariant subspaces}. An additional
difference in our work is that the invariance of the subspaces is not forced by
symmetry, but instead by the invariance of extinction in continuous-time
population models.

Despite these differences, we continue in the style of Krupa and
Melbourne~\cite{Krupa2004}. Let $P_j$ be an invariant subspace which contains
$\xi_j$ and $\xi_{j+1}$. Let $W_u\vert_{P_j}(\xi_j)$ and
$W_s\vert_{P_j}(\xi_{j+1})$ by the unstable manifold of $\xi_j$ and stable
manifold of $\xi_{j+1}$ for the flow restricted to $P_j$. Then, if
$\dim(W_u\vert_{P_j}(\xi_j))+\dim(W_s\vert_{P_j}(\xi_{j+1}))>\dim(P_j)$, then
\blue{a} heteroclinic connection from $\xi_j$ to
$\xi_{j+1}$ will be codimension zero, this is, it will persist under small 
changes to the \hbox{ODE} \blue{(so long as the changes preserve the invariant subspaces)}. If this is true for all $j$, then there exists a robust
heteroclinic cycle between the equilibria $\xi_1,\dots,\xi_n$, where robust 
here means codimension zero.

We further define
$L_j\equiv P_{j-1}\cap P_j$ and clearly $\xi_j\in L_j$.
Following~\cite{Krupa1997}, the eigenvalues of the linearisation of $f(x)$
  about each equilibrium can be classified according to the
subspaces in which the eigenspaces lie, as shown in
  table~\ref{tab:ecevals}.
\begin{table}
\caption{\label{tab:ecevals}Classification of eigenvalues. $P\ominus L$ denotes the
orthogonal complement in $P$ of the subspace $L$.}
\begin{indented}
\item[]\begin{tabular}{@{}ll}
\br
Eigenvalue class & Subspace \\  \mr Radial ($r$) &
$L_j\equiv P_{j-1}\cap P_j$ \\ Contracting ($c$) &
$V_j(c)=P_{j-1}\ominus L_j$ \\ Expanding ($e$) &
$V_j(e)=P_j\ominus L_j$
\\ Transverse ($s$) & $V_j(s)=(P_{j-1}+P_j)^{\perp}$ \\  
\br
\end{tabular}
\end{indented}
\end{table}
As we will discuss in the following, because we do not require that $P_j$
contains the unstable manifold of $\xi_j$ (unlike in the definition used by
Krupa and Melbourne~\cite{Krupa2004}), we are allowed to have positive radial
and/or contracting eigenvalues.


Methods for determining the stability properties of an isolated heteroclinic
cycle are \blue{in principal}
well-established~\cite{Chossat1997,Krupa1995,Krupa2004,Melbourne1991,
Postlethwaite2010a,Scheel1992, Podvigina2012, Podvigina2013,Podvigina2011,Podvigina2017,Castro2014,Lohse2015}: \blue{that is, one can construct a \Poincare map, by linearising the flow around the fixed points and the heteroclinic connections. Many examples have been investigated in lower dimensions ($\R^3$ and $\R^4$ in particular), but in higher dimensions, calculations can become quite intricate.}  A number of codimension-one
bifurcations have been identified in which the stability of robust heteroclinic
cycles changes, but issues of stability turn out to be more subtle than might
be at first thought (for \blue{several examples, see~\cite{Melbourne1991,Postlethwaite2010,Podvigina2011,Podvigina2012}}). Two
well-studied ways in which heteroclinic cycles can change stability are
\emph{resonance} and \emph{transverse} bifurcations. A resonance
bifurcation~\cite{Driesse2009,Krupa2004,Postlethwaite2010a,Scheel1992,Postlethwaite2006} occurs when an algebraic condition on the eigenvalues of the
equilibria in the cycle is satisfied. \blue{Typically}, resonance bifurcations are
accompanied by the birth or death of a long-period periodic orbit. In a
transverse bifurcation from a heteroclinic cycle~\cite{Chossat1997}, a local
bifurcation causes a transverse eigenvalue of one of the equilibria in the
cycle to change sign. This can result in a bifurcating periodic orbit or
heteroclinic cycle, depending on the specific situation.

In this paper, we use the standard methods for analysing the dynamics close to
a heteroclinic cycle, namely, we construct a \Poincare map which approximates
the flow of the differential equations close to the heteroclinic cycle, but as
mentioned in the introduction, there turn out to be several subtleties which
must be carefully navigated. 
We do not explicitly compute the stability of
the heteroclinic cycle but rather compute conditions for the existence of
nearby periodic orbits.  We find that long-period periodic orbits can exist
close to the heteroclinic cycle, and can appear from the cycle in three
different ways: at a resonance bifurcation, at a bifurcation of
Belyakov--Devaney type, and at an orbit flip bifurcation. Although resonance
bifurcations have been previously studied in the context of robust heteroclinic
cycles, the bifurcation we find is of an unusual type, in that the eigenvalues
of interest are not those that one would expect~\cite{Krupa1995}.

All three of these types of bifurcations have been previously studied in the
context of homoclinic orbits, and in many cases are associated with complicated
dynamics such as homoclinic-doubling cascades~\cite{Oldeman2000,Oldeman2001}.
Useful references for each case include the work of Chow, Deng and Fiedler for
resonant homoclinic bifurcations~\cite{Chow1990a}, the work of Homburg, Kokubu,
Krauskopf and others for orbit flip bifurcations~\cite{Kokubu1996,Homburg2000},
and the work of Belyakov~\cite{Belyakov1984} and Devaney~\cite{Devaney1976} for
the Belyakov--Devaney bifurcation. However, homoclinic orbits cannot be robust,
so each of these phenomena is at least a codimension two bifurcation (there
must be another parameter associated with the existence of the homoclinic
orbit). In the case of a robust heteroclinic bifurcation, then these phenomena
can occur as codimension one, and as such the dynamics associated with the
bifurcations may be somewhat different, and indeed, we find that this is the case.

\section{\blue{Derivation of ODEs and existence of heteroclinic cycles}}
\label{sec:setup}

In this paper, we examine the behaviour of the travelling wave solutions
in one dimension, and so we consider equations~\eref{eq:RPS_PDEs} with only one
spatial dimension, so $\nabla^2=\frac{\partial^2}{\partial x^2}$. We move to a
travelling frame with wavespeed $\gamma>0$, so define $z=x+\gamma{t}$, then
$\ddx{}\rightarrow\ddz{}$ and $\ddt{}\rightarrow\gamma\ddz{}+\ddt{}$. \blue{This results in the following set of PDEs in the travelling frame:
 \begin{eqnarray}
 \ddt{a} +\gamma\ddz{a} &= a (1 - (a+b+c) - (\sigma+\zeta)b + \zeta c) + \ddzz a, \nonumber\\
 \ddt{b}+\gamma\ddz{b} &= b (1 - (a+b+c) - (\sigma+\zeta)c + \zeta a) + \ddzz b, \nonumber \\
 \ddt{c}+\gamma\ddz{c} &= c (1 - (a+b+c) - (\sigma+\zeta)a + \zeta b) + \ddzz c. \nonumber
 \end{eqnarray} }

Travelling wave (TW) solutions in the moving frame have $\ddt{}=0$. \blue{We thus set $\ddt{}=0$, and add additional variables for the first derivative of $a$, $b$ and $c$ with respect to $z$.  Therefore,} TW
solutions of~\eref{eq:RPS_PDEs} correspond to periodic solutions of the
following set of six first-order ODEs:
 \begin{eqnarray}
 a_z  =  u,  \nonumber \\
 u_z  =  \gamma u - a(1-(a+b+c)-(\sigma+\zeta)b+\zeta c),  \nonumber \\
 b_z  =  v,  \nonumber \\
 v_z  =  \gamma v - b(1-(a+b+c)-(\sigma+\zeta)c+\zeta a), \label{eq:odes}\\
 c_z  =  w,  \nonumber \\
 w_z  =  \gamma w - c(1-(a+b+c)-(\sigma+\zeta)a+\zeta b).  \nonumber
 \end{eqnarray}
Since $a$, $b$ and $c$ are non-negative, we define a {\em positive} travelling 
wave as a periodic solution of~\eref{eq:odes} with $a,b,c>0$ for all~$z$.
In an abuse of notation, the independent variable~$z$ will be referred to as 
`time' (and denoted with a `$t$') when we construct \Poincare maps in the next section. 

Let $\vec{x}=(a,u,b,v,c,w)$, and note that the coexistence and on-axis equilibria of~\eref{eq:odes} correspond to 
$\vec{x}=\frac{1}{3+\sigma}(1,0,1,0,1,0)$, $\vec{x}=(1,0,0,0,0,0)$, $\vec{x}=(0,0,1,0,0,0)$ and
$\vec{x}=(0,0,0,0,1,0)$. We label these equilibria $\xih$, $\xia$, $\xib$ and $\xic$
respectively. Also note that the ODEs~\eref{eq:odes} are invariant under the rotation symmetry $g$:
 \begin{equation}
 g(a,u,b,v,c,w)=(b,v,c,w,a,u).
 \end{equation}

The Jacobian matrix at $\xia$ is
\begin{equation}
J_A=
\left(
\begin{array}{llllll}
0 & 1& 0 & 0 & 0 & 0 \\
1 & \gamma & 1+\sigma+\zeta  & 0 & 1-\zeta & 0 \\
0 & 0 & 0 & 1 & 0 & 0 \\
0 & 0 & -\zeta & \gamma & 0 & 0 \\
0 & 0 & 0 & 0 & 0 & 1 \\
0 & 0 & 0 & 0 & \sigma+\zeta & \gamma  \\
\end{array}\right)
\end{equation}
The eigenvalues of $J_A$ are given in table~\ref{tab:evals}. Note that we frequently refer to `the eigenvalues of $\xia$', by which of course we mean the eigenvalues of $J_A$.
 By the symmetry $g$, $\xib$ and $\xic$ have the same eigenvalues.

 Let the four-dimensional subspace $\{c=w=0\}$ be labelled $P(\xia)$. It can easily be seen that $P(\xia)$ is invariant under the flow 
of~\eref{eq:odes}. For the dynamics restricted to $P(\xia)$, $\xia$ has a three-dimensional unstable manifold, and $\xib$ has a
two-dimensional stable manifold. \blue{By dimension counting, it is reasonable to expect that these manifolds will intersect, and hence that there is a heteroclinic connection within $P(\xia)$
between $\xia$ and $\xib$, which persists under small perturbations. We are able to numerically confirm the existence of a heteroclinic connection for a wide range of parameter values.}
By symmetry, there is thus a robust heteroclinic cycle
between $\xia$, $\xib$ and $\xic$.

As discussed earlier, because our definition of robust heteroclinic cycle did
not require the unstable manifold of $\xia$ to be contained in $P(\xia)$, we
can have radial or contracting eigenvalues that have positive real part, and in
fact, this is what we find (see table~\ref{tab:evals}). Specifically, we note
that $\lambda_c^-<0<\lambda_c^+$, $\lambda_r^-<0<\lambda_r^+$, and
$0<Re(\lambda_e^-)\leq Re(\lambda_e^+)$.

\begin{table}
\begin{tabular}{ll}
\strut
Label & Eigenvalues \\ \hline
\strut
Radial &  $\lambda_r^{\pm}=\frac{1}{2}\left(\gamma\pm\sqrt{\gamma^2+4}\right)$  \\
\hline
\strut
Contracting & $\lambda_c^{\pm}=\frac{1}{2}\left(\gamma\pm\sqrt{\gamma^2+4(\sigma+\zeta)}\right)$ \\
\hline
\strut
Expanding ($\gamma^2-4\zeta>0$) & $\lambda_e^{\pm}=\frac{1}{2}\left(\gamma\pm\sqrt{\gamma^2-4\zeta}\right)$\\
\strut
Expanding ($\gamma^2-4\zeta<0$) &  $\lambda_e^{\pm}=\lambda_e^R\pm{i}\lambda_e^I=\frac{1}{2}\left(\gamma\pm{i}\sqrt{4\zeta-\gamma^2}\right)$
 \end{tabular}
 \caption{\label{tab:evals}Eigenvalues of the equilibrium $\xi_A$ in~\eref{eq:odes}.}
 \end{table}

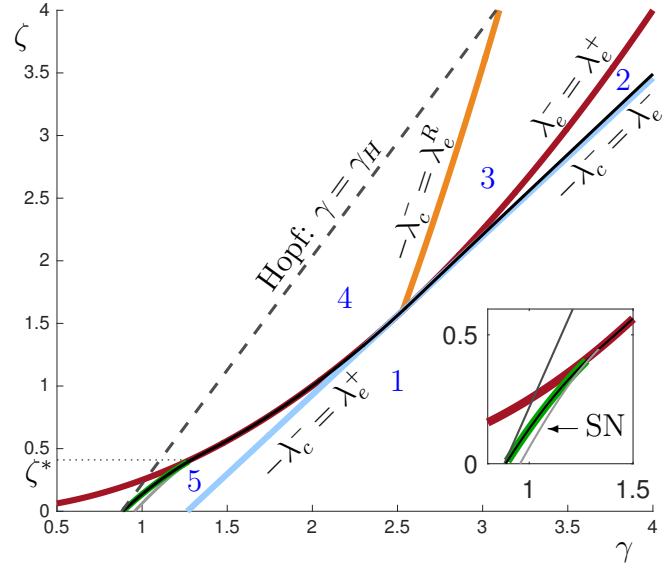
\begin{figure}
\setlength{\unitlength}{1mm}
\begin{center}
\begin{picture}(86,75)(0,0)
\put(0,0){\includegraphics[trim= 1.3cm 0cm 1.5cm 0.9cm,clip=true,width=86mm]{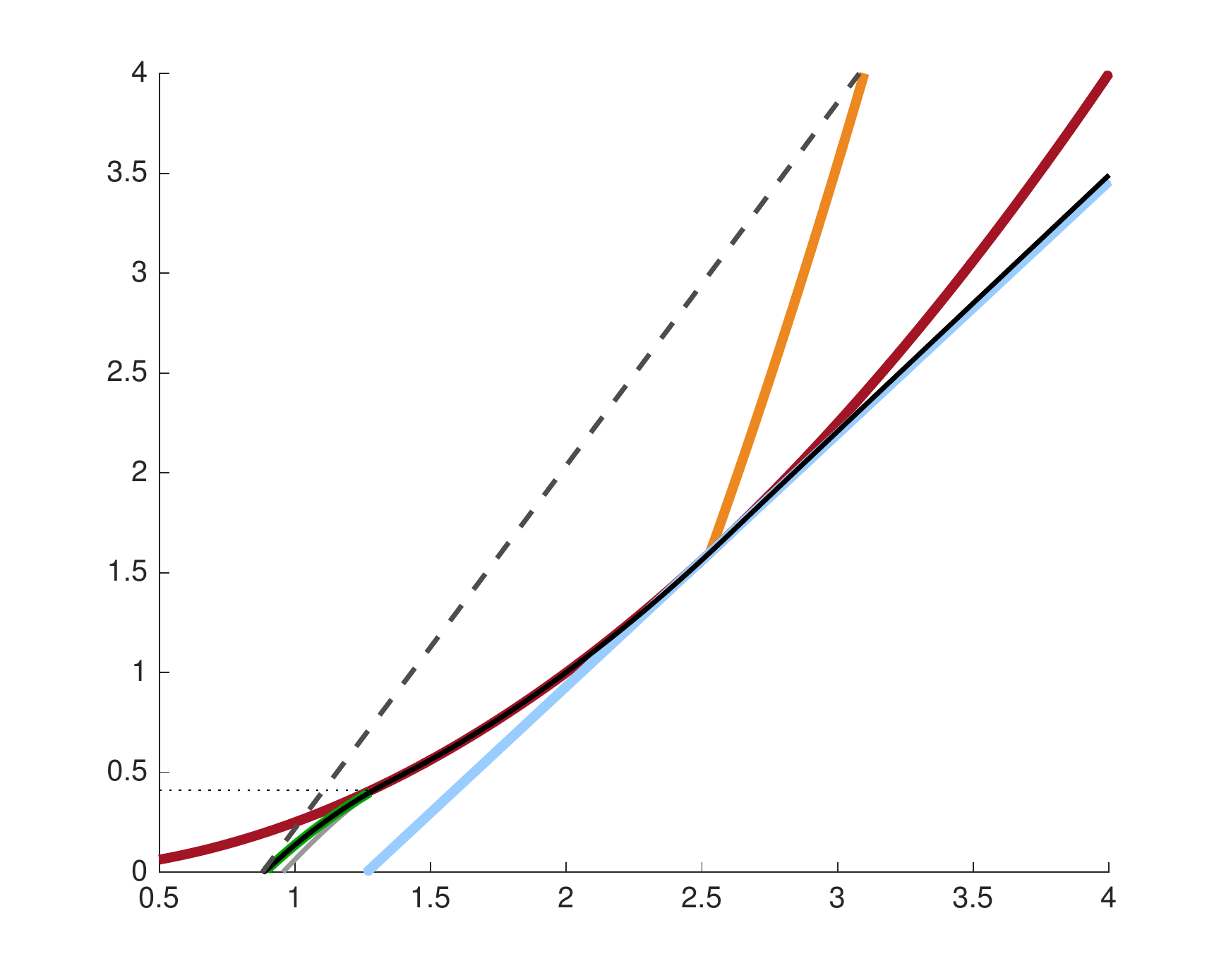}}                             
\put(80,3){{$\gamma$}}
\put(0,71){$\zeta$}
\put(32,38){\rotatebox{54}{Hopf: $\gamma=\gamma_H$}}
\put(50,25){\textcolor{blue}{$1$}}
\put(80,65){\textcolor{blue}{$2$}}
\put(62,52){\textcolor{blue}{$3$}}
\put(43,36){\textcolor{blue}{$4$}}
\put(23,11.7){\textcolor{blue}{$5$}}
\put(68,58.5){\rotatebox{55}{$\lambda_e^-=\lambda_e^{+}$}}
\put(71,50){\rotatebox{45}{$-\lambda_c^-=\lambda_e^-$}}
\put(32,13){\rotatebox{45}{$-\lambda_c^-=\lambda_e^{+}$}}
\put(49,42){\rotatebox{72}{$-\lambda_c^-=\lambda_e^{R}$}}
\put(1.5,13){$\zeta^*$}
\put(57,10){\includegraphics[trim= 0.cm 0.cm 0.cm 0.1cm,clip=true,width=28mm]{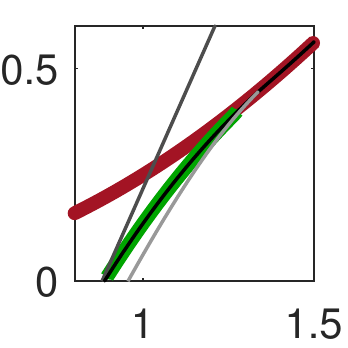}}
\put(76,19){SN}
\put(75,20){\vector(-1,0){4}}
\end{picture}
\end{center}
\caption{Bifurcation diagram for the ODEs~\eref{eq:odes}, in $(\gamma,\zeta)$
parameter space, with $\sigma=3.2$. The blue line
($\zeta=\sqrt{\frac{\sigma}{2}}\gamma-\frac{\sigma}{2}$) and red curve
($4\zeta=\gamma^2$) are tangent at $(\gamma, \zeta)=(\sqrt{2\sigma},\sigma/2)$,
where they meet the yellow curve ($4(\sigma+\zeta)=3\gamma^2$). These three
curves divide the parameter space into five regions, labelled by blue numbers,
and defined in table~\ref{tab:pspace}. The green curve is the locus of a
heteroclinic orbit flip. The dashed grey line is a curve of Hopf bifurcations 
(equation~\eref{eq:hopf}).
Periodic orbits bifurcate to the right of this line and disappear in a curve of
heteroclinic bifurcations (black). A curve of saddle-node bifurcations of
periodic orbits (light grey) exists for smaller $\zeta$. The inset shows a zoom
near the saddle-node \blue{of periodic orbits} (SN) and \blue{heteroclinic} orbit flip (green) bifurcations.
 \label{fig:pspace}} 
 \end{figure}

The Jacobian matrix at $\xih$ is:
\begin{equation}
J_H=
\left(
\begin{array}{llllll}
0 & 1& 0 & 0 & 0 & 0 \\
\frac{1}{3+\sigma} & \gamma & \frac{1+\sigma+\zeta}{3+\sigma}  & 0 & \frac{1-\zeta}{3+\sigma} & 0 \\
0 & 0 & 0 & 1 & 0 & 0 \\
\frac{1-\zeta}{3+\sigma} & 0 & \frac{1}{3+\sigma}& \gamma & \frac{1+\sigma+\zeta}{3+\sigma}  & 0 \\
0 & 0 & 0 & 0 & 0 & 1 \\
\frac{1+\sigma+\zeta}{3+\sigma}  & 0 & \frac{1-\zeta}{3+\sigma}  & 0 & \frac{1}{3+\sigma}  & \gamma  \\
\end{array}\right)
\end{equation}
$J_H$ has pure imaginary eigenvalues~$\pm i\omega_H$ when
$\gamma=\gamma_H(\sigma,\zeta)$, where
 \begin{equation}\label{eq:hopf}
 \gamma_H(\sigma,\zeta)\equiv\frac{\sqrt{3}(\sigma+2\zeta)}{\sqrt{2\sigma(\sigma+3)}},
 \quad\textrm{and}\quad
 \omega_H^2=\frac{\sigma}{2(\sigma+3)},
 \end{equation}
at which point a Hopf bifurcation creates periodic orbits of
period (in the $z$~variable)~$\Lambda_H=\frac{2\pi}{\omega_H}$. Numerical
analysis of equations~\eref{eq:odes} with AUTO~\cite{Postlethwaite2017,Doedel2001} show that the
branch of periodic orbits grows in period as $\gamma$ is increased from the
Hopf bifurcation, eventually ending in a heteroclinic bifurcation. The Hopf and
heteroclinic bifurcation curves can be seen in Figure~\ref{fig:pspace} for
$\sigma=3.2$ as the grey dashed and black solid curves respectively.
Also shown
in figure~\ref{fig:pspace} are various curves depicting relationships between
the eigenvalues (the red, yellow and blue curves), and a curve showing the
location of when the heteroclinic connection undergoes an orbit flip (green
curve). Recall that the heteroclinic cycle is of codimension zero, and so the
orbit flip curve is of codimension one. The orbit flip curve is found by
solving a boundary value problem, as explained further in
section~\ref{sec:orbflip}.

The heteroclinic bifurcation curve in figure~\ref{fig:pspace} is of three
different types, depending on the parameters $\zeta$ and $\sigma$. For the
value of $\sigma$ used to create figure~\ref{fig:pspace} ($\sigma=3.2$), we
have: (a) if $\zeta>\sigma/2=1.6$, the heteroclinic bifurcation is of resonance
type, and occurs when $-\lambda_c^-=\lambda_e^-$ (where the black curve
coincides with the blue curve in figure~\ref{fig:pspace}); (b) if $\zeta^*
<\zeta<\sigma/2=1.6$ (where $\zeta^*\approx 0.46$), then the heteroclinic
bifurcation is of Belyakov--Devaney type, and occurs when $\lambda_e^I=0$
(where the black curve coincides with the red curve in
figure~\ref{fig:pspace}), and (c) if $0<\zeta<\zeta^*$, then the heteroclinic
bifurcation is of orbit flip type (where the black curve coincides with the
green curve in figure~\ref{fig:pspace}).

 \blue{
 In the first two cases, the Hopf and heteroclinic bifurcation curves denote the existence boundaries of periodic orbits in the ODEs, and hence also of travelling waves in the PDEs. Specifically, the Hopf bifurcation curve indicates the minimum wavespeed $\gamma$ (and minimum wavelength, given by $\Lambda_H$ as written after equation~\eref{eq:hopf}), and the heteroclinic bifurcation curve indicates the maximum wavespeed. That is, for $\zeta>\sigma/2$, the allowed wavespeeds are
 \[
 \gamma_H(\sigma,\zeta)<\gamma<\sqrt{\frac{2}{\sigma}}\zeta+\sqrt{\frac{\sigma}{2}},
 \]
 where $\gamma_H$ is given in~\eref{eq:hopf}.
For $\zeta^*<\zeta<\sigma/2=1.6$, the allowed wavespeeds are
 \[
 \gamma_H(\sigma,\zeta)<\gamma<2\sqrt{\zeta}.
 \]
For $\zeta<\zeta^*$, the heteroclinic bifurcation is of orbit-flip type, and there also exists a branch of saddle-node bifurcations of periodic orbits (light grey curve). Here, the right hand boundary for existence of travelling waves is the saddle-node bifurcation curve, not the heteroclinic bifurcation curve. The location of both of these curves depends on global parameters, so here we cannot give an explicit expression for the maximum wavespeed. }

\begin{table}
\caption{Definitions of the regions of parameter space shown
in Fig.~\ref{fig:pspace} and eigenvalue properties therein.
 \label{tab:pspace}}
\begin{center}
\begin{tabular}{ccc}
Region & Definition & Eigenvalue properties \\ \hline
1 &  $\zeta<\sqrt{\frac{\sigma}{2}}\gamma-\frac{\sigma}{2}$ & $\lambda_e^{\pm}\in\mathbb{R}$, $\lambda_e^-<|\lambda_c^-|<\lambda_e^{+}$  \\
\hline
2 & $\zeta>\frac{\sigma}{2}$, $ \sqrt{\frac{\sigma}{2}}\gamma-\frac{\sigma}{2}<\zeta<\frac{\gamma^2}{4}$ &
 $\lambda_e^{\pm}\in\mathbb{R}$, $|\lambda_c^-|<\lambda_e^-<\lambda_e^{+}$ \\
\hline
3 & $\frac{\gamma^2}{4}<\zeta<\frac{3}{4}\gamma^2-\sigma$ & $\lambda_e^{\pm}\in\mathbb{C}$, $|\lambda_c^-|<\lambda_e^R$ \\
\hline
4 & $\zeta>\frac{\gamma^2}{4}$, $\frac{3}{4}\gamma^2-\sigma<\zeta$ & $\lambda_e^{\pm}\in\mathbb{C}$, $\lambda_e^R<|\lambda_c^-|$
\\
\hline
5 & $\zeta<\frac{\sigma}{2}$, $ \sqrt{\frac{\sigma}{2}}\gamma-\frac{\sigma}{2}<\zeta<\frac{\gamma^2}{4}$ & $\lambda_e^{\pm}\in\mathbb{R}$, $\lambda_e^-<\lambda_e^{+}<|\lambda_c^-|$
\end{tabular}
\end{center}
\end{table}

\begin{figure}
\setlength{\unitlength}{1mm}
\begin{center}
\begin{picture}(90,122)(0,0)
\put(5,80){\includegraphics[trim= 1.2cm 0cm 1.5cm 0cm,clip=true,width=85mm]{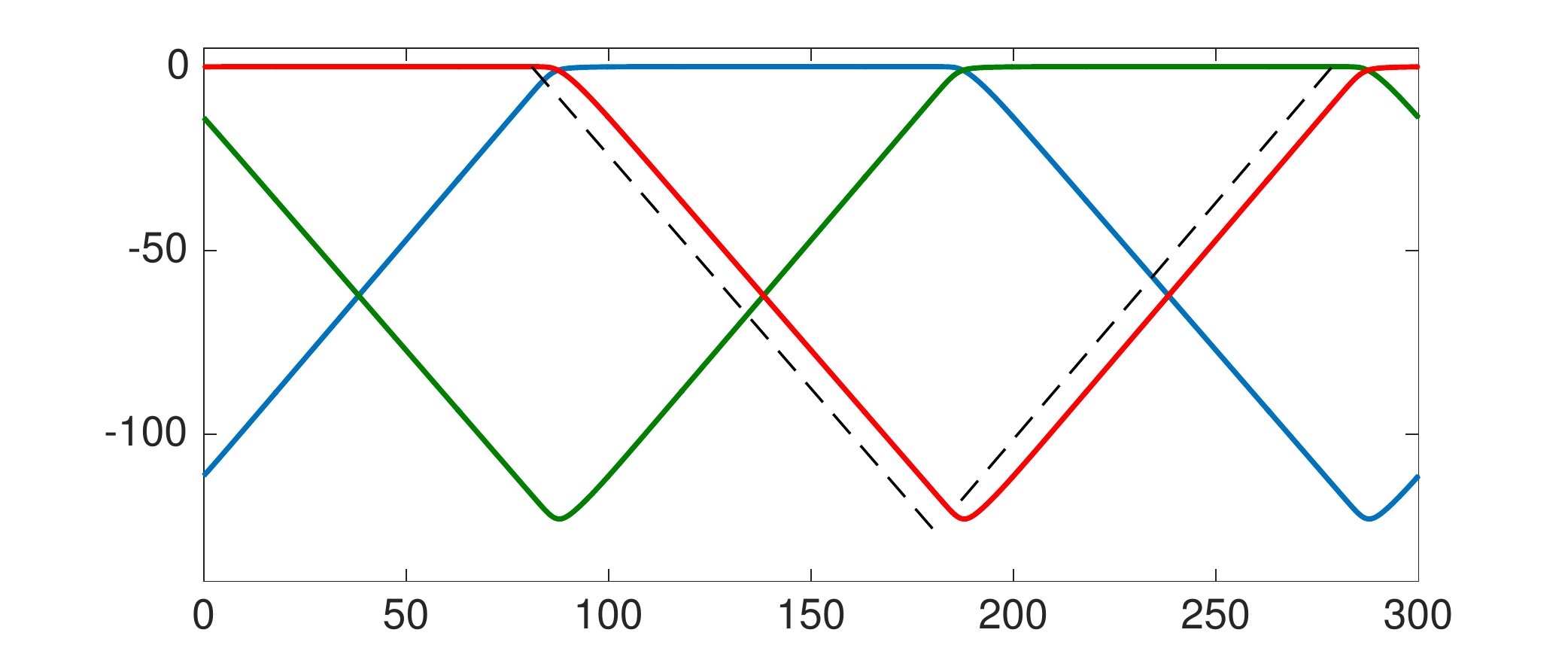}}
\put(46,92){$\lambda_c^-$}
\put(62,100){$\lambda_e^-$}
\put(5,40){\includegraphics[trim= 1.2cm 0cm 1.5cm 0cm,clip=true,width=85mm]{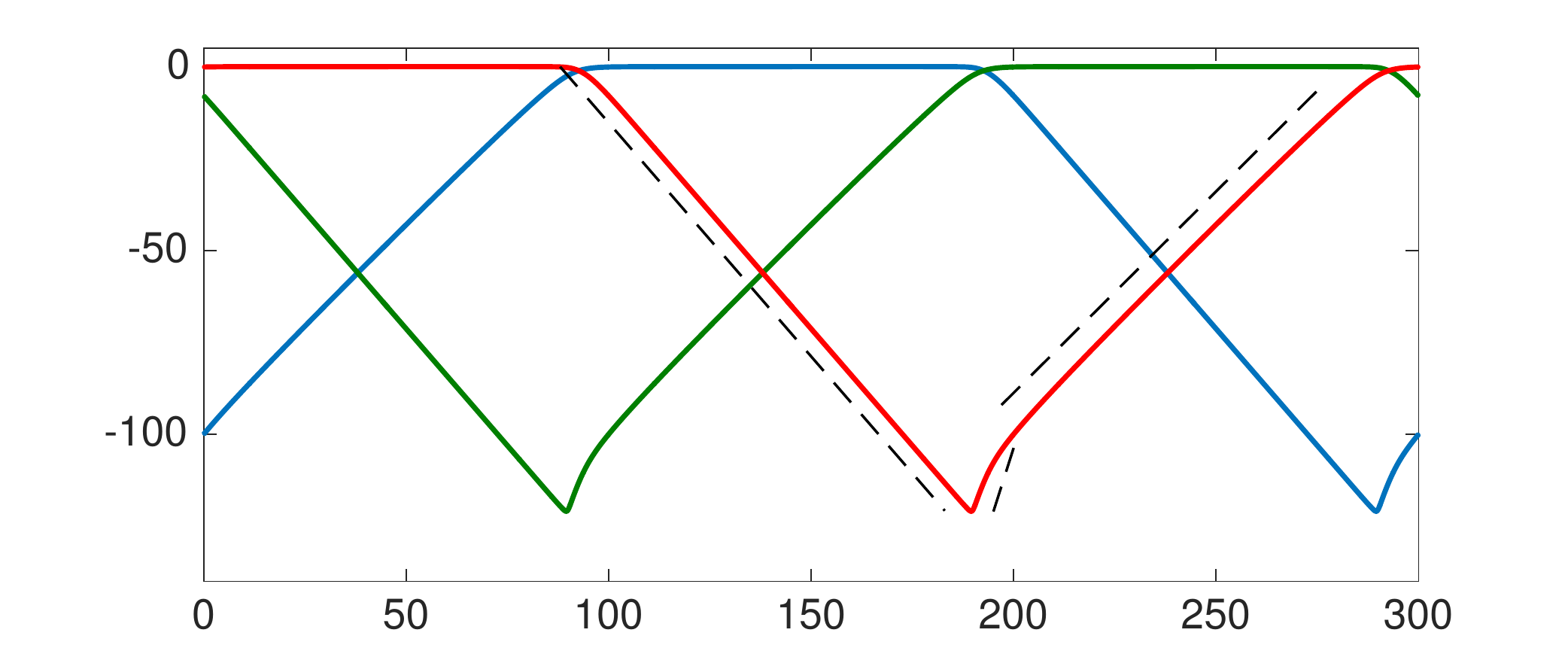}}
\put(45,55){$\lambda_c^-$}
\put(63,50){$\lambda_c^+$}
\put(62,62){$\lambda_e^R$}
\put(5,0){\includegraphics[trim= 1.2cm 0cm 1.5cm 0cm,clip=true,width=85mm]{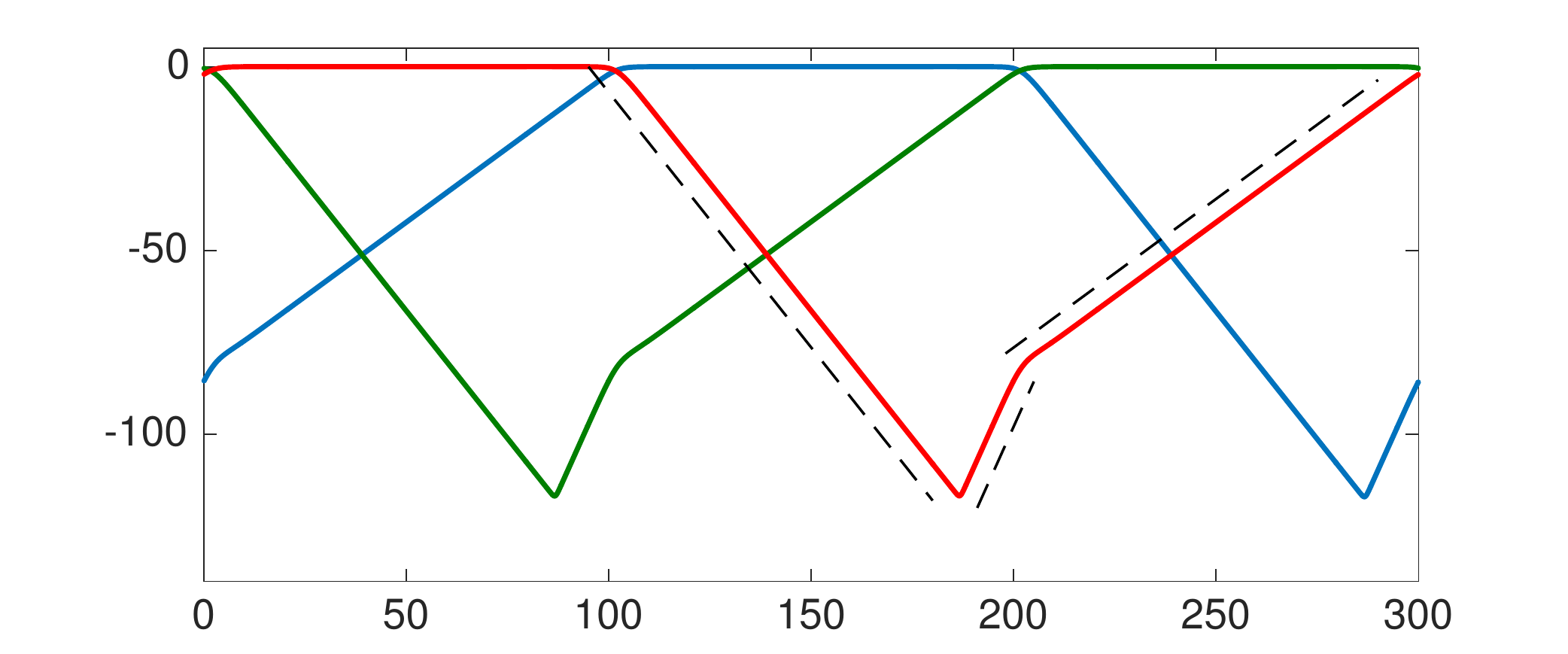}}
\put(45,15){$\lambda_c^-$}
\put(63,10){$\lambda_c^+$}
\put(63,24){$\lambda_e^+$}
\put(80,0){$z$}
\put(2,6){\rotatebox{90}{$\log a, \log b, \log c$}}
\put(2,46){\rotatebox{90}{$\log a, \log b, \log c$}}
\put(2,86){\rotatebox{90}{$\log a, \log b, \log c$}}
\put(2,118){(a)}
\put(2,78){(b)}
\put(2,38){(c)}
\end{picture}
\end{center}
\caption{The figures show time series (in logarithmic coordinates) of periodic
solutions to the ODEs~\eref{eq:odes}, computed using AUTO, near the
heteroclinic bifurcation. The coordinates $a$, $b$ and $c$ are shown in red,
blue and green respectively, all in logarithmic coordinates. Parameter values
are $\sigma=3.2$, and (a) $\zeta=3.2238$, $\gamma=3.7917$ (b) $\zeta=1.2024$,
$\gamma=2.1441$ (c) $\zeta=0.2096$, $\gamma=1.0679$. Dashed lines have
gradients indicated by the eigenvalues, which are given in
table~\ref{tab:evals}. In (a), the expanding eigenvalues are real, and the
periodic orbit is close to a heteroclinic resonance bifurcation. In (b), the
expanding eigenvalues are complex, and the periodic orbit is close to a
heteroclinic Belyakov--Devaney\blue{-type} bifurcation. In (c), the expanding eigenvalues
are real, and the periodic orbit is close to a heteroclinic orbit flip
bifurcation. In (b) and (c), the periodic orbits are kinked at the transition 
from the contracting to the expanding phase.}
 \label{fig:tseries}
 \end{figure}
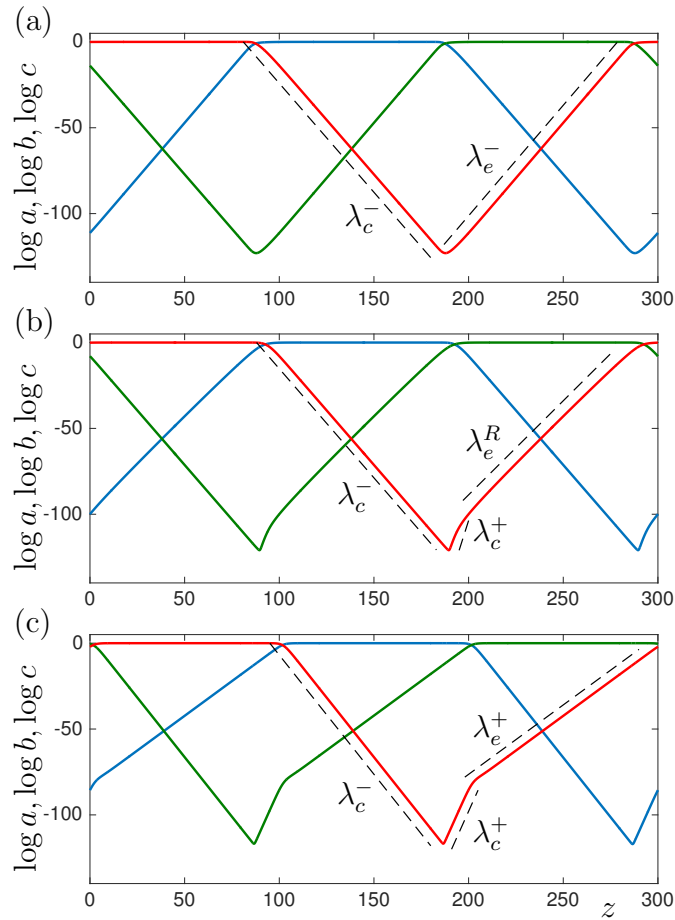

In figure~\ref{fig:tseries} we show time-series of periodic solutions
of~\eref{eq:odes} which are close to the three types of heteroclinic
bifurcations. The examples are all right-travelling waves \blue{(in the PDE setup)}; left-travelling
waves are also possible. In panel (a), we show a periodic orbit close to the
heteroclinic resonance bifurcation (near the edge of region~2, the expanding
eigenvalues are real). The slopes in the contracting and expanding phases can
be seen to be very close to $\lambda_c^-$ and $\lambda_e^-$. In panel (b), we
show a periodic orbit close to the heteroclinic Belyakov--Devaney\blue{-type} bifurcation
(near the edge of region~4, the expanding eigenvalues are complex). Here,
$\lambda_e^I$ is very close to zero, and the slope in the expanding phase is
close to $\lambda_e^R$. In the contracting phase, we see slopes equal to both
the negative contracting eigenvalue, $\lambda_c^-$, and the positive
contracting eigenvalues $\lambda_c^+$. In panel (c), we show a periodic orbit
close to the heteroclinic orbit-flip bifurcation (in region~5, the expanding
eigenvalues are real). The slope in the expanding phase is $\lambda_e^+$,
because the periodic solution lies close to the heteroclinic orbit, which is
close to tangent to the strong unstable manifold. Again, in the contracting phase
we see both the positive and negative slopes. We refer later to periodic orbits
which have both a positive and negative slope in the contracting phase as those
having a \emph{kink} -- the kink refers to the change in growth rate at the 
transition from the contracting to the expanding phase.

In long-period orbits such as in Figure~\ref{fig:tseries}, the \blue{total amount of} decay in the
contracting phase must balance the growth in the expanding phase; \blue{the contracting and expanding phases must be the same length because of the symmetry between the $a$, $b$ and $c$ coordinates in the orbit}.  Therefore,
orbits of this type cannot exist in regions~2 and~3: $\lambda_c^-$,~the only
negative non-radial eigenvalue, is less in absolute value than the (real part
of the) smaller of the two expanding eigenvalues, and so there can't be enough
decay to balance the growth.

A further point to note is that not all periodic solutions of~\eref{eq:odes}
correspond to positive travelling wave solutions of~\eref{eq:RPS_PDEs}. In
particular, because we are considering a population model, we will start with
initial conditions (of~\eref{eq:RPS_PDEs}) which have $a,b,c\geq0$, and, given
reasonable conditions on the smoothness of the initial conditions, it can be
shown that $a,b,c\geq0$ for all~$t$ (in~\eref{eq:RPS_PDEs}). 
Only periodic solutions of~\eref{eq:odes}
which have $a,b,c>0$ for all~$z$ correspond to positive travelling wave
solutions of~\eref{eq:RPS_PDEs}. This may be important, because the variables
in~\eref{eq:odes} may change sign along the heteroclinic connections ---
clearly they will in the case that the expanding eigenvalues are complex.

\section{Constuction of a \Poincare map \blue{and analysis of heteroclinic bifurcations}}
\label{sec:Pmap}

In this section we construct a \Poincare map which approximates the dynamics
close to the heteroclinic cycle of equations~\eref{eq:odes} as described in
section~\ref{sec:setup}. We treat the cases in which the expanding eigenvalues
are real and complex separately, although the computations are quite similar.
Regions of real (1, 2 and~5) and complex (3 and 4) eigenvalues are divided by
the red curve in figure~\ref{fig:pspace}. In this section, we refer to the 
independent variable of equations~\eref{eq:odes} as time ($t$) rather than~$z$.

The \Poincare map we construct here will follow a trajectory that starts on an
incoming section near $\xia$ and ends on an incoming section near $\xib$. In
both real and complex cases, we define \Poincare sections close to $\xia$ and
$\xib$, and derive a local map which approximates the flow close to $\xia$. We
combine this with a global map linearised about the location of the
heteroclinic connection from $\xia$ to $\xib$ and then use the symmetry $g$ to
map the coordinates back to a \Poincare section close to $\xia$. \blue{We are able to disregard the radial directions in our computations. This is often done because eigenvalues in the radial directions are negative, with an invocation to an invariant sphere~\cite{Field1996a}. In our case, we have a positive radial eigenvalue, and the same argument may not hold. However, because of the invariance of the subspace containing the heteroclinic cycle, the radial directions decouple (to lowest order) in the construction of the \Poincare map. Since we are looking for fixed points of the map, rather than computing stability criteria, we can thus find the fixed points, and examine their properties while ignoring the radial direction.}

 In constructing the \Poincare map, we do not explicitly compute the amount of time
$T$ that the trajectory spends close to the equilibria, but leave this as an
unknown defined implicitly in terms of the other coordinates in the map: the
map is defined in terms of three coordinates, and the time $T$.
It then becomes possible to solve the equations for fixed points of the
\Poincare map by writing each of the coordinates in term of $T$, allowing us to
construct a single equation with a single unknown,~$T$. 
Letting $T$ become large will give us the locations of the heteroclinic 
bifurcations.

In the case where the
expanding eigenvalues are real, we are able to find two different types of
solution for large~$T$, depending on which terms in this equation dominate: one
type of solution generates the resonant bifurcation, and the other generates
the orbit-flip bifurcation. In the case where the expanding eigenvalues are
complex, we find only one type of solution, corresponding to a
Belyakov--Devaney\blue{-type} bifurcation. 

The period of the bifurcating periodic orbit scales differently with the
distance from the bifurcation point, depending on the type of bifurcation.
Suppose that $\mu$ is a parameter which measures the distance from the
heteroclinic bifurcation in each of the three cases, then:
 (a) in the resonance bifurcation, $\mu\propto|\lambda_e^-+\lambda_c^-|$, and
$T$ scales like $1/|\mu|$ (see equation~\eref{eq:Tres});
 (b), in the Belyakov--Devaney\blue{-type} bifurcation $\mu\propto|\lambda_e^I|$, and $T$
scales like $1/|\mu|$ (see equation~\eref{eq:TBD}); and
 (c) in the orbit flip bifurcation, $\mu\propto A_3$, a global constant which
determines the angle at which the heteroclinic connection exits a neighbourhood
of $\xia$, and $T$ scales like $\log|\mu|$ (see equation~\eref{eq:a3of}).
 
 In each case, once we have computed an expression for the fixed points of the
\Poincare map, we also check that the corresponding periodic orbits satisfy the
condition that $a,b,c>0$ for all time. The coordinates will need to be checked
when they are close to $\xia$: during the transition between equilibria the
coordinates will be order 1 and hence will not change sign. In a neighbourhood
of $\xia$, it is clear that $a(t)$ will not change sign, as it is order 1. The
heteroclinic connection leaving $\xia$ lies in an invariant subspace which has
$c=0$, so $c$ cannot change sign during the transition from $\xia$ to $\xib$.
Thus the coordinate which will need to be checked is $b(t)$.
 
Finally, we will check whether or not we expect the solution to be `kinked'.

\subsection{Real eigenvalues}

To begin, we define new coordinates which we use when the trajectory is near
$\xia$
 \begin{equation} \label{eq:acoords}
 \xea=\lambda_e^-b-v,\ \yea=\lambda_e^{+} b-v,\ \xca=\lambda_c^-c-w,\ \yca=\lambda_c^+c-w.
 \end{equation}
Recall that we are interested in solutions which have $b(t)>0$, which in these
coordinates, means we must have $\yea>\xea$. The coordinates
in~\eref{eq:acoords} are aligned with the eigenvectors of the Jacobian matrix,
and so the linearised equations near $\xia$ can be written
 \begin{equation}
 \frac{d\xea}{dt}=\lambda_e^+ \xea,\ \frac{d\yea}{dt}=\lambda_e^- \yea,\ \frac{d\xca}{dt}=\lambda_c^+ \xca,\  \frac{d\yca}{dt}=\lambda_c^- \yca. 
 \end{equation}  
We will also make use of polar coordinates in the expanding directions, namely
$\rea$ and $\thea$, defined by
 \begin{equation}
 (\rea)^2=(\xea)^2+(\yea)^2 \quad \mathrm{and} \quad \tan\thea =\frac{\yea}{\xea}. \label{eq:rth}
 \end{equation}
The constraint  $\yea>\xea$ means that $\pi/4<\thea<5\pi/4$. We similarly
define new coordinates for use near $\xib$:
 \begin{equation}
 \xeb=\lambda_e^-c-w,\ \yeb=\lambda_e^{+} c-w,\ \xcb=\lambda_c^-a-u,\ \ycb=\lambda_c^+a-u.
 \end{equation}
We further write $\xa=(\xea,\yea,\xca,\yca)$ and $\xb=(\xeb,\yeb,\xcb,\ycb)$.

We define \Poincare sections, close to $\xia$ and $\xib$:
 \begin{eqnarray*}
 \HAin&=\{\vec{x} | \yca=h\} \\
 \HAout&=\{\vec{x} | \rea=h\} \\
 \HBin &= \{\vec{x} |\ycb=h \}
 \end{eqnarray*}
for some $h\ll 1$.

\begin{figure}
\setlength{\unitlength}{1mm}
\begin{center}
\begin{picture}(73,73)(0,0)

\put(0,0){\includegraphics[trim= 0cm 0cm 0cm 0.cm,clip=true,width=73mm]{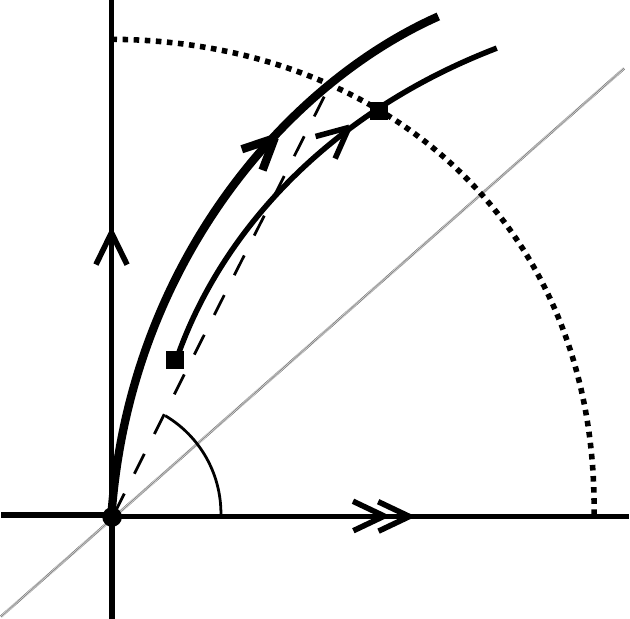}}                             
\put(7,67){{$\yea$}}
\put(70,8){$\xea$}
\put(19.5,14){$\hthea$}
\put(20,61){$\HAout$}
\put(48,58){$\xa(T)$}
\put(23.5,30.5){$\xa(0)$}
\put(17,47){$\gab$}
\put(36,64){$\hxa$}
\put(61,50){$b=0$}

\end{picture}
\end{center}

\caption{The figure shows a schematic of the expanding subspace from $\xia$, in the case when the expanding eigenvalues are real. The bold line indicates the heteroclinic connection $\gab$, and it intersects the \Poincare section $\HAout$ (shown by a dotted curve) at $\hxa$. A trajectory close to the the heteroclinic connection is shown, starting at a point $\xa(0)$ and hitting $\HAout$ at $\xa(T)$ (both points marked with black squares). The grey line indicates where $b=0$; $b>0$ above this line.
\label{fig:HAout}} 
\end{figure}
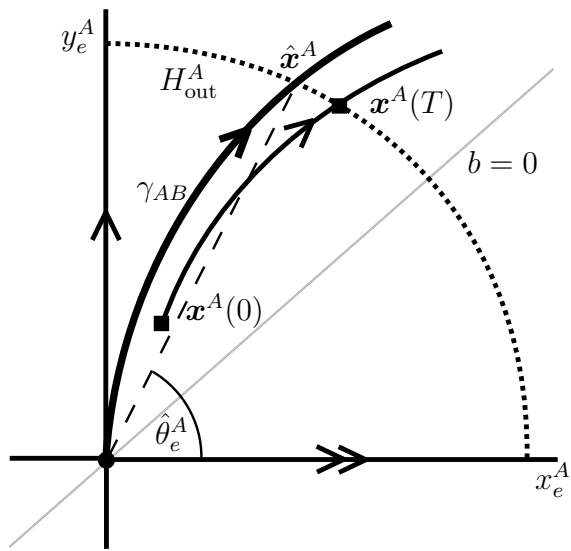

We will now construct a local map near $\xia$ and a global map from $\xia$ to $\xib$ as follows. Let the time it takes the trajectory to travel from $\HAin$ to $\HAout$ be $T$. The local map is
\begin{eqnarray*}
\pil:\HAin \rightarrow \HAout \\
\xa(T)=\pil(\xa(0)) \\
(\xea(T),\yea(T),\xca(T),\yca(T))=\pil(\xea(0),\yea(0),\xca(0),h),
\end{eqnarray*}
where $\xea(T)^2+\yea(T)^2=h^2$, and the global map is
\begin{eqnarray*}
\pig:\HAout \rightarrow \HBin \\
\xb=\pig(\xa(T)) \\
(\xeb,\yeb,\xcb,h)=\pig(\xea(T),\yea(T),\xca(T),\yca(T))
\end{eqnarray*}
where again $\xea(T)^2+\yea(T)^2=h^2$. In figure~\ref{fig:HAout} we show a schematic of the expanding dynamics near $\xia$.

We label the heteroclinic connection between $\xia$ and $\xib$ as $\gab$. Recall that the unstable manifold of $\xia$, $W^u(\xia)$, is four-dimensional. The heteroclinic connection is a one-dimensional sub-manifold of $W^u(\xia)$. In addition, we also
 know that the connection lies in the invariant subspace $P(\xia)$ (which has $c=w=0$, equivalently,  $\xca=\yca=0$ near $\xia$ or $\xeb=\yeb=0$ near $\xib$). 
We consider the points at which the heteroclinic connection intersects the \Poincare sections, and write
 \begin{equation} \label{eq:gabint}
 \gab\cap\HAout=\hxa=(\hxea,\hyea,0,0),\quad \gab\cap\HBin=\hxb=(0,0,0,h)
 \end{equation}
where 
 \begin{equation}
 (\hxea)^2+(\hyea)^2=h, \quad\frac{\hyea}{\hxea}= \tan\hthea.
 \end{equation}
The $\xcb$ coordinate of $\hxb$ is zero because $\gab$ must lie in the stable
manifold of $\xib$, and $\xcb$ is the coordinate associated with the  positive
contracting eigenvalue, $\lambda_c^+$. The angle $\hthea$ is marked in
figure~\ref{fig:HAout}. Note that generically, the heteroclinic connection
$\gab$ will be tangent to the $\yea$ axis at $\xia$, and so generically
$\hthea$ will be order one. In the orbit flip bifurcation which we consider in
section~\ref{sec:orbflip}, the heteroclinic connection is tangent to the strong
stable manifold, i.e., the $\xea$ axis, and then $\hthea$ will be very close to
$\pi$ (so $|\tan\hthea| \ll 1$).

\subsubsection{Local map}
\label{sec:local}

We consider a trajectory which starts at time $t=0$, at a point
$\xa(0)\in\HAin$, and we write the solution to the equations linearised around
$\xia$ as
 \numparts
 \begin{eqnarray}
 \xea(t)&=\xea(0)\e^{\lambda_e^+ t} \label{eq:xealin} \\
 \yea(t)&=\yea(0)\e^{\lambda_e^- t} \label{eq:yealin} \\
 \xca(t)&=\xca(0)\e^{\lambda_c^+ t} \\
 \yca(t)&=h\e^{\lambda_c^- t} 
 \end{eqnarray}
 \endnumparts
The time it takes the trajectory to travel from $\HAin$ to $\HAout$ is $T$, so $\xa(T)\in \HAout$, and $T$ is defined by
\begin{equation}
 \rea(T)^2=\xea(T)^2+\yea(T)^2=h^2.
\end{equation}
 This gives the five equations
 \numparts
\begin{eqnarray}
\xea(T)&=\xea(0)\e^{\lambda_e^+ T} \\
\yea(T)&=\yea(0)\e^{\lambda_e^- T} \\
\xca(T)&=\xca(0)\e^{\lambda_c^+ T} \\
\yca(T)&=h\e^{\lambda_c^- T} \\
 h^2&=\xea(0)^2\e^{2\lambda_e^+ T}+\yea(0)^2\e^{2\lambda_e^- T}
\end{eqnarray}
\endnumparts
which define $\xea(T), \yea(T), \xca(T), \yca(T)$ and (implicitly) $T$ in terms
of $\xea(0), \yea(0)$ and $\xca(0)$, thus defining the local map from $\HAin$
to $\HAout$. Note that we do not attempt to solve for $T$ at this stage.

\subsubsection{Global map}
\label{sec:global}

We next construct the global map from $\HAout$ to $\HBin$. We only consider trajectories which lie close to the heteroclinic connection from $\xia$ to $\xib$, so $\thea(T)$ will be close to $\hthea$ (see figure~\ref{fig:HAout}). Then we write
\begin{equation}
\thea(T)=\arctan\left(\frac{\yea(T)}{\xea(T)}\right)
\end{equation}
and Taylor expand the right hand side around $\hxa$ to get
\begin{eqnarray}
\thea(T)& = \arctan\left(\frac{\hyea+(\yea(T)-\hyea)}{\hxea+(\xea(T)-\hxea)}\right) \nonumber \\
&=\hthea-\frac{\hyea}{({\hxea})^2+({\hyea})^2}(x_e(T)-\hxea) +\frac{\hxea}{({\hxea})^2+({\hyea})^2}(y_e(T)-\hyea) \nonumber \\
& = \hthea-\frac{\hyea}{h^2}x_e(T) +\frac{\hxea}{h^2}y_e(T) \label{eq:thea}
\end{eqnarray}
where we are assuming $(\yea(T)-\hyea)$ and $(\xea(T)-\hxea)$ are small and have used the fact that $({\hxea})^2+({\hyea})^2=h^2$.

Recall that a point on $\HAout$ can be defined by the coordinates $\xca(T),
\yca(T)$ and $\thea(T)$. For a trajectory close to the heteroclinic connection,
$\xca(T)$ and $\yca(T)$ are small (since the heteroclinic connection lies in
$P(\xia)$ which has $\xca=\yca=0$), and  $(\thea(T)-\hthea)$ is also small. A
point on $\HBin$ is defined by the coordinates $\xcb$, $\xeb$ and $\yeb$, which
are also all small for a trajectory close to $\gab$ (see
equation~\eref{eq:gabint}). Thus, in the global map, to first order, $\xcb$,
$\xeb$ and $\yeb$ can be written as a linear combination of $\xca(T)$,
$\yca(T)$ and $(\thea(T)-\hthea)$. In addition, the global map must preserve
the invariance of $P(\xia)$. The global map can thus be written to first order
as:
 \numparts
 \begin{eqnarray}
 \xcb&=F_1 (\thea(T)-\hthea) +F_2 \xca(T) +F_3 \yca(T) \\
 \xeb&=F_4 \xca(T) +F_5 \yca(T) \label{eq:pmapf2} \\
 \yeb& = F_6 \xca(T)+F_7 \yca(T) \label{eq:pmapf3}
 \end{eqnarray}
 \endnumparts
where the $F_j$ are order one constants.

Using equation~\eref{eq:thea}, we replace $\thea(T)$, and renaming the constants gives
\numparts
\begin{eqnarray}
\xcb&=A_1 \xca(0) \e^{\lambda_c^+ T}  +A_2 h \e^{\lambda_c^- T} +A_3 \xea(0) \e^{\lambda_e^+ T} +A_4 \yea(0) \e^{\lambda_e^- T}  \label{eq:pmapab1} \\
\xeb&=B_1 \xca(0) \e^{\lambda_c^+ T}  +B_2 h \e^{\lambda_c^- T} \\
\yeb& = C_1 \xca(0) \e^{\lambda_c^+ T}  +C_2 h \e^{\lambda_c^- T} \label{eq:pmapab3}
\end{eqnarray}
\endnumparts
Note that $A_3=-F_1\hyea/h^2$ and $A_4=F_1\hxea/h^2$, so $\tan\hthea=-A_3/A_4$. 

Usually in these sorts of calculations, it is assumed that the order one
constants which arise in the global map are not functions of the eigenvalues.
This is not entirely true, as they will be dependent on the global dynamics,
but to leading order, if we are only considering small changes of the
eigenvalues (such as near a bifurcation point), then the constants will be
close enough to constant that it doesn't matter. However, in this case, we note
that the constants $B_1$, $B_2$, $C_1$ and $C_2$ do have a degeneracy near a
particular degeneracy of the eigenvalues, arising because of the way we have
defined our coordinates.

Specifically, consider the trajectory of $c$ and $w$ during the passage from $\HAout$ to $\HBin$. Both $c$ and $w$ are assumed small, and write $(c_A,w_A)$ for the coordinates on $\HAout$ and $(c_B,w_B)$ for the coordinates on $\HBin$. Then to lowest order, the global map can be written
 \begin{equation}
 \left( \begin{array}{l}
 c_B \\ w_B
 \end{array} \right)
 =
 \left( \begin{array}{ll}
  G_1 & G_2 \\ G_3 & G_4 
 \end{array} \right)
 \left( \begin{array}{l}
 c_A \\ w_A
 \end{array} \right)
 \end{equation}
where the $G_j$ are indeed generically order one constants. When we rewrite this in terms of $\xeb$, $\yeb$, $\xca$ and $\xcb$, we have
 \begin{equation}
 \left( \begin{array}{l}
 \xeb \\ \yeb
 \end{array} \right)
 =
 \left( \begin{array}{ll}
 \lambda_e^- & -1 \\ \lambda_e^+ & -1 
 \end{array} \right)
 \left( \begin{array}{ll}
  G_1 & G_2 \\ G_3 & G_4 
 \end{array} \right)
 \left( \begin{array}{ll}
 \lambda_c^- & -1 \\ \lambda_c^+ & -1 
 \end{array} \right)^{-1}
 \left( \begin{array}{l}
 \xca(T) \\ \yca(T)
 \end{array} \right)
 \end{equation}
That is, (referring to~\eref{eq:pmapf2} and~\eref{eq:pmapf3})
 \begin{equation}
 \left( \begin{array}{ll}
  B_1 & B_2 \\ C_1 & C_2 
 \end{array} \right)
 =
 \left( \begin{array}{ll}
 \lambda_e^- & -1 \\ \lambda_e^+ & -1 
 \end{array} \right)
 \left( \begin{array}{ll}
 G_1 & G_2 \\ G_3 & G_4 
 \end{array} \right)
 \left( \begin{array}{ll}
 \lambda_c^- & -1 \\ \lambda_c^+ & -1 
 \end{array} \right)^{-1}
 \label{eq:realevscoordtransform}
 \end{equation}
There are thus degeneracies in $B_1$, $B_2$, $C_1$ and $C_2$ when either
$\lambda_c^-=\lambda_c^+$ or $\lambda_e^-=\lambda_e^+$. The former case doesn't
occur in our system, because we assume that $\sigma,\zeta>0$ (see
table~\ref{tab:evals}), but the latter can occur, when $4\zeta=\gamma^2$: where
the expanding eigenvalues change from being real to complex. In this case, when
$\lambda_e^-=\lambda_e^+$, then $B_1=C_1$, and $B_2=C_2$, and the determinant 
of the matrix on the left hand side of~\eref{eq:realevscoordtransform} is
$\Delta_{BC}=B_1C_2-C_1B_2=0$. \blue{We assume in this section that we are away from the point where the expanding eigenvalues are equal. In section~\ref{sec:complex}, we consider the case where the expanding eigenvalues are complex, but use a coordinate change which limits to the repeating eigenvalues case when the imaginary part of the complex pair vanishes.}

\subsubsection{Fixed point of the \Poincare map}

Equations~\eref{eq:pmapab1} to~\eref{eq:pmapab3} map a point on $\HAin$ to a
point on $\HBin$. Due to the symmetry $g$ in equation~\eref{eq:odes}, a fixed
point of a full \Poincare return map will also be a fixed point
of~\eref{eq:pmapab1} to~\eref{eq:pmapab3}. Fixed points of a full \Poincare
return map can thus be found by dropping the $A$ and $B$ superscripts, and the
dependence on $0$ on the right hand side, to give the following four nonlinear
equations, with four unknowns, $x_c$, $x_e$, $y_e$ and $T$:
 \numparts
 \begin{eqnarray}
 x_c&=A_1 x_c \e^{\lambda_c^+ T}  +A_2 h \e^{\lambda_c^- T} +A_3 x_e \e^{\lambda_e^+ T} +A_4 y_e \e^{\lambda_e^- T} \label{eq:fpxc} \\
 x_e&=B_1 x_c \e^{\lambda_c^+ T}  +B_2 h \e^{\lambda_c^- T}  \label{eq:fpxe} \\
 y_e& = C_1 x_c \e^{\lambda_c^+ T}  +C_2 h \e^{\lambda_c^- T}  \label{eq:fpye} \\
 h^2&=x_e^2\e^{2\lambda_e^- T}+y_e^2\e^{2\lambda_e^+ T}  \label{eq:fpT}
 \end{eqnarray}
 \endnumparts
We substitute equations~\eref{eq:fpxe} and~\eref{eq:fpye} into~\eref{eq:fpxc}
to eliminate $x_e$ and $y_e$, which upon rearranging gives:
 \begin{equation}
 \fl x_c(1- \e^{\lambda_c^+ T}( A_1  +A_3 B_1  \e^{ \lambda_e^+ T}  +A_4  C_1  \e^{ \lambda_e^- T} ))= h\e^{\lambda_c^- T}( A_2   +A_3 B_2 \e^{ \lambda_e^+ T}  +A_4 C_2  \e^{\lambda_e^- T})
 \end{equation}
Recall that $\lambda_c^+>0>\lambda_c^-$ and  $\lambda_e^+> \lambda_e^->0$.
Since $T$ is assumed to be large, and is certainly positive, we can neglect the
first ($1$) and second ($A_1$) terms on the left hand side, and the first
($A_2$) term on the right hand side, to get:
 \begin{equation}
 \label{eq:xc}
 x_c = -h \e^{(\lambda_c^- -\lambda_c^+ )T}
   \frac{A_3 B_2 \e^{ \lambda_e^+ T}  +A_4 C_2  \e^{\lambda_e^-T}}
        {A_3 B_1 \e^{ \lambda_e^+ T}  +A_4  C_1  \e^{ \lambda_e^- T} }
 \end{equation}


We next substitute~\eref{eq:xc} into~\eref{eq:fpxe} and~\eref{eq:fpye} and then finally into the expression for $T$~\eref{eq:fpT}, which we will then solve for $T$.
This gives us:
\numparts
\begin{eqnarray}
x_e&=B_1 x_c \e^{\lambda_c^+ T}  +B_2 h \e^{\lambda_c^- T} \nonumber \\
 & = - B_1h\e^{\lambda_c^-  T}\frac{ A_3 B_2 \e^{ \lambda_e^+ T}  +A_4 C_2  \e^{\lambda_e^- T}}{A_3 B_1  \e^{ \lambda_e^+ T}  +A_4  C_1  \e^{ \lambda_e^- T} }
+B_2 h \e^{\lambda_c^- T} \label{eq:xe_2} \\
& = -hA_4\Delta_{BC}  \left(  \frac{   \e^{ (\lambda_e^- + \lambda_c^-) T}  }{A_3 B_1  \e^{ \lambda_e^+ T}  +A_4  C_1  \e^{ \lambda_e^- T} }   \right) \label{eq:xe}
\end{eqnarray}
\endnumparts
where $\Delta_{BC}=B_1C_2-C_1B_2$, and 
\numparts
\begin{eqnarray}
y_e&=C_1 x_c \e^{\lambda_c^+ T}  +C_2 h \e^{\lambda_c^- T}  \nonumber \\
 & = - C_1h\e^{\lambda_c^-  T}\frac{ A_3 B_2 \e^{ \lambda_e^+ T}  +A_4 C_2  \e^{\lambda_e^- T}}{A_3 B_1  \e^{ \lambda_e^+ T}  +A_4  C_1  \e^{ \lambda_e^- T} }
+C_2 h \e^{\lambda_c^- T}   \label{eq:ye_2} \\
& = hA_3\Delta_{BC}  \left(  \frac{\e^{( \lambda_e^+ +\lambda_c^-) T}    }{A_3 B_1  \e^{ \lambda_e^+ T}  +A_4  C_1  \e^{ \lambda_e^- T} }   \right) \label{eq:ye}
\end{eqnarray}
\endnumparts
Note that when simplifying~\eref{eq:xe_2} to get~\eref{eq:xe}
and~\eref{eq:ye_2} to get~\eref{eq:ye}, terms in the numerator in
$\e^{(\lambda_e^+ +\lambda_c^-)T}$ and $\e^{(\lambda_e^- +\lambda_c^-)T}$,
respectively, cancel out.

We substitute~\eref{eq:xe} and~\eref{eq:ye} into~\eref{eq:fpT} to get:
\begin{eqnarray}
 h^2=x_e^2\e^{2\lambda_e^+ T}+y_e^2\e^{2\lambda_e^- T}  \nonumber \\
1=|\Delta_{BC}|\sqrt{A_4^2+A_3^2}\frac{ \e^{ ( \lambda_e^- + \lambda_e^++\lambda_c^-) T }}{ A_3 B_1  \e^{ \lambda_e^+ T}  +A_4  C_1  \e^{ \lambda_e^- T}} \nonumber \\
A_3 B_1  \e^{ \lambda_e^+ T}  +A_4  C_1  \e^{ \lambda_e^- T}
=|\Delta_{BC}|\sqrt{A_4^2+A_3^2} \e^{ ( \lambda_e^- + \lambda_e^++\lambda_c^-) T }  \label{eq:T}
\end{eqnarray}
The final task is to solve~\eref{eq:T}, which gives the period  of a periodic
orbit in the flow (the actual period is~$3T$), close to
the heteroclinic cycle, which corresponds to a fixed point in the map. 
For large~$T$, the periodic orbit will be close to the heteroclinic cycle.
We will
do this in two different cases in sections~\ref{sec:res} and~\ref{sec:orbflip}.

Note that the left hand side of equation~\eref{eq:T} is the denominator in the equations for $x_c$, $x_e$ and $y_e$ (equations~\eref{eq:xc},~\eref{eq:xe} and~\eref{eq:ye} respectively) so we substitute~\eref{eq:T} into these equations to  simplify them, to get
\numparts
\begin{eqnarray}
x_c & = - h\e^{ -\lambda_c^+ T}\frac{ A_3 B_2 \e^{ -\lambda_e^- T}  +A_4 C_2  \e^{-\lambda_e^+ T}}{|\Delta_{BC}|\sqrt{A_4^2+A_3^2}  } \label{eq:xcr} \\
x_e & = -h\frac{A_4\mathrm{sgn}(\Delta_{BC} ) }{\sqrt{A_4^2+A_3^2} } \e^{- \lambda_e^+T  } \label{eq:xer} \\
y_e & = h\frac{A_3\mathrm{sgn}(\Delta_{BC} ) }{\sqrt{A_4^2+A_3^2} } \e^{- \lambda_e^- T  } \label{eq:yer}
\end{eqnarray}
\endnumparts
These three equations give the coordinates of the fixed point in terms of $T$.
Note that in all three co-ordinates, the coefficient(s) of $T$ in the exponential is (are) negative, meaning the coordinates (of the fixed point) get smaller as $T$ gets larger, as would be expected.

We now check that $b(t)>0$ for all $t\in[0,T]$ for these solutions, namely that 
the periodic orbit corresponds to a positive travelling wave solution 
of~\eref{eq:RPS_PDEs}. Note
from~\eref{eq:xealin} and~\eref{eq:yealin} that $\xea(t)$ and $\yea(t)$ do not
change sign, and in order to have $b(t)>0$ we require that $\yea(t)>\xea(t)$
for all $t\in[0,T]$. Writing~\eref{eq:xealin} and~\eref{eq:yealin} with initial
conditions from~\eref{eq:xer} and~\eref{eq:yer} gives us
 \begin{eqnarray}
 \xea(t)= - A_4 E_1 \e^{- \lambda_e^+(T-t)  } \\
 \yea(t)= A_3 E_1 \e^{- \lambda_e^- (T-t)  }
 \end{eqnarray}
where $E_1=h\frac{\mathrm{sgn}(\Delta_{BC} ) }{\sqrt{A_4^2+A_3^2} }$, and so we require
\begin{equation} \label{eq:b0t}
A_3 E_1 \e^{- \lambda_e^- (T-t)}>- A_4 E_1 \e^{- \lambda_e^+(T-t)}
\end{equation}

There are four cases to consider depending on the signs of $A_4E_1$ and
$A_3E_1$, and the corresponding quadrant in $\xea$-$\yea$ space in which the
solutions lie. Since we are only considering solutions that lie close to the
heteroclinic connection, we assume in each case that the solutions $\xea(t)$
and $\yea(t)$ lie in the same quadrant as~$\hthea$ .

If $A_4E_1,A_3E_1>0$, then $\yea(t)>0$ and $\xea(t)<0$ and we are done. If
$A_3E_1,A_4E_1<0$, then $\yea(t)>0$ and $\xea(t)<0$ so solutions have $b(t)<0$
for all $t$ and this is not a positive travelling wave. If $A_3E_1<0<A_4E_1$,
then $\xea,\yea<0$ and~\eref{eq:b0t} gives us:
 \begin{equation}
 -\frac{A_3}{A_4}=\tan\hthea< \e^{ (\lambda_e^--\lambda_e^+)(T-t)}
 \end{equation}
Note that the left-hand side of the inequality is positive and the
right-hand side is 
between $0$ and~$1$, and so we require $0<\tan\hthea<1$ for the solution to be 
positive. 
Since $\xea,\yea<0$, putting these together means that $\pi<\hthea<5\pi/4$.
Furthermore, solutions must satisfy 
 \begin{equation}
 T<\frac{-1}{\lambda_e^+ -\lambda_e^-} \log(\tan\hthea)
 \end{equation}
which implies that $\lambda_e^+ -\lambda_e^-$ must decrease to~$0$ as 
$T\rightarrow\infty$, as the heteroclinic bifurcation is approached.

Finally, suppose  $A_4E_1<0<A_3E_1$, then $\xea,\yea>0$ and~\eref{eq:b0t}
gives:
 \begin{equation} \label{eq:b_xyg0}
 -\frac{A_3}{A_4}=\tan\hthea> \e^{ (\lambda_e^--\lambda_e^+)(T-t)}
 \end{equation}
Again, the left-hand side of the inequality is positive and the
right-hand side is 
between $0$ and~$1$, so $\tan\hthea>1$. With 
$\xea,\yea>0$, this means that $\pi/4<\hthea<\pi/2$.

In summary, solutions will have $b(t)>0$ for all $t$ if the heteroclinic
connection is such that $\pi/4<\hthea<\pi$. If $\pi<\hthea<5\pi/4$, then we can
also find solutions with large $T$ with $b(t)>0$, so long as $\lambda_e^+-\lambda_e^-$ decreases
to zero as $T$ tends to infinity. For other values of $\hthea$, periodic solutions close to the
heteroclinic cycle will not correspond to positive travelling wave solutions
of the PDEs~\eref{eq:RPS_PDEs}.
 
 In the following two sections, we consider two different cases depending on the
relative size of the the two terms on the left-hand side of
equation~\eref{eq:T}.

\subsubsection{Resonant bifurcation at $\lambda_c^-+\lambda_e^-=0$}
\label{sec:res}

In this section, we will show that a resonant-type heteroclinic bifurcation
occurs when $\lambda_c^-+\lambda_e^-=0$.

Suppose that $A_3 B_1  \e^{ \lambda_e^+ T} \gg A_4  C_1  \e^{ \lambda_e^- T}$. This will be the case if $A_3, B_1, A_4$ and $C_1$ are order 1, since $T$ is large and $\lambda_e^+>\lambda_e^-$. Then equation~\eref{eq:T} simplifies to
\begin{equation}
1=\frac{|\Delta_{BC}|\sqrt{A_4^2+A_3^2}}{A_3B_1}  \e^{ (\lambda_c^- + \lambda_e^- ) T }
\end{equation}
or 
\begin{equation} \label{eq:Tres}
T=\frac{1}{\lambda_c^- + \lambda_e^-}\log(D_1)
\end{equation}
for $D_1=\frac{A_3B_1}{|\Delta_{BC}|\sqrt{A_4^2+A_3^2}}$. If $D_1<1$, then we see a branch of long-period periodic orbits emerging from the curve $\lambda_c^- + \lambda_e^-=0$ into the region where $\lambda_c^- + \lambda_e^-<0$. If $D_1>1$ then the solutions branch into $\lambda_c^- + \lambda_e^->0$. This bifurcation curve can be seen in figure~\ref{fig:pspace}, where the black curve of heteroclinic bifurcations coincides with the light blue curve at $-\lambda_c^-=\lambda_e^-$.
At this fixed point, taking the leading order term for $x_c$ in~\eref{eq:xcr} gives
\begin{equation}
x_c  = - h\frac{ A_3 B_2  }{|\Delta_{BC}|\sqrt{A_4^2+A_3^2}  } \e^{ -(\lambda_e^- +\lambda_c^+) T}
\end{equation}

This resonant bifurcation is unusual: usually you expect to see a resonant
bifurcation when the contracting eigenvalue is equal to the leading expanding
eigenvalue, that is, when $-\lambda_c^-=\lambda_e^+$~\cite{Krupa1995},
but here it is $-\lambda_c^-=\lambda_e^-$.

Numerical simulations of periodic orbits close to the resonance bifurcation
indicate that $x_e$ and $y_e$ are both positive ($\hthea\approx\pi/2$), and so
from~\eref{eq:b_xyg0}, we must have $\hthea>\pi/4$ in order for solutions to
have $b(t)>0$ for all $t$. Indeed, this is what we see in the numerical
simulations.

We next assess whether we expect to see a `kink' in the shape of the profile of
the long-period solutions as the bifurcation point is approached. As can be
seen in the time-series plots in figure~\ref{fig:tseries}, a kink is observed
when there is a period of time during which the solution grows exponentially
with rate $\lambda_c^+$. When the trajectory is near $\xia$, the contracting
components are $c$ and $w$, which are linear combinations of $\xca$ and $\yca$,
which  grow/decay exponentially at rates $\lambda_c^+$ and $\lambda_c^-$
respectively. Observing a kink corresponds to having $\xca(t)>\yca(t)$ for some
range of time $t$. Since $\yca$ is decaying and $\xca$ is growing, we will
observe a kink if $|\yca(T)|\ll |\xca(T)|$. We have that
 \begin{eqnarray}
 \xca(T)&=x_c \e^{\lambda_c^+ T} = - h\frac{ A_3 B_2  }{|\Delta_{BC}|\sqrt{A_4^2+A_3^2}  } \e^{ -\lambda_e^-  T}, \\
 \yca(T)&=h \e^{\lambda_c^- T}. 
 \end{eqnarray}
At the resonant bifurcation, $\lambda_c^-=-\lambda_e^-$ and so $\xca(T)$ and
$\yca(T)$ are the same order and so a kink won't be observed in solutions. This
is indeed what is observed, see panel (a) of figure~\ref{fig:tseries}.

In summary, we expect to find a resonant heteroclinic bifurcation with 
$-\lambda_c^-=\lambda_e^-$, that is, on the blue line in figure~\ref{fig:pspace}
with $\zeta>\sigma/2$, at the boundary between regions~1 and~2.

\subsubsection{Orbit flip bifurcation at $A_3=0$}
\label{sec:orbflip}

In this section we show that a branch of long-period periodic orbits can emerge
when the heteroclinic cycle undergoes an orbit flip: that is, in the case when
the heteroclinic connection is tangent to the strong unstable manifold. Recall
that $\hthea$ gives the position at which the heteroclinic connection
intersects $\HAout$. We have that $\tan(\hthea)=-A_3/A_4$, and so as $A_3$ goes
to zero, $\hthea$ goes to~$\pi$, which corresponds to the heteroclinic
connection being tangent to the strong unstable manifold (the $\xea$ axis; see
figure~\ref{fig:HAout}), that is, a point of heteroclinic orbit flip.

We suppose that $A_3$ is small enough that the two terms on the right hand side
of~\eref{eq:T} are of the same order, that is, neither can be discarded. We
then rewrite equation~\eref{eq:T} as
 \begin{equation}\label{eq:a3of}
 A_3=-\frac{A_4  C_1 }{B_1} \e^{ (\lambda_e^--  \lambda_e^+ )T} +\frac{|\Delta_{BC}|A_4}{B_1}\e^{ (\lambda_e^- + \lambda_c^-) T }
 \end{equation}
where we have assumed $A_3\ll A_4$ and so it can be dropped from the square root. Note that $A_3$ will only be small if the expressions in both exponentials are negative, namely if $\lambda_c^-+\lambda_e^-<0$, and then as $T$ goes to infinity, $A_3$ goes to zero.
 This holds in regions 1 and 5 of figure~\ref{fig:pspace}.

For fixed points in this case, we find the leading order term in $x_c$ to be
\begin{eqnarray}
x_c& = - h\e^{ -\lambda_c^+ T}\frac{ A_3 B_2 \e^{ -\lambda_e^- T}  +A_4 C_2  \e^{-\lambda_e^+ T}}{|\Delta_{BC}|\sqrt{A_4^2+A_3^2}  }  \\
& = \frac{h\e^{ -\lambda_c^+ T}}{B_1}\left(\mathrm{sgn}(\Delta_{BC})  \e^{-\lambda_e^+ T} +B_2  \e^{\lambda_c^- T}  \right)
\end{eqnarray}

Numerical simulations of periodic orbits close to the heteroclinic orbit flip
bifurcation indicate that $x_e<0<y_e$, $\hthea$ is very close to (but just less
than) $\pi$, and so we automatically satisfy the condition that $b(t)$ remains
positive for all time.

To get a kinked solution, we again require that $|\yca(T)|\ll|\xca(T)|$. For
solutions which start at the fixed point, this gives
 \begin{eqnarray}
 \xca(T)&=x_c \e^{\lambda_c^+ T} = \frac{h}{B_1}\left(\mathrm{sgn}(\Delta_{BC})  \e^{-\lambda_e^+ T} +B_2  \e^{\lambda_c^- T}  \right)  \\
 \yca(T)&=h \e^{\lambda_c^- T} 
 \end{eqnarray}
If $\lambda_c^-<-\lambda_e^+$, then $|\yca(T)|<|\xca(T)|$, and we will see a
kinked solution. If $\lambda_c^->-\lambda_e^+$, then $|\yca(T)|$ and
$|\xca(T)|$ will be of the same order, and we will not observe a kink. However, we note that in order for solutions in this region to expand as much as they contract, we would instead observe a kink in the expanding phase, that is, a change in growth rate from $\lambda_e^-$ to $\lambda_e^+$.

 The
location of the orbit flip (if it exists at all) is determined by the global
dynamics (that is, it can not be predicted by the eigenvalues). For
equations~\eref{eq:odes}, we find the location of the orbit flip numerically,
by solving a boundary value problem to locate the heteroclinic orbit between
$\xia$ and $\xib$, and insisting that the heteroclinic orbit is tangent to the
strong unstable manifold at $\xia$. The location of the orbit flip is shown by a
green curve in region 5 of figure~\ref{fig:pspace}. This green curve coincides
with the black curve of heteroclinic bifurcations. In region 5,
$\lambda_c^-<-\lambda_e^+$, and the periodic orbits close to this heteroclinic
bifurcation do indeed show a kinked solution --- see panel (c) of
figure~\ref{fig:tseries}. We note that the orbit flip curve terminates on the
curve where $\lambda_e^-=\lambda_e^+$ (the red curve in
figure~\ref{fig:pspace}), which is to be expected, as equation~\eref{eq:a3of}
clearly does not generate large $T$ solutions at this point.

\subsubsection{Saddle-node bifurcation of periodic orbits}
\label{sec:snpo}

Equation~\eref{eq:a3of} gives the possibility of a saddle-node bifurcation
between periodic orbits near the orbit-flip bifurcation. We compute:
 \begin{equation}
 \frac{d A_3}{dT}=-(\lambda_e^--  \lambda_e^+ )\frac{A_4  C_1 }{B_1} \e^{ (\lambda_e^--  \lambda_e^+ )T} + (\lambda_e^- + \lambda_c^-)\frac{|\Delta_{BC}|A_4}{B_1}\e^{ (\lambda_e^- + \lambda_c^-) T }
 \end{equation}
and set $\frac{d A_3}{dT}=0$ to find
 \begin{equation} \label{eq:Tsn}
 \frac{  C_1 (\lambda_e^--  \lambda_e^+ )}{|\Delta_{BC}|(\lambda_e^- + \lambda_c^-)}=\e^{ (\lambda_e^+ + \lambda_c^-) T }
 \end{equation}
giving a branch of saddle-node bifurcations of periodic orbits at
 \begin{equation}\label{eq:a3sn}
 \fl A_3=-\frac{A_4  C_1 }{B_1} \left(\frac{  C_1 (\lambda_e^--  \lambda_e^+ )}{|\Delta_{BC}|(\lambda_e^- + \lambda_c^-)}\right)^{ \frac{(\lambda_e^--  \lambda_e^+ )}{(\lambda_e^+ + \lambda_c^-)}} +\frac{|\Delta_{BC}|A_4}{B_1}\left(\frac{  C_1 (\lambda_e^--  \lambda_e^+ )}{|\Delta_{BC}|(\lambda_e^- + \lambda_c^-)}\right)^{ \frac{(\lambda_e^-+  \lambda_c^- )}{(\lambda_c^- + \lambda_e^+)}}
 \end{equation}

Recall that the orbit flip bifurcations only occur if
$\lambda_c^-+\lambda_e^-<0$, so for the left hand side of
equation~\eref{eq:Tsn} to be positive, we require $C_1>0$. The branch of
saddle-node bifurcations can terminate in the branch of orbit flip bifurcations
if the right hand side of~\eref{eq:a3sn} becomes equal to zero. This can happen
in a number of different ways, for instance, by the eigenvalue
condition $-\lambda_c^-=\lambda_e^+$, or if one of the constants $A_4$ or $C_1$
become equal to zero. In figure~\ref{fig:pspace} it appears that the first of
these does not happen, and since $A_4$ and $C_1$ do not depend on the
eigenvalues in an obvious way, we cannot say for sure what happens at the end
of the branch of saddle-node bifurcations.

\subsection{Complex eigenvalues}
\label{sec:complex}

\begin{figure}
\setlength{\unitlength}{1mm}
\begin{center}
\begin{picture}(73,73)(0,0)

\put(0,0){\includegraphics[trim= 0cm 0cm 0cm 0.cm,clip=true,width=73mm]{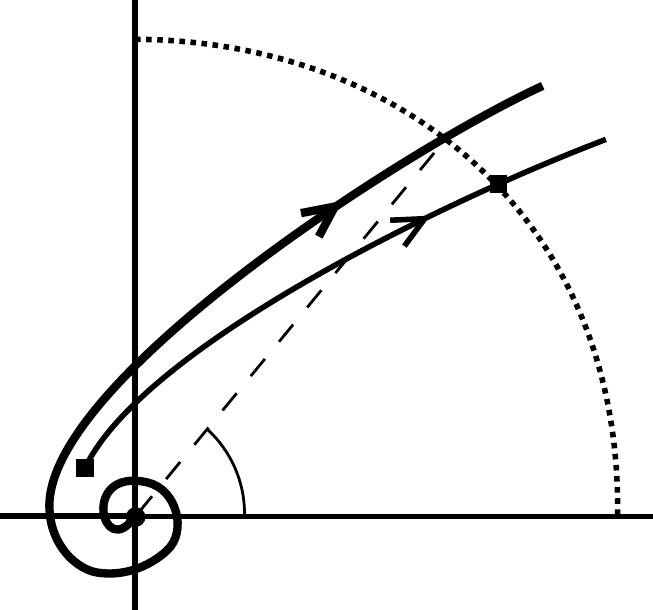}}                             
\put(71,6){{$\xea$}}
\put(10,68){$\yea$}
\put(21,13){$\hthea$}
\put(58,27){$\HAout$}
\put(0,23){$\xa(0)$}
\put(59,46){$\xa(T)$}
\put(22,40){$\gab$}
\put(48,55){$\hxa$}


\end{picture}
\end{center}

\caption{The figure shows a schematic of the expanding subspace from $\xia$, \blue{in the case where the expanding eigenvalues are complex}. 
The bold line is the heteroclinic connection $\gab$, and it intersects the
\Poincare section $\HAout$ (shown by a dotted curve) at $\hxa$. A trajectory
close to the the heteroclinic connection is shown, starting at a point $\xa(0)$
and hitting $\HAout$ at $\xa(T)$ \blue{(both start and end points are marked by squares)}.
Note that we have positive travelling wave solutions to~\eref{eq:RPS_PDEs} 
($b>0$) when \blue{$\yea>0$}.}
 \label{fig:HAoutc}
\end{figure}

We now repeat the \Poincare map calculations in the region where the expanding
eigenvalues are complex (regions~3 and~4). We make a different change
of coordinates near $\xia$, and instead write
 \begin{equation}
 \xea=\lambda_e^R b-v,\ \yea= b,\ \xca=\lambda_c^-c-w,\ \yca=\lambda_c^+c-w.
 \end{equation}
 
 In the new $\xea, \yea$ coordinates, the local part of the flow becomes
 \[
 \frac{d}{dt}\vectortwo{\xea}{\yea}  = \matrixtwo{\lambda_e^R}{(\lambda_e^I)^2}{-1}{\lambda_e^R} \vectortwo{\xea}{\yea}.
 \]
 Note that in the limit as $\lambda_e^I\rightarrow 0$, the Jordan form of the linear part here becomes what one would use in the case of repeated eigenvalues. The solution to the local flow is
 \begin{eqnarray*}
 \xea(t)&=\e^{\lambda_e^R t}\left( \xea(0)\cos(\lambda_e^I t)+  \yea(0)\lambda_e^I\sin(\lambda_e^I t)\right) \\
 \yea(t)& = \e^{\lambda_e^R t}\left(- \xea(0)\frac{\sin(\lambda_e^I t)}{\lambda_e^I}+\yea(0)\cos(\lambda_e^I t)\right)
 \end{eqnarray*}
 We note again that in the limit $\lambda_e^I\rightarrow 0$, these solutions are exactly those that one would expect for the case with two repeated eigenvalues (in particular, the term $\sin(\lambda_e^I t)/\lambda_e^I$ limits to $t$).
 
We also define new coordinates for use near $\xib$:
 \begin{equation}
\xeb=\lambda_e^Rc-w,\
\yeb=c,\
\xcb=\lambda_c^-a-u,\
\ycb=\lambda_c^+a-u.
\end{equation}
We define $\rea$ and $\thea$ as before, as in equation~\eref{eq:rth}.

We use the same \Poincare sections, close to $\xia$ and $\xib$:
\begin{eqnarray*}
\HAin&=\{\vec{x} | \yca=h\} \\
\HAout&=\{\vec{x} | \rea=h\} \\
\HBin &= \{\vec{x} |\ycb=h \}
\end{eqnarray*}
for some $h\ll 1$.

The solution to equations~\eref{eq:odes} linearised around $\xia$ is now
\begin{eqnarray*}
\xea(t)&=\e^{\lambda_e^R t}\left( \xea(0)\cos(\lambda_e^I t)+  \yea(0)\lambda_e^I\sin(\lambda_e^I t)\right) \\
\yea(t)& = \e^{\lambda_e^R t}\left(- \xea(0)\frac{\sin(\lambda_e^I t)}{\lambda_e^I}+\yea(0)\cos(\lambda_e^I t)\right)\\
\xca(t)&=\xca(0)\e^{\lambda_c^+ t} \\
\yca(t)&=\yca(0)\e^{\lambda_c^- t} 
\end{eqnarray*}
the local map then gives us
\begin{eqnarray}
\xea(T)&=\e^{\lambda_e^R T}\left( \xea(0)\cos(\lambda_e^I T)+  \yea(0)\lambda_e^I\sin(\lambda_e^I T)\right) \label{eq:locc1} \\
\yea(T)& = \e^{\lambda_e^R T}\left(- \xea(0)\frac{\sin(\lambda_e^I T)}{\lambda_e^I}+\yea(0)\cos(\lambda_e^I T)\right)  \label{eq:locc2}  \\
\xca(T)&=\xca(0)\e^{\lambda_c^+ T}  \label{eq:locc3}   \\
\yca(T)&=h\e^{\lambda_c^- T}   \label{eq:locc4} 
\end{eqnarray}
where $T$ is again defined by
\[
\xea(T)^2+\yea(T)^2=h^2.
\]

We now note that in these coordinates, to ensure that $b(t)>0$ for all~$t$, we will require that
$\yea>0$ for all~$t\in (0,T)$. We can write
\begin{eqnarray}
b(t)& =  \e^{\lambda_e^R t}\left(-\xea(0) \frac{\sin(\lambda_e^I t)}{\lambda_e^I}+\yea(0) \cos(\lambda_e^I t)\right)  \label{eq:btc2} \\
& = K \e^{\lambda_e^R t}\sin(\lambda_e^I t+\phi)
\end{eqnarray}
where
\[ K^2=\frac{\xea(0)^2}{{\lambda_e^I}^2}+\yea(0)^2,\quad\mathrm{and}\quad \tan\phi=-\lambda_e^I\frac{\yea(0)}{\xea(0)}
\]

Thus, in order for $b(t)$ to remain positive for all $t\in[0,T]$, a clear upper bound on $\lambda_e^I T$ is $\pi$ (because the $\sin$ changes sign with frequency $\pi$). Thus, $\lambda_e^I<\pi/T$, and since we are interested in solutions for which $T$ is large, $\lambda_e^I$ will be small.

The global part of the map doesn't change, that is, we still have
\begin{eqnarray}
\xcb&=F_1 (\thea(T)-\hthea) +F_2 \xca(T) +F_3 \yca(T) \nonumber \\
\xeb&=F_4 \xca(T) +F_5 \yca(T) \label{eq:gloc} \\
\yeb& = F_6 \xca(T)+F_7 \yca(T) \nonumber
\end{eqnarray}
for some order one constants $F_j$.

Using~\eref{eq:thea}, we can again write
\begin{eqnarray}
 F_1 (\thea(T)-\hthea)&= A_3x_e(T) +A_4y_e(T) \nonumber 
\end{eqnarray}
where $A_3$ and $A_4$ have the same \purple{expression} as in the real part, namely $\tan\hthea=-A_3/A_4$ \purple{(although note that the values are different because the angle $\hthea$ is defined differently because of the different coordinate transformations made)}.

Substituting in for the right-hand side, we get
\purple{
\begin{eqnarray}
\fl F_1 (\thea(T)-\hthea)& = \e^{\lambda_e^R T}  \left(A_3\left( \xea(0)\cos(\lambda_e^I T)+  \yea(0)\lambda_e^I\sin(\lambda_e^I T)\right)
\right. \nonumber \\
& \quad \left.+ A_4\left(- \xea(0)\frac{\sin(\lambda_e^I T)}{\lambda_e^I}+\yea(0)\cos(\lambda_e^I T)\right)   \right) 
\label{eq:theac}
\end{eqnarray}
}

Putting the global map~\eref{eq:gloc} together with the local map~\eref{eq:locc1} to~\eref{eq:locc4}, using~\eref{eq:theac}, renaming the constants, and finallydropping the superscripts and the dependence on $0$, gives us the following equations for the fixed points:
\purple{
\begin{eqnarray}
\fl x_c=A_1 x_c \e^{\lambda_c^+ T}  +A_2 h \e^{\lambda_c^- T} +  \e^{\lambda_e^R T} \left(x_e \left(A_3\cos(\lambda_e^I T)-\frac{A_4}{\lambda_e^I}\sin(\lambda_e^I T)  \right) \right.  \nonumber \\ \quad+
\left. y_e \left(A_3 \lambda_e^I\sin(\lambda_e^I T)+A_4  \cos(\lambda_e^I T) \right)   \right) 
 \label{eq:xcc} \\
\fl x_e=B_1 x_c \e^{\lambda_c^+ T}  +B_2 h \e^{\lambda_c^- T} \label{eq:xec}  \\
\fl y_e = C_1 x_c \e^{\lambda_c^+ T}  +C_2 h \e^{\lambda_c^- T} \label{eq:yec} \\
\fl h^2=(x_e^2+y_e^2)\e^{2\lambda_e^R T} \label{eq:Tc}
\end{eqnarray}
}
There are again four unknowns, $x_c$, $x_e$, $y_e$ and $T$.

As in the case for real expanding eigenvalues, we again consider the magnitudes of the constants $B_1$, $B_2$, $C_1$, and $C_2$. Here, we find
\begin{equation}
\left( \begin{array}{ll}
 B_1 & B_2 \\ C_1 & C_2 
\end{array} \right)
=
\left( \begin{array}{ll}
\lambda_e^R & -1 \\ 1 & 0 
\end{array} \right)
\left( \begin{array}{ll}
 G_1 & G_2 \\ G_3 & G_4 
\end{array} \right)
\left( \begin{array}{ll}
\lambda_c^- & -1 \\ \lambda_c^+ & -1 
\end{array} \right)^{-1}
\end{equation}
So, in this case, there are no degeneracies in these constants.

We now continue to find the fixed points of the return map.  \purple{As noted earlier, we are interested in the limit when $T$ is large and hence $\lambda_e^I$ is small. To make the notation clear, we write $\epsilon=\lambda_e^I$. Recall that $0<\epsilon T<\pi$, and in particular, we make the ansatz
\[
\epsilon T= \pi -K\epsilon+O(\epsilon^2)
\]
for some order-one unknown $K$. We demonstrate below that this ansatz is correct. We can then write
\[
\sin\epsilon T = K\epsilon +O(\epsilon^3),\quad \cos\epsilon T=-1+O(\epsilon^2)
\]
}

Again, we substitute the expressions~\eref{eq:xec} and~\eref{eq:yec} for $x_e$ and $y_e$ into the expression~\eref{eq:xcc} for $x_c$. This gives
\purple{
\begin{eqnarray} \nonumber
x_c&= A_1 x_c \e^{\lambda_c^+ T}  +A_2 h \e^{\lambda_c^- T}-\e^{\lambda_e^R T}(B_1 x_c \e^{\lambda_c^+ T}  +B_2 h \e^{\lambda_c^- T})(A_3+A_4K) \\ 
 &\quad \e^{\lambda_e^R T}( C_1 x_c \e^{\lambda_c^+ T}  +C_2 h \e^{\lambda_c^- T} )(\epsilon^2A_3K-A_4).
\end{eqnarray}
Rearranging gives
\begin{eqnarray} \nonumber
\fl x_c\left(1-A_1 \e^{\lambda_c^+ T}  - \e^{(\lambda_c^++\lambda_e^R) T}\left(-B_1(A_3+A_4K)+
C_1(\epsilon^2 A_3K-A_4) \right)\right)\\
  =h\left(A_2  \e^{\lambda_c^- T}+ \e^{(\lambda_c^++\lambda_e^R) T}\left(-B_2(A_3+A_4K)+
C_2(\epsilon^2 A_3K-A_4)   \right)\right) \label{eq:xcc2}
\end{eqnarray}
}

The first term in the parentheses on the left hand side of equation~\eref{eq:xcc2} is clearly smaller than the others. \purple{Dropping this term, and the terms of $O(\epsilon^2)$ gives
\begin{equation}
 x_c=-h\e^{(\lambda_c^--\lambda_c^+ )T}\frac
{ A_2  - \e^{\lambda_e^R T}\left(B_2(A_3+A_4K)+A_4C_2  \right) }
{A_1  -\e^{\lambda_e^R T}\left(B_1(A_3+A_4K)+A_4C_1  \right)}  \label{eq:fpxe}
\end{equation}
}
Substituting this into the expressions~\eref{eq:xec} and~\eref{eq:yec} for $x_e$ and $y_e$ gives, after some \purple{cancellation},
\purple{
\begin{eqnarray}
x_e&=B_1 x_c \e^{\lambda_c^+ T}  +B_2 h \e^{\lambda_c^- T} \nonumber \\
& =h\e^{\lambda_c^- T} \frac{\Delta_{AB}+\e^{\lambda_e^RT}\Delta_{BC}A_4   }
{A_1  -\e^{\lambda_e^R T}\left(B_1(A_3+A_4K)+A_4C_1  \right)} \label{eq:fpxec}
\end{eqnarray}
}
where  $\Delta_{AB}=A_1B_2-A_2B_1$.
Similarly,
\purple{
\begin{eqnarray}
y_e& = C_1 x_c \e^{\lambda_c^+ T}  +C_2 h \e^{\lambda_c^- T} \nonumber \\
& =  h\e^{\lambda_c^- T}\frac{\Delta_{AC}-\e^{\lambda_e^RT}\Delta_{BC}(A_3+A_4K) }
{A_1  -\e^{\lambda_e^R T}\left(B_1(A_3+A_4K)+A_4C_1  \right)} \label{eq:fpyec}
\end{eqnarray}}
where $\Delta_{AC}=A_1C_2-A_2C_1$.
\purple{At this point, we note that the numerators and denominators of the expressions in~\eref{eq:fpxe},~\eref{eq:fpxec} and~\eref{eq:fpyec} all contain one term which is multiplied by $\e^{\lambda_e^R T}$, and one which is not. Since $\lambda_e^R >0$ and $T$ is large, we might think that the latter term is much smaller than the former and at lowest order, can be ignored. This is true for the numerators, since the term multiplying $\e^{\lambda_e^R T}$ consists of $O(1)$ global constants which generically are non-zero. However, the expression multiplying $\e^{\lambda_e^R T}$ in the denominators of these fractions contains the unknown constant $K$. It turns out that this expression is very small, and in fact, in the calculations below, we approximate it to lowest order by zero when finding $K$. Thus, both terms in the denominators must be kept.

Following this observation, we use equation~\eref{eq:Tc}
to compute an expression for $T$, where we will ignore the terms not multiplied by $\e^{\lambda_e^R T}$ in the numerators of both~\eref{eq:fpxec} and~\eref{eq:fpyec}. First, we use~\eref{eq:fpxec} and~\eref{eq:fpyec} to compute
\begin{eqnarray*}
x_e^2+y_e^2& = h^2 \e^{2(\lambda_c^- +\lambda_e^R)T}\frac{(\Delta_{BC})^2(A_4^2+(A_3+A_4K)^2)}
 {\left(A_1  -\e^{\lambda_e^R T}\left(B_1(A_3+A_4K)+A_4C_1  \right)\right)^2} \end{eqnarray*}
Substituting into~\eref{eq:Tc} and rearranging, we get
\begin{eqnarray}\label{eq:Tfpc}
\fl B_1(A_3+A_4K)+A_4C_1  =A_1 \e^{-\lambda_e^RT} -|\Delta_{BC}|\e^{(\lambda_e^R+\lambda_c^-) T}  \sqrt{ A_4^2+  (A_3+A_4K)^2 }.
\end{eqnarray}
Both terms on the right-hand side are non-zero, but exponentially small (as $T$ is large), if $\lambda_e^R+\lambda_c^-<0$. If $\lambda_e^R+\lambda_c^->0$, there are no solutions to this equation as there is nothing to balance the second term on the right-hand side, which would be exponentially large. Thus, we require $\lambda_e^R+\lambda_c^-<0$
 (that is, we can't be  in region~3 of figure~\ref{fig:pspace} and so must be in region~4), and so to lowest order, solutions will have
\[
B_1(A_3+A_4K)+A_4C_1=0
\]
or
\[
K=-\frac{A_3}{A_4}-\frac{C_1}{B_1},
\]
confirming that $K$ is order $1$. Thus our solution for $T$ is given by 
\begin{equation}
T=\frac{\pi}{\lambda_e^I}+\left(\frac{A_3}{A_4}+\frac{C_1}{B_1}\right)+O({\lambda_e^I})  \label{eq:TBD}
\end{equation}
}

As noted above, in order for $b(t)$ to remain positive for all $t\in(0,T)$, we must have $T<\pi/\lambda_e^I$. For large $T$ solutions then, we require \purple{ $\frac{A_3}{A_4}+\frac{C_1}{B_1}<0$}.
Since $A_3, A_4, B_1$ and $C_1$ are functions of the global dynamics (that is, the are not
solely dependent on the eigenvalues of the equilibria), we cannot say where in
parameter space this condition holds (apart from being within region~4,
close to the boundary between regions~4 and~5).

\purple{
We note that $K$ can change sign when $A_4$ passes through zero: this occurs when the heteroclinic connection is tangent to the positive $y_e$-axis. Since the coordinate changes we have used in the real and complex cases are different, this corresponds to the heteroclinic connection in the real case being tangent to the negative $x_e$-axis, which is exactly the point where the orbit-flip bifurcation curve terminates (see section~\ref{sec:orbflip}). We thus expect a transition between orbit-flip and Belyakov--Devany-type bifurcation to occur, at a location determined by the global constants. This is consistent with what is observed in figure~\ref{fig:pspace}.
}

Again, we check to see whether we expect to see kinked solutions. Recall that to get a kinked solution, we require that $|\yca(T)|\ll|\xca(T)|$. \purple{Using equation~\eref{eq:Tfpc} in the denominator in the $x_c$ equation~\eref{eq:fpxe} we can see that the denominator scales like $\e^{(2\lambda_e^R+\lambda_c^- )T}$}. The numerator will be order \purple{$\e^{\lambda_e^R T}$}. Thus
\begin{eqnarray*}
\xca(T) &= E_1 h \e^{(-\lambda_c^+-\lambda_e^R )T } \e^{\lambda_c^+ T} =  E_1 h \e^{-\lambda_e^R T } 
\end{eqnarray*}
for some \purple{$O(1)$ constant  $E_1$} and
\begin{eqnarray*}
\yca(T) &= h \e^{\lambda_c^- T}
\end{eqnarray*}
As noted above, $\lambda_c^-<-\lambda_e^R$, so $|\yca(T)|\ll|\xca(T)|$, and we expect to see a kinked solution, as observed (see panel (b) of figure~\ref{fig:tseries}).

In summary, we expect to find a Belyakov--Devaney\blue{-type} heteroclinic bifurcation with
$\lambda_e^-=\lambda_e^+$, that is, on the red curve in
figure~\ref{fig:pspace} with $\zeta<\sigma/2$, at the boundary between
regions~4 and~5.

\subsection{Summary}
\label{sec:Pmapsummary}

In summary, we conclude that heteroclinic bifurcations can only occur on the
boundary between regions~1 and~2 (see figure~\ref{fig:pspace}), or on the
boundary between regions~4 and~5, or within regions~1 or~5.  All our numerical results (detailed below) point
to all periodic orbits created in the Hopf bifurcation ending in heteroclinic bifurcations. If $\zeta>\sigma/2$,
the heteroclinic bifurcation is of resonance type ($-\lambda_c^-=\lambda_e^-$),
on the boundary between regions~2 and~1. If $\zeta<\sigma/2$, there are two
possibilities: the heteroclinic bifurcation can either be of Belyakov--Devaney
type (expanding eigenvalues changing from real to a complex-conjugate pair), on
the boundary between regions~4 and~5, or of orbit flip type (when a constant in
the global part of the map vanishes as the way in which the trajectories
between equilibria change their orientation), within region~5 or region~1. The transition 
between the resonance and Belyakov--Devaney\blue{-type} bifurcations occurs at 
$\zeta=\sigma/2$. The transition from Belyakov--Devaney\blue{-type} to orbit flip occurs 
when a global coefficient changes sign and so cannot be deduced only from 
considerations of eigenvalues.

\section{Further PDE simulations}
\label{sec:moresims}

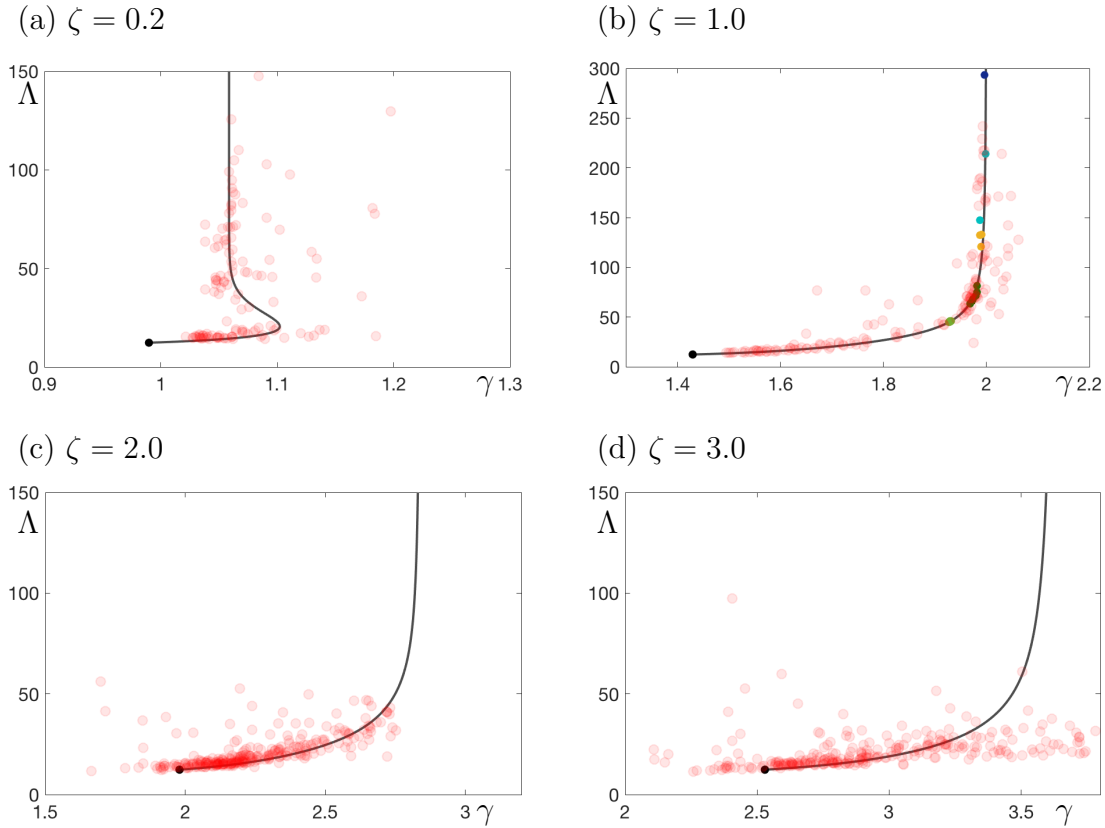
\begin{figure}
\setlength{\unitlength}{1mm}
\begin{center}
\begin{picture}(150,115)(0,0)
\put(0,57){\includegraphics[trim= 0cm 0cm 0cm 0cm,clip=true,width=68mm]{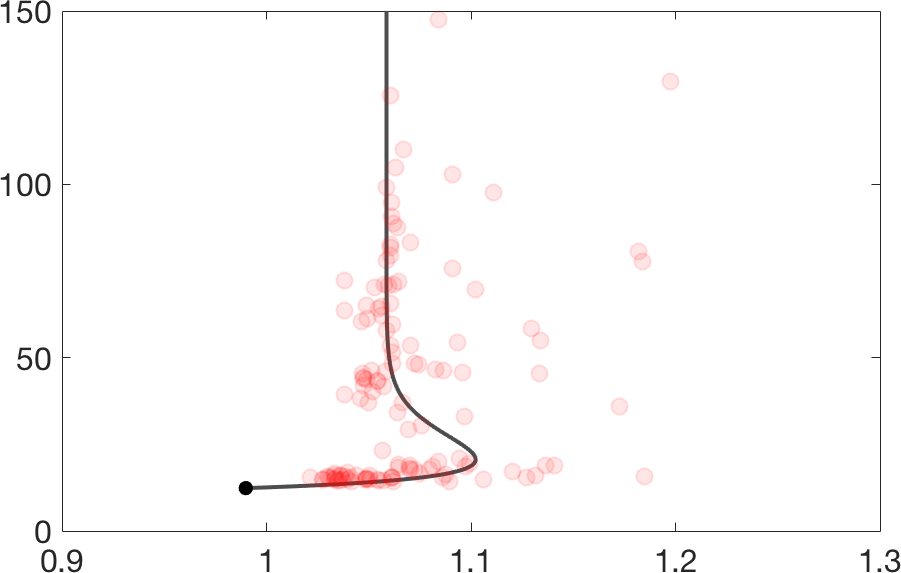}}
\put(77,57){\includegraphics[trim= 0cm 0cm 0cm 0cm,clip=true,width=68mm]{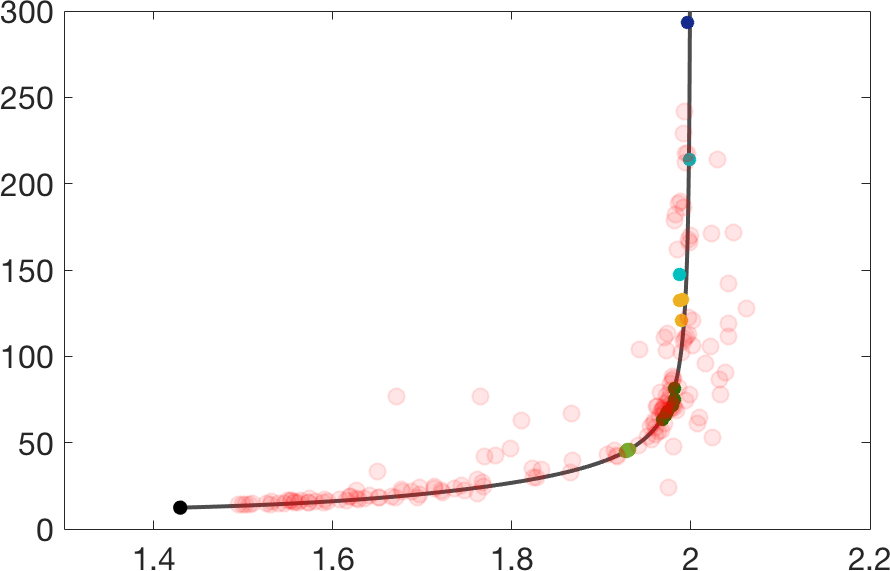}}

\put(0,0){\includegraphics[trim= 0cm 0cm 0cm 0cm,clip=true,width=68mm]{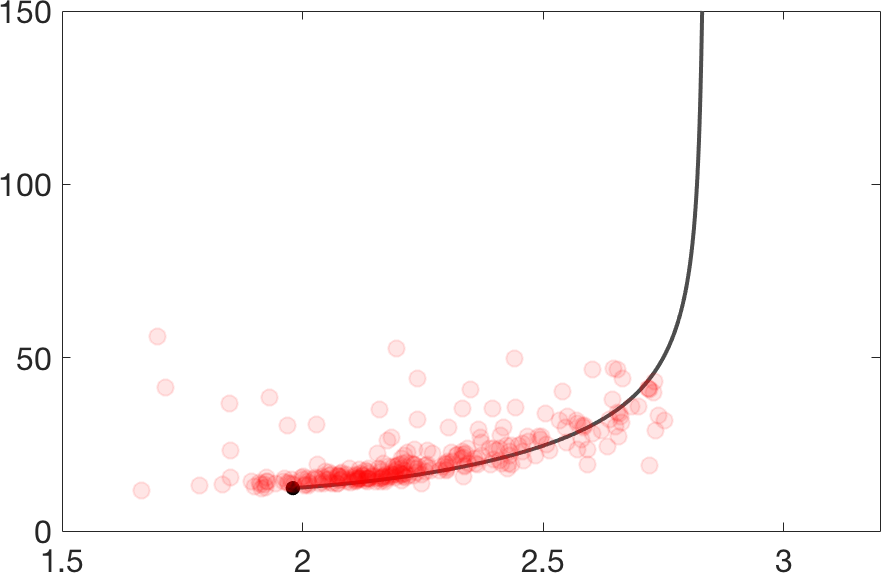}}
\put(77,0){\includegraphics[trim= 0cm 0cm 0cm 0cm,clip=true,width=68mm]{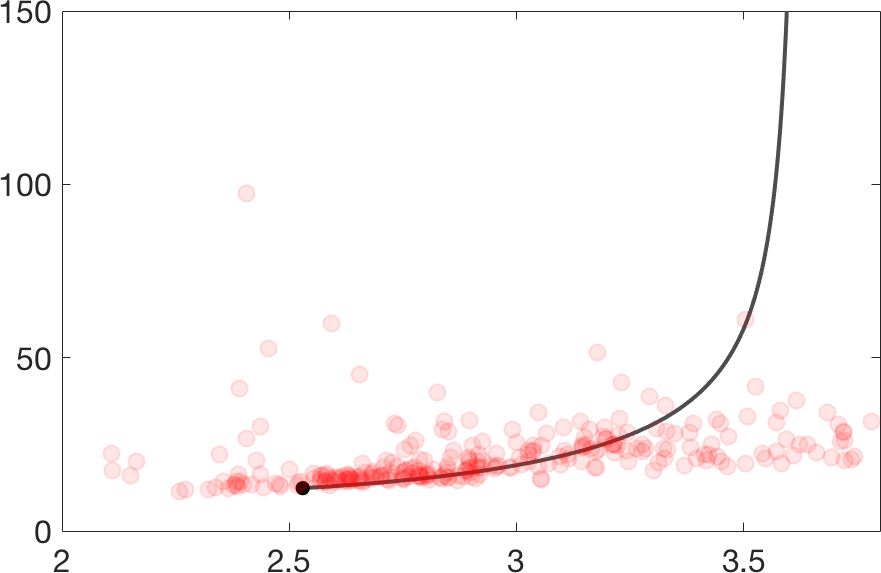}}

\put(1,105){(a) $\zeta=0.2$}
\put(62,57){{$\gamma$}}
\put(1,95){$\Lambda$}

\put(78,105){(b) $\zeta=1.0$}
\put(139,57){{$\gamma$}}
\put(78,95){$\Lambda$}

\put(1,48){(c) $\zeta=2.0$}
\put(62,0){{$\gamma$}}
\put(1,38){$\Lambda$}

\put(78,48){(d) $\zeta=3.0$}
\put(139,0){{$\gamma$}}
\put(78,38){$\Lambda$}

%
%
\end{picture}
\end{center}


\caption{In each panel, the solid curve shows the wavelength (period in~$\xi$)
$\Lambda$, as $\gamma$ is varied, of periodic orbits in the
ODEs~\eref{eq:odes}, computed using AUTO, with $\sigma=3.2$ and
values of $\zeta$ as indicated. Each curve of periodic
orbits arises in a Hopf bifurcation on the left (black dot), and
ends in a heteroclinic (long-period) bifurcation on the right. Effectively
these curves are nonlinear dispersion relations for travelling wave solutions in the
PDEs~\eref{eq:RPS_PDEs}. We additionally show estimated wavespeeds and wavelengths from PDE simulations as red points; the points are transparent, so darker areas indicate an accumulation of points. In (b), in addition, different coloured dots correspond to estimated wavespeeds and wavelengths in long-time behaviour for different initial conditions. Note that the scale on the $y$-axes is different in (b) so that these points can be seen. Further details can be found in the text.
\label{fig:per_orbs}}
\end{figure}

In this section, we continue the numerical PDE simulations first discussed in section~\ref{sec:sims}, and relate the results of these to the results of our calculations of the heteroclinic bifurcations. We begin by showing dispersion relations computed from the ODEs~\eref{eq:odes}, which relate the period of the orbit to the parameter $\gamma$ (in the terminology of the ODEs), or equivalently, the wavelength of the travelling wave $\Lambda$, to the wavespeed $\gamma$ (in the terminology of the PDEs). The dispersion relations are computed in AUTO, by following periodic orbits created the Hopf bifurcation given in~\eref{eq:hopf}, as done in~\cite{Postlethwaite2017}. The resulting curves are shown in figure~\ref{fig:per_orbs}. 

We ran numerical simulations of the PDEs~\eref{eq:RPS_PDEs} for values of $\zeta\in\{0.2, 1.0, 2.0, 3.0\}$, on a periodic domain of size $500$. For each simulation, we started with small, randomly generated initial conditions, and integrated for a time period of $10,000$ to remove any transient behaviour. We then sample the solution at timepoints $t=10,000+100k$, for $k=1,\dots,40$. At each sample point, we compute the wavelengths and wavespeeds of the current solution profile. The wavelengths are computed by calculating the distances (in $x$) between points which have both $\log(a)=-1$, and $\frac{da}{dx}>0$. Wavespeeds at each of these points are computed by locally calculating $\frac{da}{dx}$ and $\frac{da}{dt}$ and using $\gamma=\frac{da}{dt}/\frac{da}{dx}$. \purple{Waves are only included in the analysis if each of the three variables $\log a$, $\log b$ and $\log c$ goes both above and below $-1$, over the wavelength of the wave.} The results are plotted in figure~\ref{fig:per_orbs}(a).

The first thing to notice about these results is that there is a lot of scatter. This is for two main reasons. Firstly, although we can compute a `local' wavespeed (i.e. a wavespeed for some specific point $(x,t)$), we cannot reliably compute a `local' value of the wavelength. Secondly, in the simulations there are many different waves travelling both left and right (see figure~\ref{fig:1dsims}); whenever the waves collide there is a region of time and space for which the wavespeed and wavelength are not well-defined, and our computations do not take account of this. However, for each of the values of $\zeta$ shown, it can be seen that there is a concentration of points along the AUTO-computed dispersion relation curve.

Further observations of numerical experiments indicate that for $\zeta=1.0$, solutions will often converge to a single travelling wave after sufficient time has passed (sometimes in excess of $t=50,000$). For the other values of $\zeta$ used in these experiments, we do not observe this convergence. In figure~\ref{fig:per_orbs}(b) we show the results of further similar computations for $\zeta=1.0$, but now we run multiple simulations from randomly chosen initial conditions, and sample the wavelengths and wavespeeds of the solution at a single timepoint, after the solutions have become close to a single travelling wave. Different initial conditions converge to travelling waves with different numbers of waves fitting into the box, but all of these lie very close to the AUTO-computed dispersion relation curves.

The values of $\zeta$ used above, together with $\sigma=3.2$, correspond to observing each of the three types of heteroclinic bifurcation discussed in section~\ref{sec:Pmap}: the curve of heteroclinic bifurcations at $\zeta=0.2$ is of orbit flip type, at $\zeta=1.0$ is of Belyakov--Devaney type, and at $\zeta=2.0$ is of resonance type (see figure~\ref{fig:pspace}). We also include $\zeta=3.0$ to match with the data shown in figures~\ref{fig:2d} and~\ref{fig:1dsims}. The heteroclinic bifurcation calculations we have done tell us about existence criteria for periodic orbits in the ODEs, which correspond to existence criteria for travelling waves in the PDEs. For instance, we can say that travelling wave solutions exist to the left of the heteroclinic bifurcation curve, and to the right of the Hopf bifurcation curve in figure~\ref{fig:pspace}, and this in turn gives us a maximum and minimum wavespeed for observed travelling waves. In order to be able to give firm predictions about whether these travelling waves would be observed in simulations, we would also need to understand the stability of the travelling waves, which is beyond the scope of this paper (but the subject of future work).

\section{\blue{Bifurcation diagrams for varying $\sigma$}}
\label{sec:smallsig}

In this section we give some numerical results showing different bifurcation diagrams in
the $(\gamma,\zeta)$ plane as $\sigma$ is varied. Most of the bifurcation curves were
computed using AUTO~\cite{Doedel2001}. Maintaining computational accuracy for periodic
orbits close to heteroclinic cycles can be difficult for two reasons. Firstly, because
the periodic orbits are of very long period, it is necessary to have a large number of
mesh points defining the periodic orbit. Secondly, the heteroclinic connections lie in
invariant planes where some of the coordinates are zero. The nearby periodic orbits thus
will have coordinates which are very close to zero. In order to overcome the numerical
issues associated with small numbers we make the following change of coordinates:
 \begin{equation}
 \fl A=\log(a),\quad U=\frac{u}{a},\quad B=\log(b),\quad V=\frac{v}{b},\quad C=\log(c),\quad W=\frac{w}{c},
 \end{equation}
and use differential equations for $A,B,C,U,V$ and $W$ in our numerical computations
instead of the original equations. Since the periodic orbits which we are interested in
exist entirely in the positive orthant (they correspond to positive travelling waves of~\eref{eq:RPS_PDEs}), we have no issues with taking the logarithm of a
negative number. In AUTO, we compute a curve of periodic orbits which has a large, fixed,
period ($T=300$ in the following calculations), and say that this curve well approximates
the curve of heteroclinic bifurcations.

\begin{figure}
\setlength{\unitlength}{1mm}
\begin{center}
\begin{picture}(155,75)(0,0)
\put(2,37){\includegraphics[trim= 0.9cm 0cm 1.2cm 0.8cm,clip=true,width=35mm]{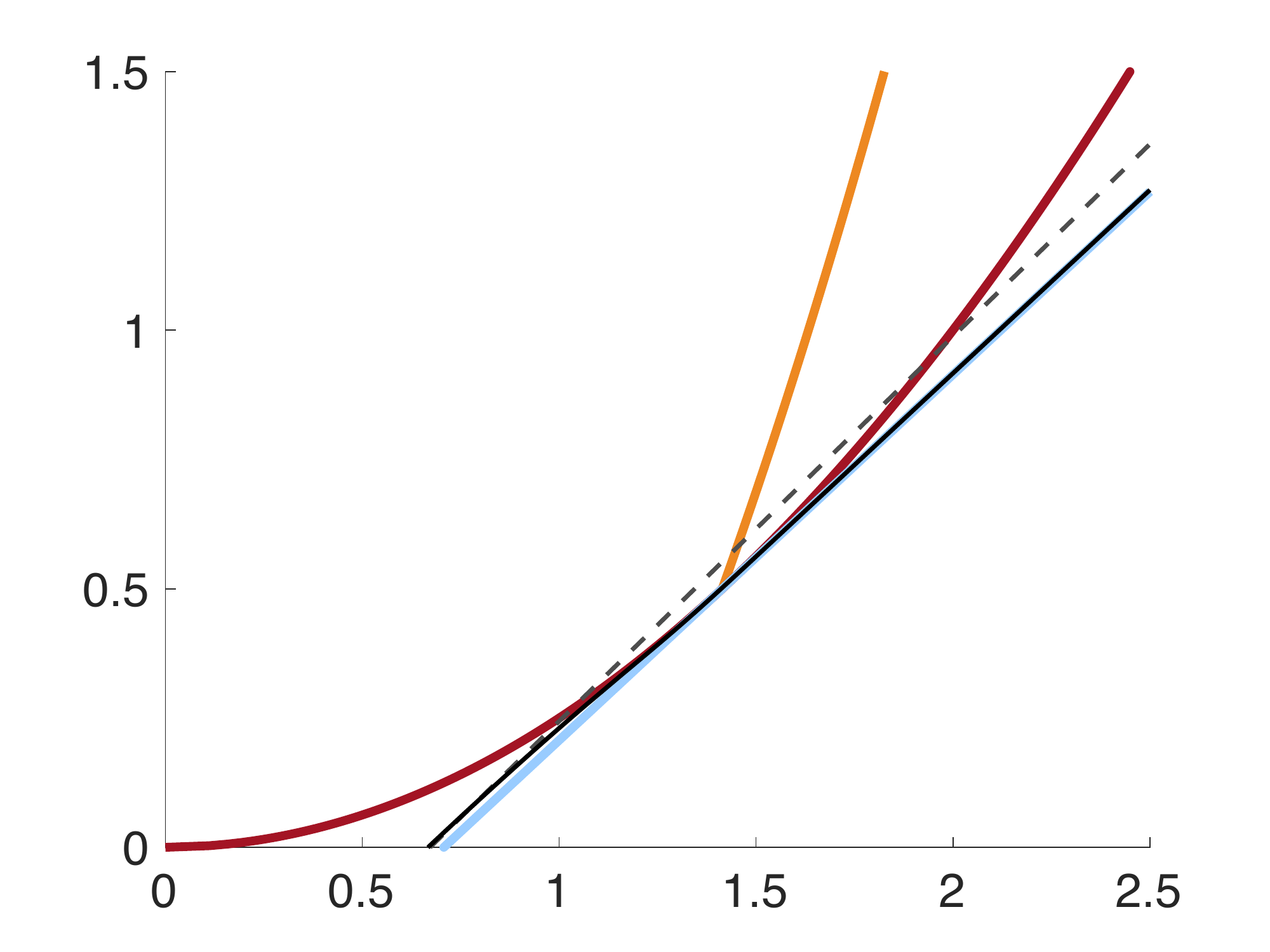}} 
\put(38,37){\includegraphics[trim= 0.9cm 0cm 1.2cm 0.8cm,clip=true,width=35mm]{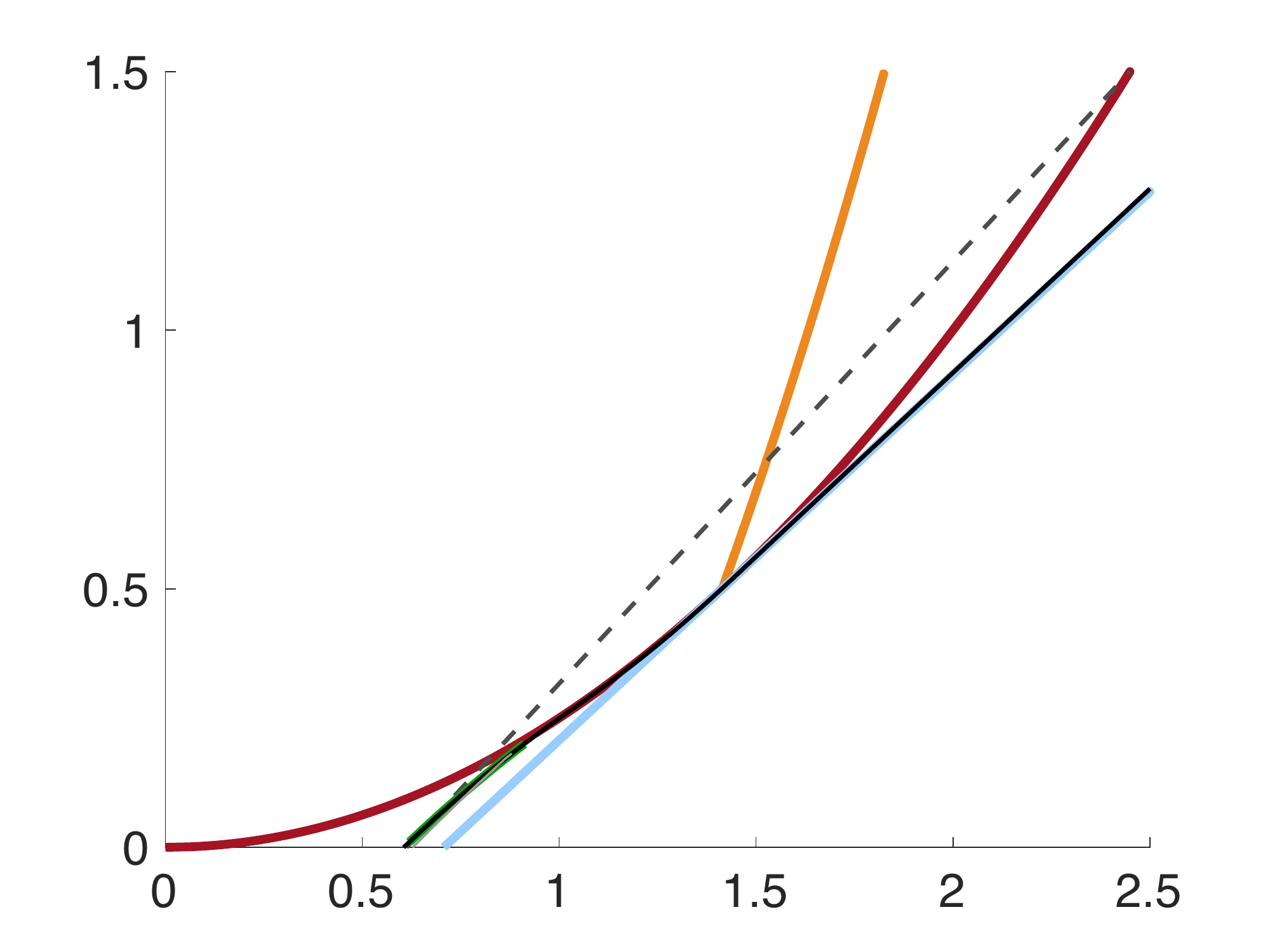}}        
\put(74,37){\includegraphics[trim= 0.9cm 0cm 1.2cm 0.8cm,clip=true,width=35mm]{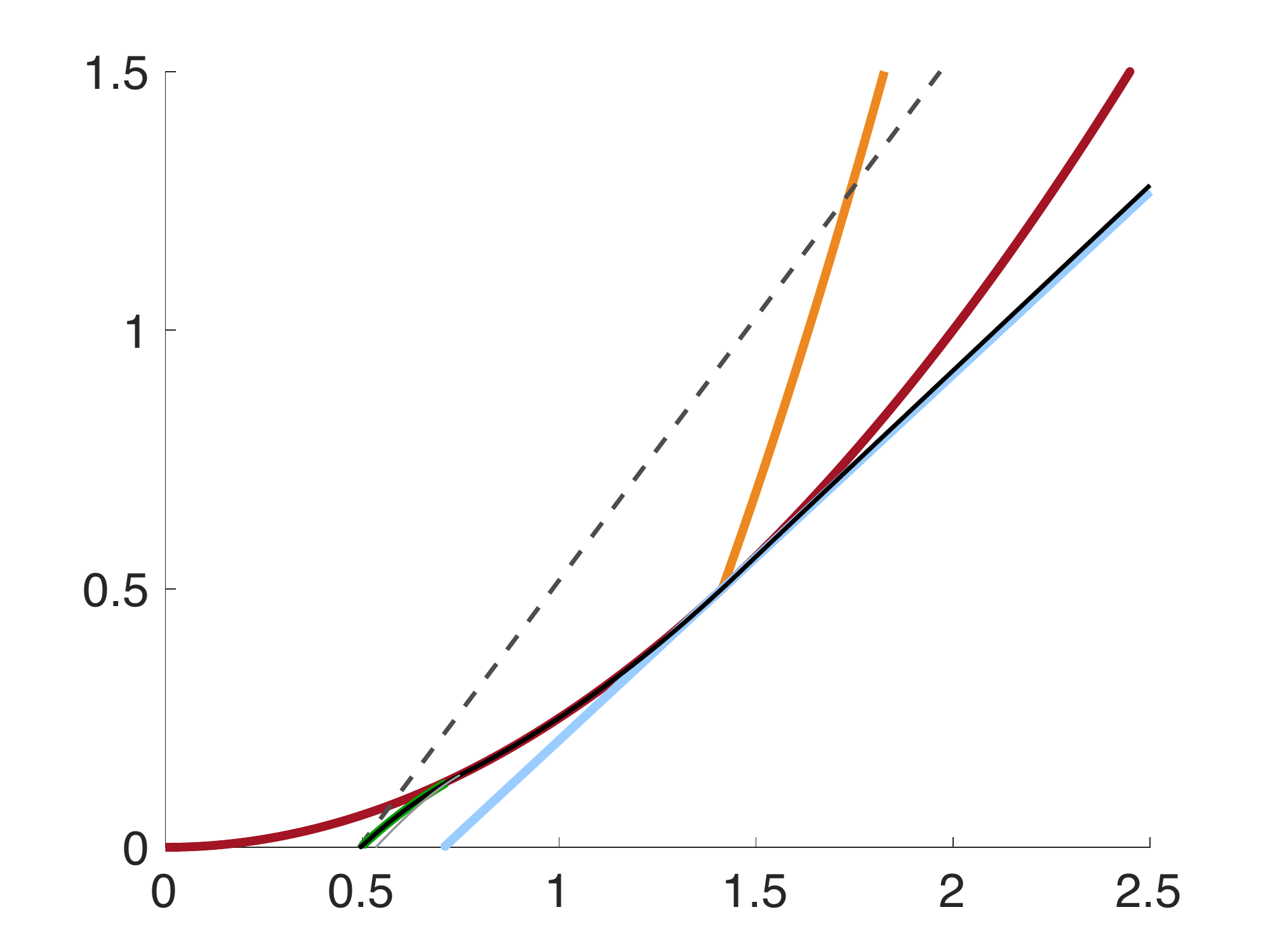}}        
\put(110,37){\includegraphics[trim= 0.9cm 0cm 1.2cm 0.8cm,clip=true,width=35mm]{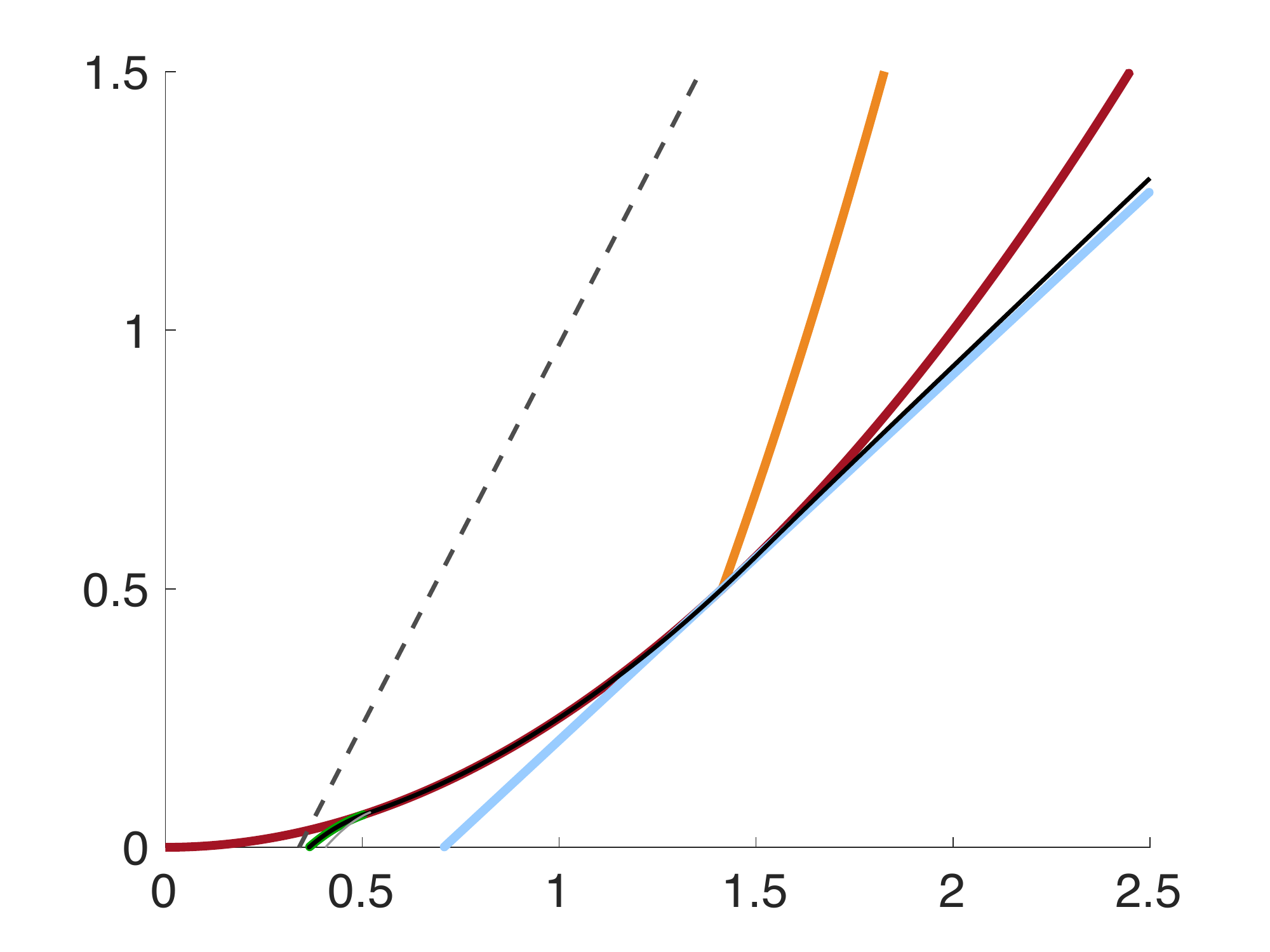}}     
\put(2,3){\includegraphics[trim= 0.9cm 0cm 1.2cm 0.8cm,clip=true,width=35mm]{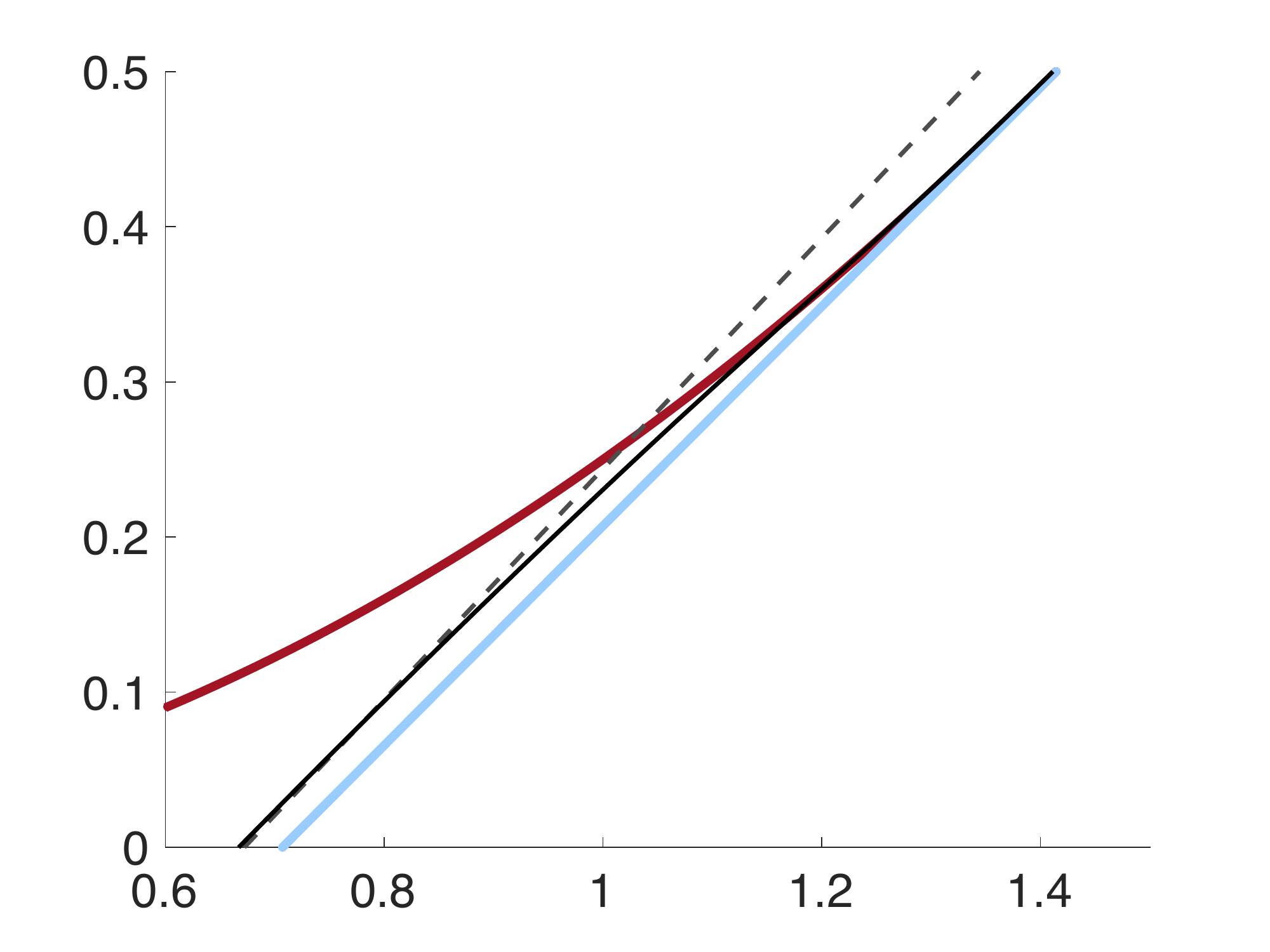}} 
\put(38,3){\includegraphics[trim= 0.9cm 0cm 1.2cm 0.8cm,clip=true,width=35mm]{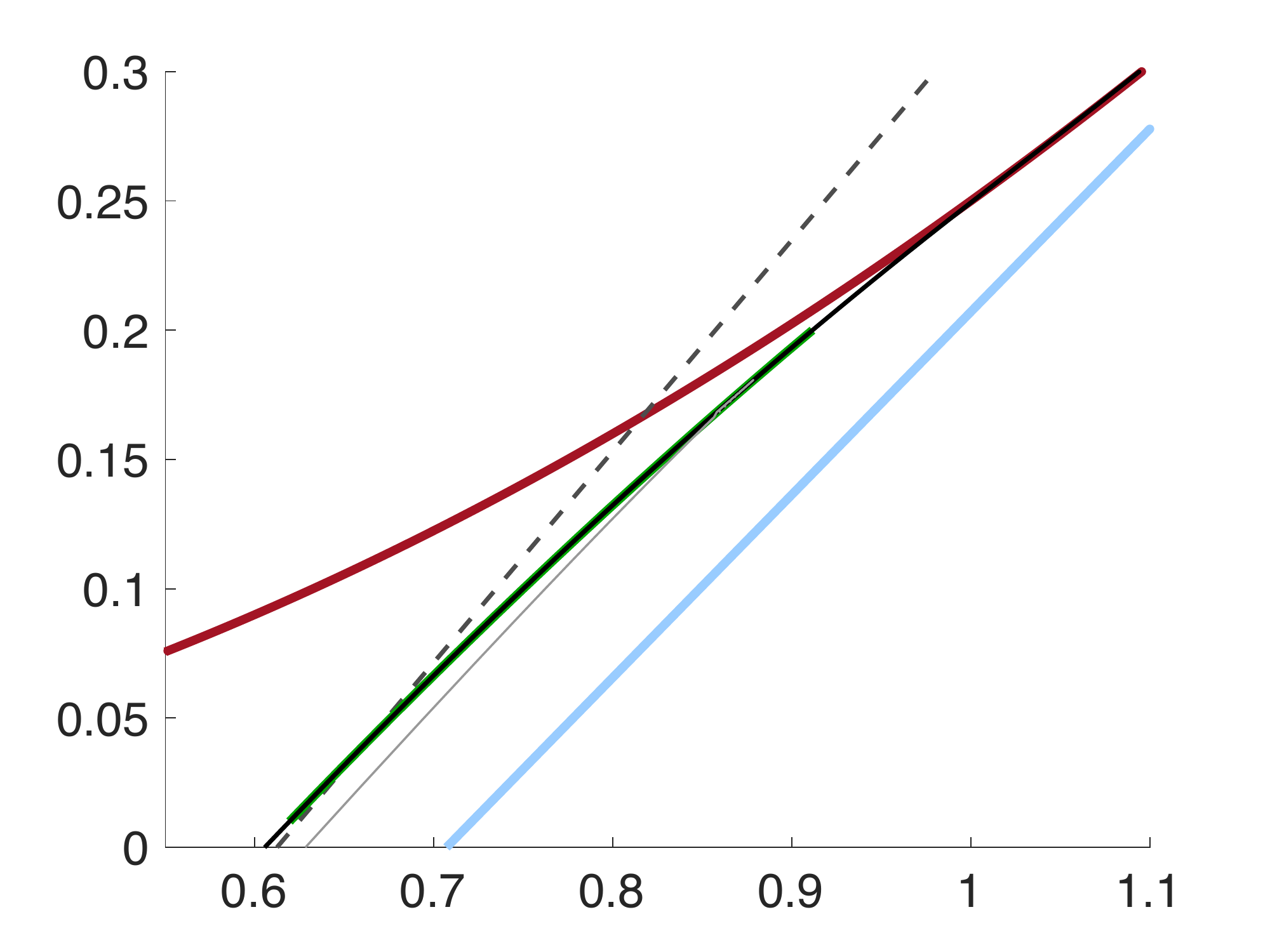}}        
\put(74,3){\includegraphics[trim= 0.9cm 0cm 1.2cm 0.8cm,clip=true,width=35mm]{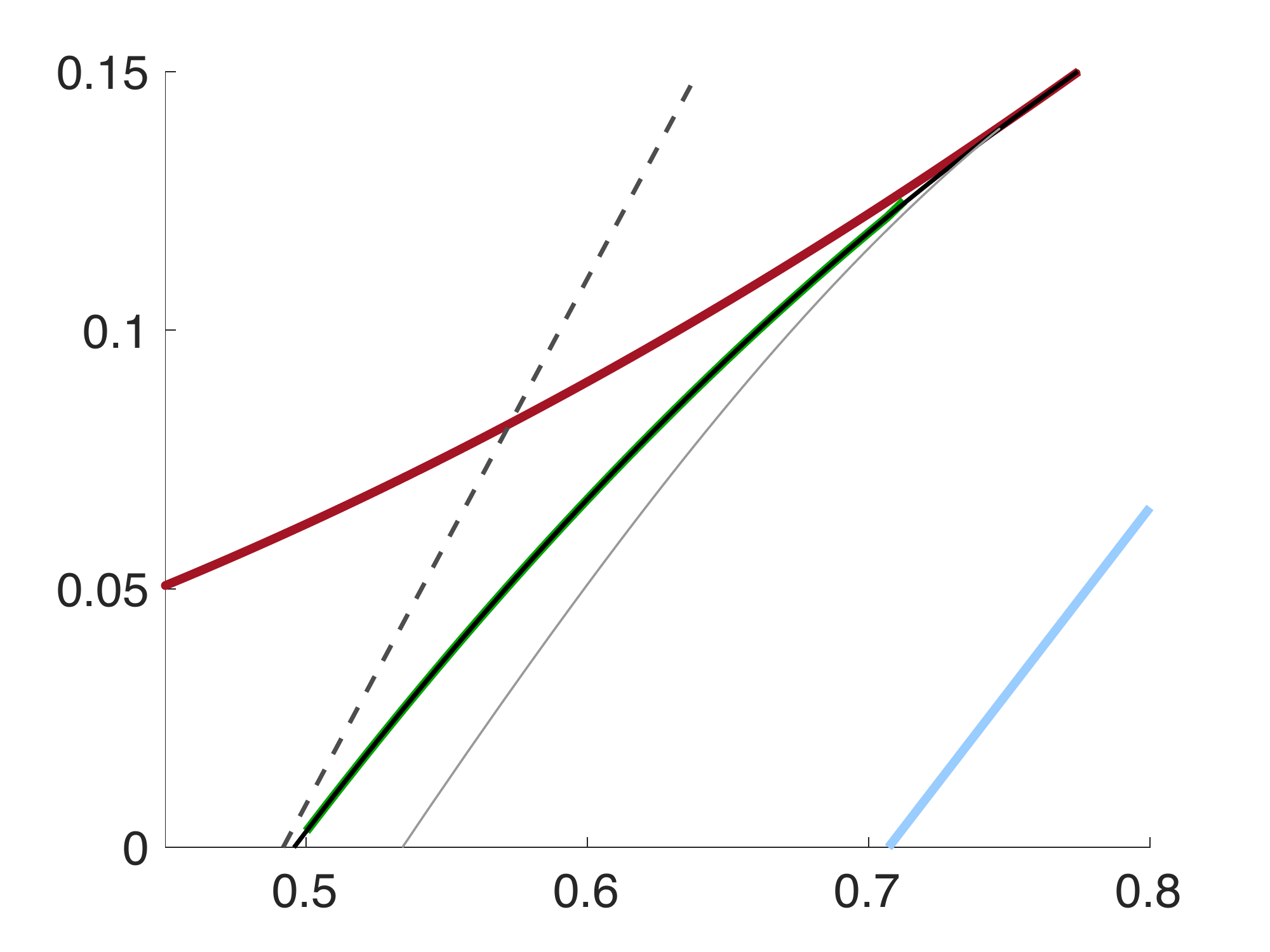}}        
\put(110,3){\includegraphics[trim= 0.9cm 0cm 1.2cm 0.8cm,clip=true,width=35mm]{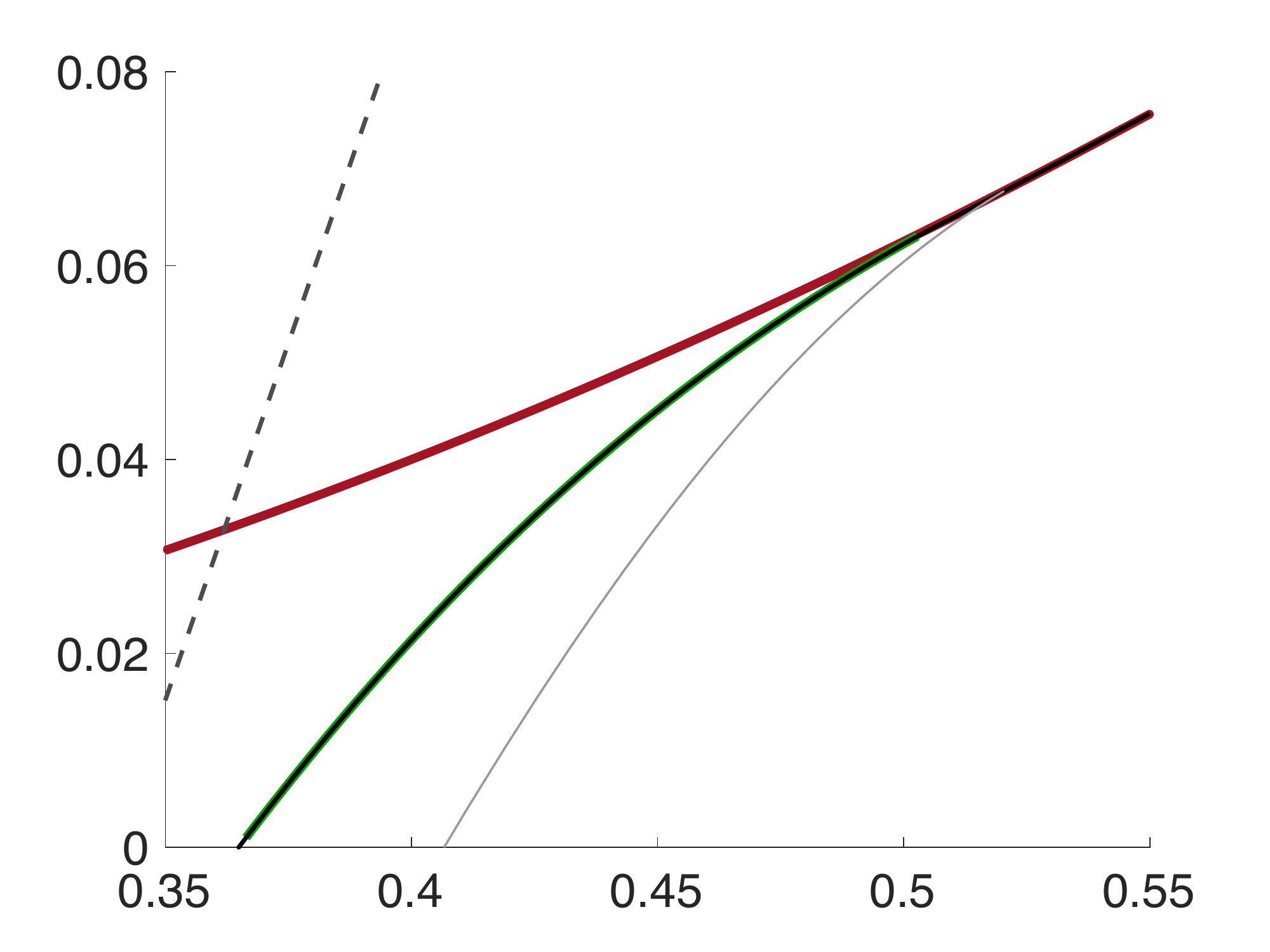}}  

\put(2,71){(a) $\sigma=0.32$}
\put(38,71){(b) $\sigma=1.0$}
\put(74,71){(c) $\sigma=3.2$}
\put(110,71){(d)  $\sigma=10.0$}

\put(35,1){$\hat{\gamma}$}
\put(71,1){$\hat{\gamma}$}
\put(107,1){$\hat{\gamma}$}
\put(143,1){$\hat{\gamma}$}

\put(0,59){$\hat{\zeta}$}
\put(0,30){$\hat{\zeta}$}


\end{picture}
\end{center}
\caption{Bifurcation diagrams for the
ODEs~\eref{eq:odes}, in $(\hat{\gamma},\hat{\zeta})$ parameter space, with $\sigma$ as indicated for each column. 
The blue line ($\zeta=\sqrt{\frac{\sigma}{2}}\gamma-\frac{\sigma}{2}$)
and red curve ($4\zeta=\gamma^2$) are tangent at $(\gamma, \zeta)=(\sqrt{2\sigma},\sigma/2)$, where they meet the yellow curve ($4(\sigma+\zeta)=3\gamma^2$). The purple curve ($\sigma+\zeta=2\gamma^2$). The green curve is the locus of a heteroclinic orbit flip.
The dark grey line is a curve of Hopf bifurcations. Periodic orbits bifurcate to the
right of this line and disappear in a curve of heteroclinic bifurcations
(black). A curve of saddle-node bifurcations of periodic orbits (light
grey) exists for smaller $\zeta$. The lower panels show zooms of the upper panels near the orbit-flip bifurcation.
\label{fig:varsig}} 
\end{figure}
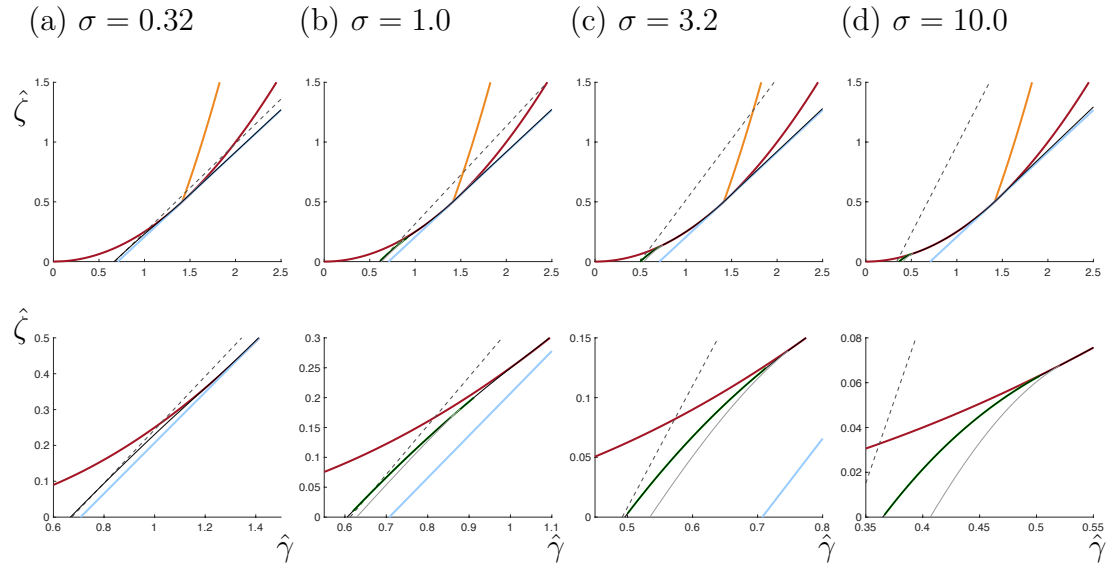

Figure~\ref{fig:varsig} shows bifurcation diagrams of system~\eref{eq:odes} for various
values of $\sigma$. For ease of comparison, we rescale $\zeta$ and $\gamma$ by writing
$\hat\zeta=\zeta/\sigma$ and $\hat\gamma=\gamma/\sqrt{\sigma}$. Note that with this
rescaling, all of the coloured lines (given by equations involving eigenvalues) in the
bifurcation diagrams do not depend on~$\sigma$. However, the location of the Hopf curve
changes. The grey curve shows the Hopf bifurcation curve, as given by~\eref{eq:hopf}, and
the black curve shows the heteroclinic bifurcation curve, as computed by AUTO. In (b),
(c) and (d), a light grey curve, also computed by AUTO, shows a curve of saddle-node
bifurcations of periodic orbits. The green curve is a curve of orbit-flip heteroclinic
orbits, computed by solving a boundary value problem, as explained in
section~\ref{sec:orbflip}.

For all four values of $\sigma$ shown, the heteroclinic curve coincides with the curve $\lambda_c^-=\lambda_e^-$ (the light blue curve) for values of $\zeta$ greater than $\sigma/2$. For values of $\zeta$ below $\sigma/2$, there is a range of $\zeta=[\zeta^*,\sigma/2)$ for which the heteroclinic curve coincides with the red curve, where the expanding eigenvalues are equal: the expanding eigenvalues are real to the right of this curve and complex to the left of this curve. Then for $\zeta<\zeta^*$, the heteroclinic curve coincides with the green curve: the curve of orbit flip heteroclinic orbits. We note that the transition point $\zeta^*$ is dependent on the global dynamics, and varies as $\sigma$ is varied. The curve of saddle-node of periodic orbits also appears to terminates at $\zeta=\zeta^*$.

In the lower panels of figure~\ref{fig:varsig}, we show zooms of each set of curves near to
$\hat\zeta=0$, showing the orbit flip and saddle-node curves more clearly. We note that
the numerical calculations become more difficult as $\sigma$ decreases, and for this
reason we do not show the orbit flip or saddle-node curves on the panel for
$\sigma=0.32$. In particular, we note that for the original Rock--Paper--Scissor
equations with no diffusion~\eref{eq:RPS_ODEs}, there is a degeneracy when $\sigma=0$,
namely that the Hopf and heteroclinic resonant bifurcations are degenerate (the Jacobian
matrix at the coexistence point has imaginary eigenvalues for all values of $\zeta$, and
the heteroclinic orbit is at resonance for all values of $\zeta$). \blue{Something similar happens in this six-dimensional system: it is simple to shown that the Hopf bifurcation curve and the resonance bifurcation curve collapse onto one another as $\sigma$ is reduced to zero, and it appears numerically that in fact the whole heteroclinic bifurcation curve approaches the Hopf curve in this limit}.

\section{Discussion}
\label{sec:disc}

We have summarised the results of our \Poincare map construction in
section~\ref{sec:Pmapsummary}: the PDEs~\eref{eq:RPS_PDEs} have been shifted
into a travelling frame of reference moving at speed~$\gamma$~\eref{eq:odes}.
In these sixth-order ODEs, travelling waves correspond to periodic solutions
that are created in a Hopf bifurcation and, with increasing wavespeed, are
destroyed in one of three types of heteroclinic bifurcation. Although the
construction of the \Poincare map follows reasonably standard lines, there are
technical issues: the unstable manifolds of the equilibria are
four-dimensional, and we restated some standard definitions of heteroclinic
cycles in order to accommodate (for example) positive contracting eigenvalues.
We find it advantageous to delay solving for the period~$T$ of the orbit
until the very end, since due to cancellation of some exponential terms in the calculations, it isn't obvious which terms can be safely neglected.

The periodic orbits we find can be kinked because one of the contracting
eigenvalues is positive and the growth rate changes in magnitude but not sign 
at the transition from
the contracting phase to the expanding phase. In addition, each of the three
heteroclinic bifurcations is non-standard or new in some way. The resonance
bifurcation, with $-\lambda_c^-=\lambda_e^-$, involves the leading
expanding (that is, smallest positive) eigenvalue; usually it would be the non-leading (largest positive) expanding eigenvalue,
that is, $-\lambda_c^-=\lambda_e^+$~\cite{Krupa1995}. The
Belyakov--Devaney\blue{-type} bifurcation, with the expanding eigenvalues changing from
real to a complex-conjugate pair, and the orbit flip heteroclinic bifurcation,
where the trajectories between equilibria change their orientation, are both 
new because they involve a robust (codimension zero) heteroclinic
cycle, rather than a higher codimension homoclinic 
orbit~\cite{Belyakov1984,Devaney1976,Homburg2000}.

It seems to be the case that stability conditions of heteroclinic cycles can be
much more complicated than perhaps was thought several decades ago when the
study of robust heteroclinic cycles was in its infancy. Much of this complexity
perhaps arises in cases where unstable manifolds have dimensions greater than
one. In the case in this paper, we have a positive contracting eigenvalue, and
other types of stability are often seen when cycles have positive transverse
eigenvalues (see, e.g.~\cite{Chossat1997,Postlethwaite2010,Podvigina2011,Lohse2015}). Recently, the
study of heteroclinic networks is receiving increasing attention in the
literature: by definition, such networks must have at least one equilibrium
with an unstable manifold of dimension greater than one. The stability of
heteroclinic networks is almost certainly very subtle~\cite{Kirk2012,Castro2014}, and we
expect many interesting results in this area in the future.

Viewing the ODEs~\eref{eq:odes} as an Initial Value Problem, the heteroclinic
cycles we describe are hopelessly unstable. However, the fixed points of the
map correspond to travelling waves in the PDEs~\eref{eq:RPS_PDEs}, and these
may (viewed as a Boundary Value Problem on an appropriate periodic domain) be
stable. We plan in future to use the results in this paper to address the
stability of the travelling waves within the PDEs\blue{: intriguing preliminary results have shown that long-wavelength travelling wave solutions are stable when they bifurcate from the Belyakov--Devaney-type bifurcation curve, but unstable otherwise.} In further work we will
address the problem of existence and stability of spiral waves in the
two-dimensional problem. Speculating further, it seems plausible that these
ideas can be used to examine spiral waves in other systems, such as
reaction--diffusion systems, or other spatially extended population models
(e.g., Lokta--Volterra).

%

\section*{Acknowledgements}

We are grateful to Graham Donovan, \blue{Cris Hasan,} Mauro Mobilia and Hinke Osinga for helpful
conversations, to Beverley White for reminding us to ``think about the eigenvalues'', and to two anonymous referees for some very helpful reports. We acknowledge the London Mathematical Society for financial support through
a Research in Pairs (Scheme~4) grant, and the hospitality of and financial 
support from the Department
of Mathematics, University of Auckland.

\section*{References}

\bibliographystyle{unsrt}
\bibliography{allrefs}

%
%
%
%
%
%
%
%
%
%
%
%

\end{document}